%% file: arxivMain.tex
\documentclass[a4paper]{article}

\usepackage[%
  paperwidth=192mm,
  paperheight=262mm,
  vmargin={19mm,19mm},
  hmargin={13.7mm,13.7mm},
  headsep=12pt,
  footskip=12pt,
]{geometry}

\usepackage[authoryear,longnamesfirst]{natbib}
\usepackage{hyperref}
\usepackage{subcaption}
\usepackage{graphicx}
\usepackage{fancyhdr}
\usepackage{amsmath}
\usepackage{url}
\usepackage{amssymb}
\usepackage{lineno}
\usepackage{setspace}

\begin{document}
\onehalfspacing 




\title{diffSPH: Differentiable Smoothed Particle Hydrodynamics for Adjoint Optimization and Machine Learning}  



%







\author{Rene Winchenbach\\Technical University Munich \and Nils Thuerey\\Technical University Munich}










\maketitle
\begin{abstract}
We present \emph{diffSPH}, a novel open-source differentiable Smoothed Particle Hydrodynamics (SPH) framework developed entirely in PyTorch with GPU acceleration. 
\emph{diffSPH} is designed centrally around differentiation to facilitate optimization and machine learning (ML) applications in Computational Fluid Dynamics~(CFD), including training neural networks and the development of hybrid models.
Its differentiable SPH core, and schemes for compressible (with shock capturing and multi-phase flows), weakly compressible (with boundary handling and free-surface flows), and incompressible physics, enable a broad range of application areas.
We demonstrate the framework's unique capabilities through several applications, including addressing particle shifting via a novel, target-oriented approach by minimizing physical and regularization loss terms, a task often intractable in traditional solvers. 
Further examples include optimizing initial conditions and physical parameters to match target trajectories, shape optimization, implementing a solver-in-the-loop setup to emulate higher-order integration, and demonstrating gradient propagation through hundreds of full simulation steps. 
Prioritizing readability, usability, and extensibility, this work offers a foundational platform for the CFD community to develop and deploy novel neural networks and adjoint optimization applications.
\end{abstract}





\input{01-introduction}
\input{02-governing}
\input{03-framework}
\input{04-00-application}

\input{05-discussion}

\appendix
\input{A0-appendix}


\section*{Acknowledgments}
This work was supported by the German Research Foundation (DFG) under the Individual Research Grant TH 2034/1-2.

\bibliographystyle{unsrt}

\bibliography{cas-refs}



\end{document}

%% file: 01-introduction.tex
\section{Introduction}\label{sec:introduction}

Computational Fluid Dynamics~(CFD) is used across a broad range of scientific and engineering applications, especially when involving complex phenomena, and has been an active area of research for decades.
In this field, Smoothed Particle Hydrodynamics~(SPH) is an example of a powerful and versatile Lagrangian scheme, capable of simulating a broad range of problems.
Initially proposed in an astrophysical context~\cite{gingold1977smoothed} for the simulation of celestial bodies, SPH's capabilities to handle large deformations, free surfaces, and complex geometries lead to its adoption in multiple fields, e.g., astrophysics~\cite{springel2010smoothed}, engineering~\cite{ye2019smoothed}, and animation~\cite{koschier2022survey}. 
Despite these successes, many open challenges remain~\cite{Vacondio2021Grand}, among others, regarding convergence, boundary conditions, adaptivity, and coupling to other solvers.

Concurrently with the development of traditional numerical methods in scientific computing, there is a strong transformative drive by recent machine learning techniques~(ML), e.g., for image recognition~\cite{deng2009imagenet}, Large Language Models~\cite{radford2018improving}, and unsupervised learning~\cite{silver2017mastering}.
This influence goes beyond image processing and language and includes significant progress in data-driven methods~\cite{goodfellow2016deep}, as well as scientific applications~\cite{jumper2021highly}.
More specifically, within the scientific and engineering domains, differentiable solvers~\cite{holl2020phiflow,hu2019difftaichi,koehler2024apebench} have seen significant adoption to a variety of phenomena due to their capabilities in solving inverse and adjoint problems.
Automatic Differentiation~(AD), the algorithmic engine underpinning modern ML frameworks like PyTorch~\cite{pytorch} and JAX~\cite{jax2018github}, is not only useful for training complex neural networks, e.g., surrogate emulators~\cite{alkin2024upt,Winchenbach2024SFBC}, but also provides powerful mechanisms to compute exact gradients of numerical problems. 
Prominent examples using these gradients include  Physics Informed Neural Networks~(PINNs) that differentiate through physical loss terms~\cite{cai2021physicsinformed} and solver-in-the-loop approaches that combine numerical schemes with learned correctors~\cite{um2021solverintheloop}, which require gradients not just for the neural networks but also for the numerical solver components.
Crucially, the effective development and deployment of such hybrid schemes necessitate numerical codes designed for differentiability from the ground up to enable the seamless integration of physics with ML. 
Furthermore, differentiable numerical frameworks also empower the direct optimization of adjoint problems and shape optimization tasks, offering new avenues for design and control~\cite{courtais2019adjoint,thuerey2020deep}.

This confluence of mature simulation techniques like SPH and the rise of powerful differentiable programming paradigms presents a compelling opportunity, as there is a growing specific need for CFD frameworks that are natively differentiable. 
Such frameworks allow researchers to integrate gradient-based optimization techniques directly into the simulation loop, learn effective closure models from data by backpropagating through the solver, or even discover optimized particle configurations or interactions. 
By enabling these capabilities, a differentiable SPH framework serves as a crucial stepping stone towards flexible applications of machine learning within the CFD community, while also empowering novel solutions to longstanding challenges through the lens of optimization. 

Reflecting SPH's origins in astrophysics, numerous codes have been developed for applications in astrophysics, often emphasizing gravitational effects, complex multi-physics, and scalability.
Notable examples include Gadget-4~\cite{springel2021simulating} and Gizmo~\cite{hopkins2015new}, both aimed at parallel cosmological n-body simulations, as well as Swift~\cite{schaller2018swift} and SPH-Exa~\cite{cavelan2020smoothed}, which are designed for extreme scalability into Exa scale.
However, these codes are not designed for ease of implementation but to achieve specific targets that dictate the code design with highly customized parallelization.

While these astrophysical codes achieve remarkable performance for their specific cosmological or stellar phenomena, the requirements in engineering and broader CFD applications often differ.
Consequently, there exists a broad set of codes in the engineering community with the two most noteworthy examples being DualSPHysics~\cite{dominguez2022dualsphysics} and SPHinXsys~\cite{zhang2021sphinxsys}. 
These codes are often designed with a strong focus on validating and simulating real-world engineering problems, but generally only implement weakly compressible or incompressible physics.
There also exist many other codes, with two notable examples being PySPH~\cite{ramachandran2021a}, a python based code implementing a broad range of physics, including compressible dynamics, with manually implemented C++ optimization, as well as  SPlisHSPlasH~\cite{SPlisHSPlasH_Library}, 
which focuses on algorithms for computer graphics applications.

Despite the wide range of capabilities of the solvers discussed above, none natively support differentiability, so they were not designed for gradient-based optimizations with the solver.
This significantly limits their direct applicability in ML research beyond serving as data generators, as incorporating inverse problem solvers or end-to-end model training requires substantial reimplementation efforts, especially when requiring physical loss terms and long backpropagation chains.
A noteworthy exception is the Jax-SPH code~\cite{toshev2024jax} that is implemented in Jax, as well as the DiffFR code~\cite{10.1145/3618318}, which are designed with differentiability in mind. 
However, these codes only implement a limited scope of SPH features and are not general-purpose frameworks.

While general differentiable frameworks for physical problems exist, e.g., PhiFlow~\cite{holl2020phiflow} and Warp~\cite{warp2022}, these either have no, or only rudimentary, support for Lagrangian schemes, such as SPH.
These limitations make it difficult to apply them in scenarios requiring high-fidelity physical simulation as a baseline, or where the interaction between the learned components and solver needs to be rigorously evaluated, e.g., 
using them as data generators~\cite{toshev2023lagrangebench,koehler2024apebench}.

This situation underlines significant potential and the need for an open-source SPH framework that synergistically combines robust, comprehensively validated physics modules for diverse flow regimes (compressible, weakly compressible, and incompressible) with native, end-to-end differentiability in the context of machine learning. 
Such a platform should prioritize usability and extensibility, facilitating the development of more robust, verifiable physics-informed ML models and novel adjoint-based optimization techniques for CFD and ML researchers.

To address this need, we introduce \emph{diffSPH}, a novel Smoothed Particle Hydrodynamics framework developed entirely in PyTorch. 
\emph{diffSPH} is built around core principles of differentiability, usability, and extensibility. 
It supports a broad range of physical problems, including shock-capturing, multi-phase, weakly compressible, and free-surface flows, all validated against standard numerical benchmarks. 
The framework features efficient, differentiable SPH operators, a scalable neighbor search, and allows for straightforward extension using both pure PyTorch and a custom Domain-Specific Language~(DSL) for integrating performance-critical C++ and CUDA components. 
We demonstrate the capabilities of \emph{diffSPH} as a foundational platform for research at the intersection of machine learning and CFD by 
\begin{itemize}
    \item Solving inverse problems by optimizing initial conditions to match target shock-capturing trajectories (Sec.~\ref{sec:application:inverse})
    \item Performing shape optimization of a wave source to maximize interference at a target location (Sec.~\ref{sec:application:targeted})
    \item Training neural network correctors to match higher-order integration schemes for the wave equation (Sec.~\ref{sec:application:sitl})
    \item Developing a novel approach to address particle shifting by directly minimizing physical loss terms and regularizers using the capabilities of our differentiable SPH framework (Sec.~\ref{sec:application:shifting})
\end{itemize}

%% file: 02-governing.tex
\section{Foundations}\label{sec:governing}

This section provides the theoretical groundwork for the \emph{diffSPH} framework, detailing the implemented SPH methodologies. 
We begin by briefly revisiting the SPH fundamentals, including the integral representation and particle approximation. 
Subsequently, we present the continuous governing equations for the diverse fluid physics regimes supported—weakly compressible, incompressible, and compressible flows- and a simplified wave equation. 
For each problem, the specific, discretized SPH formulations employed in \emph{diffSPH} are then elaborated, highlighting key variants for differential operators, corrective schemes, and shock-capturing techniques. 
Finally, we describe various treatments for boundary conditions implemented to handle interactions with domain limits and solid obstacles. 

\subsection{SPH Fundamentals}
SPH is based on a vector calculus identity that is then discretized using particles~\cite{price2012smoothed,monaghan2005smoothed}. 
The basic discretization of SPH starts with an initial vector calculus identity given as
\begin{equation}
A(\mathbf{x})=\int_{\mathbb{R}^d}A(\mathbf{x}^\prime)\delta(\mathbf{x}-\mathbf{x}^\prime)d\mathbf{x}^{\prime},
\end{equation}
where $A$ is a continuous field in $d$ dimensions, $\mathbf{x}$ an arbitrary position and $\delta$ being the Dirac delta function.
The Dirac delta function is then approximated by a Gaussian kernel $W$, i.e., $A(\mathbf{x})=\int_{\mathbb{R}^d}A(\mathbf{x}^\prime)W(\mathbf{x}-\mathbf{x}^\prime)d\mathbf{x}^\prime$, which is then discretized using a summation over particles (the volume integral $\mathbf{x}^\prime$ becomes $m_j/\rho_j$ by multiplying with$\rho/\rho$), as
\begin{equation}
A(\mathbf{x}) \approx \sum_j \frac{m_j}{\rho_j}A_j W(\mathbf{x} - \mathbf{x}_j).
\end{equation}
As this summation is of complexity $\mathcal{O}(n^2)$, and 
Gaussians become vanishingly small for large distances, the Gaussian term is then replaced with a compact Kernel function with cut-off radius $h$.
Note that there is some difference in the notation using $h$, as some SPH research, e.g., the seminal work by Monaghan~\cite{monaghan2005smoothed}, uses $h$ as the smoothing scale and $H$ as the cut-off radius with a scaling value dependent on the kernel $H/h$~\cite{dehnen2012improving}. In contrast, other work denotes $h$ as the cut-off radius directly~\cite{ihmsen2013implicit}, where we chose to adopt the latter.
There are several further requirements for the compact kernel, but they are beyond our scope here, see~\cite{dehnen2012improving}.
An important distinction that needs to be made at this point is that with varying resolution in space each particle can be assigned a different support radius, i.e., for a particle pair $i,j$ there is the choice of using $h_i$ (Gather-SPH), $h_j$ (Scatter-SPH), $\frac{1}{2}(h_i+h_j)$ (Symmetric-SPH) or averaging the kernels $\frac{1}{2}(W(\mathbf{x}_i-\mathbf{x}_j,h_i)+W(\mathbf{x}_i-\mathbf{x}_j,h_j))$, which we refer to as \textit{super symmetric}.
For brevity, we denote the kernel for an interaction as $W_{ij} = W(\mathbf{x}_{ij}, h_i, h_j)$ with $\mathbf{x}_{ij} = \mathbf{x}_i - \mathbf{x}_j$, and $\mathcal{N}_i$ being the neighbors of particle $i$, giving
\begin{equation}
A_i \approx \sum_{j\in\mathcal{N}_i} \frac{m_j}{\rho_j}A_j W_ij.
\end{equation}    
Setting the support radius $h$ can be done in many ways, e.g., based on the particle resolution~\cite{DBLP:journals/cgf/WinchenbachK20}, based on the current density~\cite{monaghan2005smoothed}, or kernel moment~\cite{owen2010asph}, where an update of the support radius in every timestep necessitates the inclusion of \emph{grad-H} correction terms~\cite{price2012smoothed} that implement the derivatives of SPH operators with respect to the support radius.
These corrective terms are usually only derived for a gather formulation, even if the overall simulation is not using a gather formulation due to the difficulties in tracking the gradients of density with respect to the support radii in other schemes.
Note that this is tracked automatically by AD approaches, without requiring manual bookkeeping of all terms.

Computing gradients of quantities in SPH is generally not done using the naive gradient of the SPH operator, but instead using a variety of schemes dependent on the requirements of the operation, see~\cite{price2012smoothed} and Appendix~\ref{appendix:operators}.
Similarly, for Laplacian operators, generally required for viscous forces, several approximation schemes exist~\cite{price2012smoothed}, with the most common approach being a combination of an analytical first-order derivative combined with a finite difference approximation of the second, see~\cite{brookshaw1985method} and Appendix~\ref{appendix:operators}.
In addition to the correction for varying support radii, SPH can also be corrected to enforce consistent gradients, i.e., $\nabla 1 = \mathbf{0}$, using either explicit~\cite{sun2019consistent} or implicit~\cite{rastelli2022implicit} particle-shifting approaches. Gradients can be corrected in the vicinity of free surfaces~\cite{marrone2011delta} and to have exact reconstruction properties~\cite{frontiere2017crksph}, which are all supported by our framework.

\subsection{Differentiable Solvers}
\label{sec:foundations:differentiable}
A differentiable solver is a numerical method that allows for the differentiation of its output with respect to its input parameters.
This capability enables a wide range of applications, such as computing the derivative of a final state's energy spectrum with respect to a physical parameter~\cite{holl2020learning}, optimizing the shape of an obstacle by differentiating a velocity field~\cite{chen2021numerical}, or performing sensitivity analysis via the gradient of the final state with respect to initial conditions~\cite{cao2003adjoint}. 
While one could manually derive these gradients using symbolic differentiation, modern differentiable solvers rely on Automatic Differentiation (AD), which constructs a computational graph of the forward pass to exactly compute derivatives~\cite{holl2020phiflow}.
In the differentiable physics context, a Lagrangian PDE is represented by a function $\mathcal{P}(\mathbf{x},\mathbf{f},\sigma)$, which takes as input the current particle positions $\mathbf{x}$, per particle attributes $\mathbf{f}$ and a global state $\sigma$, summarized as a state $\mathbf{u}$, which is analogous to the theoretical framework of Pahlke and Sbalzarini~\cite{pahlke2023unifying}.
This function is then applied recurrently on a set of initial conditions $\mathbf{x}_0$, $\mathbf{f}_0$, and $\nu_0$, to result in a state at time $n$ as a sequence of operations~\cite{thuerey2021pbdl}
\begin{equation}
    \mathbf{u}_n = \mathcal{P}_n\underbrace{\circ\dots\circ\mathcal{P}_1}_{\times n}(\mathbf{u}_0),
\end{equation}
with $\circ$ denoting function decomposition.
The goal of a differentiable solver is then to compute the partial derivatives of any input element $d$, i.e., $\partial \mathcal{P}_{i,d}$, for any solver step $\mathcal{P}_i$ as 
\begin{equation}
    \begin{split} \begin{aligned}
    \frac{ \partial \mathcal P_i }{ \partial \mathbf{u} } = 
    \begin{bmatrix} 
    \partial \mathcal P_{i,1} / \partial u_{1} 
    & \  \cdots \ &
    \partial \mathcal P_{i,1} / \partial u_{d} 
    \\
    \vdots & \ & \ 
    \\
    \partial \mathcal P_{i,d} / \partial u_{1} 
    & \  \cdots \ &
    \partial \mathcal P_{i,d} / \partial u_{d} 
    \end{bmatrix} 
\end{aligned} \end{split},
\end{equation}
where $u$ includes all inputs, i.e., positions, per-particle attributes, and the global state.
However, this matrix is a Jacobian that would require excessive amounts of memory to compute and store, and instead, a deep learning framework, such as PyTorch, does not compute the explicit Jacobian.
Instead, they compute a matrix vector product $\big( \frac{\partial \mathcal P_i }{ \partial \mathbf{u} } \big)^T \mathbf{a}$ of the Jacobian with a vector $\mathbf{a}$, where $\mathbf{a}$ is referred to as the incoming gradients.
In addition to the individual steps, the overall solver also needs to resolve the chain rule implied by the recurrent invocation of the solver.
In practice, the \emph{right hand} term of this process is a scalar value $L$, commonly computed as a loss term, which results automatically in the loss computation itself yielding a vector $\mathbf{a}$ that is then propagated backwards.

While AD provides a powerful and general mechanism for obtaining gradients, a naive application to complex, iterative simulations like SPH presents significant challenges in terms of memory consumption, computational overhead, and numerical stability. 
Naive reverse-mode AD typically retains all intermediate variables from the forward pass in memory to construct the computational graph for the backward pass. 
For SPH, this would entail storing numerous pair-wise interaction terms for every particle at every timestep, leading to a memory footprint of $\mathcal{O}(T n m_\text{avg})$, where $T$ is the number of timesteps, $n$ is the number of particles, and $m_\text{avg}$ is the average number of neighbors. 
This memory requirement quickly becomes prohibitive for simulations of practical scale and duration, as one of the major limitations of scaling SPH is the size of the neighborhood, which requires storing only a single integer list of $\mathcal{O}(n m_\text{avg})$; naive AD would necessitate storing several floating-point values for each of these interactions and retaining them for all $T$ timesteps, leading to excessive memory consumption.

\subsection{Governing Equations}
While we implement several simulation schemes, i.e., compressible, weakly-compressible, and incompressible, using our framework, they have a significant shared but distinct foundation.
At this point, it is important to note that the design of \emph{diffSPH} does not just facilitate fluid simulations using SPH, but that its underlying foundation allows for the implementation of any particle-based scheme~\cite{pahlke2023unifying}, see Sec.~\ref{appendix:timestepping}.
All fluid mechanical schemes relevant for \emph{diffSPH} are built around the same underlying Navier-Stokes equation,  written as
\begin{equation}
    \frac{d\mathbf{v}}{dt} = \frac{1}{\rho}\nabla p + \nu \nabla^2 \mathbf{v} + \frac{1}{\rho}\mathbf{f},
\end{equation}
with $\mathbf{v},p,\nu$ denoting velocity, pressure, and kinematic viscosity, respectively, while $\mathbf{f}$ collects external forcing terms such as gravity. 
To discretize this continuous form, many different approaches have been proposed in the past, e.g., using a symmetric~\cite{ihmsen2013implicit}, difference~\cite{marrone2011delta}, or case-dependent~\cite{antuono2021delta} gradient formulations for the pressure term and several distinct~\cite{monaghan2005smoothed, price2012smoothed, marrone2011delta} formulations for the viscosity term. 
We utilize the symmetric pressure term for compressible schemes~\cite{frontiere2017crksph}, the formulation by Antuono et al.~\cite{antuono2012numerical} for weakly-compressible, and the summation formulation for incompressible~\cite{bender2015divergence} schemes.
To discretize the viscosity term, we generally use the formulation by Brookshaw~\cite{brookshaw1985method} with viscosity terms being represented by a term modelling the pair-wise viscosity contributions as
\begin{equation}
    \left(\frac{d\mathbf{v}_i}{dt}\right)_\text{diss} = -\sum_j \frac{m_j}{\rho_j} \Pi_{ij} \mu_{ij} \nabla_iW_{ij};\;\mu_{ij}= \frac{h_i \mathbf{v}_{ij}\cdot\mathbf{x}_{ij}}{|\mathbf{x}_{ij}| +\epsilon h_i^2}
\end{equation}
with $\mathbf{v}_{ij} = \mathbf{v}_i - \mathbf{v}_j$, $\epsilon$ denoting a small constant to avoid singularities, and $\Pi_{ij}$ depending on the specific viscosity scheme.
Note that this term is commonly written using $\mathbf{v}_{ij}$, whereas the difference formulation of the gradient usually employs $\mathbf{v}_{ji}$, thus requiring the minus sign for the dissipation term.
In addition to the Navier-Stokes equation as given above, for incompressible SPH the flow field needs to be either divergence-free~\cite{shao2003incompressible}, $\nabla\cdot\mathbf{v}=0$, equal to rest-density everywhere~\cite{ihmsen2013implicit}, $\rho_i=\rho_0$, or both~\cite{bender2015divergence}. 
We implement a relaxed Jacobi solver~\cite{ihmsen2013implicit}, treating divergence-freedom as part of the PDE evolution and the rest-density constraint as a form of particle shifting~\cite{cornelis2019optimized}. 
To evaluate the density of a particle, mainly in compressible schemes~\cite{michael2014compatibly,hopkins2015new}, a summation is utilized as
\begin{equation}
\rho_i = \sum_j m_j W_{ij},
\end{equation}
whereas weakly-compressible schemes~\cite{marrone2011delta} tend to prefer an integration of density over time using momentum
\begin{equation}
\frac{d\rho_i}{dt} = \frac{\rho_i}{\Omega_i}\sum_j \frac{m_j}{\rho_j}\left(\mathbf{v}_j - \mathbf{v}_i\right)\cdot \nabla_i W_{ij},
\end{equation}
with $\Omega$, see Eqn.~\ref{eqn:omega}, being the corrective grad-H terms and $\nabla_i$ denoting the gradient with respect to the position of particle $i$.
Modelling the evolution of energy over time is done based on the first law of thermodynamics, i.e.~\cite{price2012smoothed},
\begin{equation}
\frac{du}{dt} = T\frac{ds}{dt} + \frac{p}{\rho^2} \frac{d\rho}{dt},
\end{equation}
with $u$ being the internal energy of a particle, $s$ as the specific entropy and temperature $T$.
Generally, entropy is assumed to be constant in the system, thus $T\frac{ds}{dt}=0$, see Price~\cite{price2012smoothed} for more detail, and the density derivative being implemented as above. 
Several additional schemes exist, e.g., a pressure entropy formulation~\cite{hopkins2015new,springel2002cosmological}, but these are just a change of reference frame and part of the specific implementation of a scheme~\cite{frontiere2017crksph}.
To exactly conserve energy over time, compSPH~\cite{michael2014compatibly} and CRKSPH~\cite{frontiere2017crksph} track the specific work exchange between particle pairs $\Delta u_{ij}$ and compute a term $f_{ij}$ to distribute the work at the end of a simulation step, instead of explicitly evolving energy directly.
This has some practical considerations on the implementation of time-stepping schemes~\cite{frontiere2017crksph} that will be discussed in Sec.~\ref{sec:framework:timestepping}.

In addition to physical viscosity $\nu\nabla^2\mathbf{v}$, most SPH schemes introduce dissipation terms to stabilize the numerical behavior of SPH, generally referred to as artificial viscosity.
For more details on these terms for energy dissipation, see the works by Cleary and Monaghan~\cite{cleary1999conduction} and Frontiere et al.~\cite{frontiere2017crksph}, for more details regarding artificial viscosity see Monaghan~\cite{monaghan2002sph} and Maronne et al.~\cite{marrone2011delta}, and for details on the density dissipation, see Maronne et al.~\cite{marrone2011delta} and Antuono et al.~\cite{antuono2021delta}.
Especially within compressible simulations, artificial viscosity strength evolves both in time and space to apply dissipation terms only in the vicinity of shocks to improve the overall convergence.
A broad range of these viscosity switches have been proposed~\cite{balsara1995neumann, cullen2010inviscid,hopkins2013general,frontiere2017crksph}, with their specific tradeoffs beyond our scope here.
However, it bears pointing out that these terms often rely on shock detection approaches based on heuristics with several hyperparameters and that learning such a shock detection scheme could be realized with~\emph{diffSPH}.
These switches can also be utilized in different contexts, e.g., stabilizing spatially adaptive SPH schemes~\cite{DBLP:journals/tog/WinchenbachA020}.

To evaluate the pressure and state of particles, we utilize Equations of State~(EoS) formulations. 
For compressible fluids, we utilize the standard compressible set of equations as~\cite{price2012smoothed}
\begin{equation}
p = (\gamma - 1) u \rho;\quad c_s = \sqrt{u\gamma(\gamma-1)},
\end{equation}
where $c_s$ denotes the speed of sound of a particle and $\gamma$ is the adiabatic index.
For weakly-compressible simulations, internal energy is generally assumed to be constant, and instead of the ideal gas EoS, either the isothermal~\cite{marrone2011delta} or the Tait-Murnaghan~\cite{english2022modified} equation is used as
\begin{equation}
p_{\text{isothermal}} = c_s^2 \left(\rho - \rho_0\right);\quad p_\text{Tait} = \frac{\rho_0 c_s^2}{n} \left[\left(\frac{\rho}{\rho_0}\right)^n - 1\right],
\end{equation}
where $n$ is the polytropic exponent, usually chosen to be 7~\cite{muller2003particle}, and $c_s$ being a constant (for each fluid-phase) value set to ensure a limit on the compressibility~\cite{marrone2011delta}, generally set to $10\mathbf{v}_\text{max}$, with $\mathbf{v}_\text{max}$ being the maximum instantaneous velocity across the entire simulation run.
For incompressible simulations we use the same value for the speed of sound, as this term is required for evaluations of artificial viscosity, whereas the pressure is a result of a pressure projection problem using an advection velocity $v_\text{adv} = v + \Delta t \left(\nu \nabla^2 \mathbf{v} + \frac{1}{\rho}\mathbf{f}\right)$ solving for a divergence free corrected velocity field.

For our examples using the wave equation, we use a simple wave equation $\frac{d^2u}{dt^2} = c^2 \Delta u$, transformed into a set of first-order differential equations as
\begin{equation}
\frac{du}{dt} = v;\quad \frac{dv}{dt} = c^2\Delta u,
\end{equation}
with $\Delta$ being the Laplacian operator and $c$ being a, potentially spatially and temporally varying, speed of sound value.
While SPH is Lagrangian in nature, we solve this problem using SPH as a tool to evolve a PDE on a set of fixed, and potentially unstructured, points, similar to FEM approaches.
Note that this is different from Eulerian-SPH approaches~\cite{o2022eulerian} that are beyond our scope here, but these can also be realized within \emph{diffSPH}.

Boundary conditions for solid objects are implemented using the mDBC approach~\cite{english2022modified} for weakly-compressible fluids and the Moving Least Squares approach of Band et al.~\cite{band2018mls} for incompressible simulations, with inlet and outlet buffer regions based on Negi et al.~\cite{negi2020improved}, which all use (with slight differences) a first order Taylor expansion of the flow field into the boundary or buffer region either directly or using ghost particles projected into the fluid.
These approaches generally work by assembling a linear system of size $d+1$ for an arbitrary quantity $q$ as 
\begin{equation}
\begin{bmatrix}
\langle1\rangle & \langle\mathbf{x}_{ij}\rangle^T\\
\langle \nabla 1\rangle & \langle\nabla\otimes\mathbf{x}_{ij}\rangle
\end{bmatrix} \cdot \begin{bmatrix}
f\\f^\prime
\end{bmatrix}= \begin{bmatrix}
\langle q\rangle\\
\langle \nabla q\rangle
\end{bmatrix},
\end{equation}
with $\langle\cdot\rangle$ indicating an SPH interpolation and $\otimes$ denoting the dyadic product, which is then solved using a pseudo-inverse of the system matrix.
This approach can also be utilized to implement Neumann boundary conditions, whereas Dirichlet boundary conditions are directly implemented by setting particle quantities.

%% file: 03-framework.tex
\section{Framework Architecture}\label{sec:framework}

Sec.~\ref{sec:foundations:differentiable} highlighted some of the common limitations of using AD to develop a differentiable solver, e.g., excessive memory consumption, that would make it impractical to use for any complex problem.
To overcome these limitations, \text{diffSPH} employs a strategy of creating optimized building blocks for fundamental SPH operators, each with a manually defined and optimized gradient flow. 
Instead of storing all intermediate values, this approach uses a form of gradient checkpointing where only essential particle data is retained. 
Intermediate quantities required for the backward pass are recomputed on-the-fly, trading a manageable increase in computation for a massive reduction in memory usage. 
Furthermore, this approach allows us to address potential numerical instabilities that naive AD can introduce. 
Consequently, these optimizations are necessary to enable the application of a differentiable solver to practical problems, such as geometry optimization, and implementing such a solver does not solely require solving computational and software engineering problems, but also requires careful realization of the numerical aspects, not just for the forward pass, but also for the backward pass.

To realize the potential of differentiable programming in Smoothed Particle Hydrodynamics, the \emph{diffSPH} framework has been designed focusing on accuracy, differentiability, usability, and performance. 
This section describes the key architectural and numerical implementation details that enable these capabilities. 
First, we outline the fundamental software design and PyTorch integration (Sec.~\ref{sec:framework:core}), detailing how \emph{diffSPH} structures particle data and leverages PyTorch's automatic differentiation for SPH computations on a high level.
We then provide details 
of implementing the core SPH interactions as differentiable operators, with particular attention to kernel functions and their gradient computations (Sec.~\ref{sec:framework:operators}), which form the core of the differentiable engine. 
To address the practical demands of SPH simulations, the mechanisms for efficient neighborhood searching and considerations for framework scalability (Sec.~\ref{sec:framework:neighborhoods}) are discussed. 
Finally, we describe the temporal integration methods and the overall simulation process (Section~\ref{sec:framework:timestepping}) employed in diffSPH. 
These implementation choices are foundational to the framework's ability to perform both as a forward SPH solver and for advanced differentiable applications explored here.

\subsection{Core Framework Architecture and PyTorch Integration}\label{sec:framework:core}
\begin{figure}
    \centering
    \includegraphics[width=\linewidth]{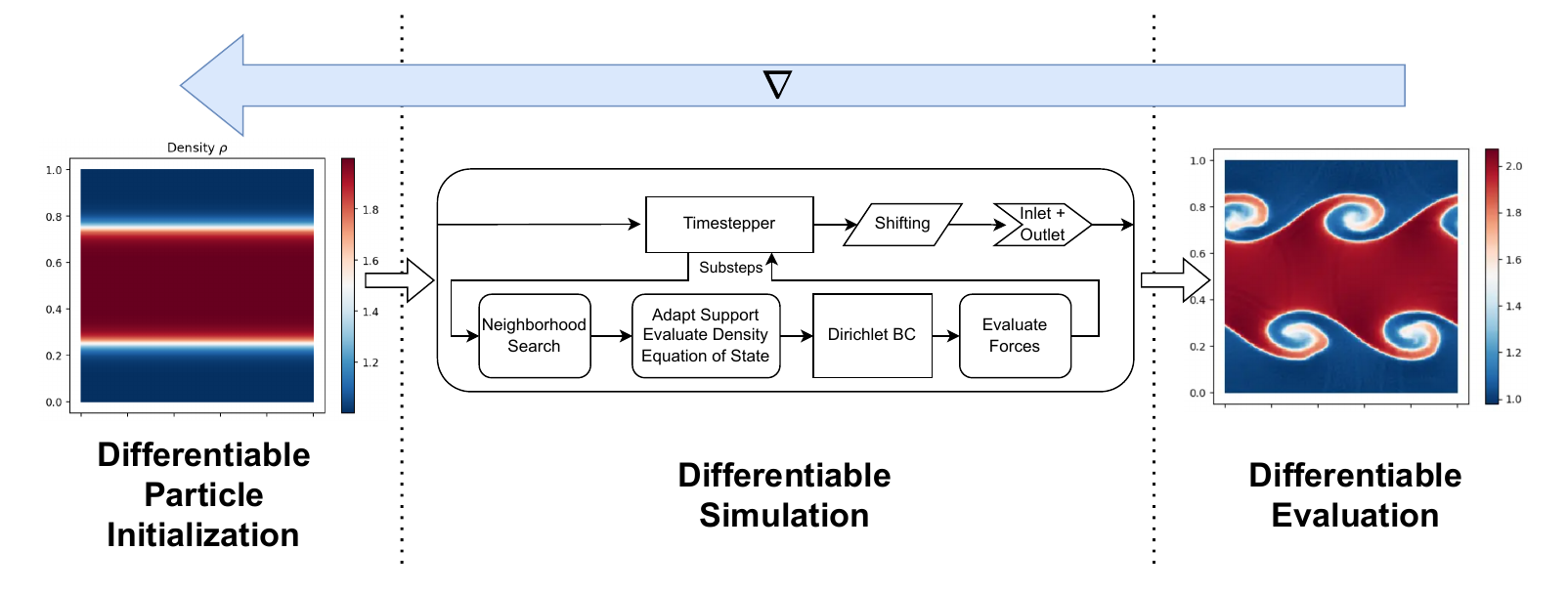}
    \caption{A schematic overview of the \emph{diffSPH} framework showcasing how all aspects of our simulation are differentiable, from the initialization using CSG, to the time stepping and boundary handling, to the evaluation of the final simulation results, allowing us to close the loop around any and all parts of the simulation for gradient computations.}
    \label{fig:flowchart}
\end{figure}

The \emph{diffSPH} framework is a modular system within PyTorch, designed to map the complexities of SPH simulations into a differentiable computational graph. 
Utilizing tensors in PyTorch for particle quantities inherently enables GPU acceleration and automatic differentiation; however, achieving seamless and efficient end-to-end differentiability for full SPH simulations requires several considerations to avoid prohibitive memory and compute costs. 
Furthermore, a primary objective is to provide a usable and extensible framework without requiring users to grasp every nuance of the simulation control flow.
Figure~\ref{fig:flowchart} shows a schematic overview of our framework.

To this end, our codebase is structured around individual modules, each expecting a standardized set of inputs: a \emph{ParticleState} object (being a Structure of Arrays container for all particle data in the simulation), simulation configuration variables, neighborhood information, and the current timestep size.
These modules, in turn, return tensors representing particle information in a Structure of Array fashion, i.e., one tensor per computed quantity.
This design facilitates the development of modules as being side-effect-free and functional implementations of individual SPH steps (e.g., the computation of pressure forces), which allows leveraging PyTorch's JIT script engine to reduce the Python overhead.
However, doing so requires strict typing of function parameters, vectorized operations on particle sets, tensor slicing and summation instead of explicit loops. This design also allows for easy extensibility~\cite{pahlke2023unifying}.

Should more fine-grained control or higher performance be necessary for individual components, or if certain algorithms are challenging to express vectorially, our framework supports the straightforward integration of custom C++ and CUDA functions.
This is facilitated by a custom DSL that simplifies the creation of Python bindings, see Appendix~\ref{appendix:DSL}.
To maintain end-to-end differentiability, these custom C++/CUDA components require explicitly implementing corresponding custom backwards passes (i.e., by wrapping them in torch.autograd.Function subclasses) to integrate seamlessly into the compute graph.
It is worth noting that using torch.compile is overly restrictive in many cases and not usable in practice.
Utilizing compilation within PyTorch requires constant tensor shapes to be efficient, which is not possible with varying neighborhoods over time.
Furthermore, the common workaround of masking and padding significantly increases memory consumption to a point of not being viable for large scales~\cite{DBLP:conf/vmv/Winchenbach019}.

Simulation domains in \emph{diffSPH} are initialized using a Constructive Solid Geometry (CSG) approach, leveraging Signed Distance Functions (SDFs), allowing the definition of complex geometries in a modular and hierarchical manner. 
A key advantage of this SDF-based CSG approach is the inherent parameterization of the initial particle distribution with respect to geometric parameters. 
For instance, the location of a region of high potential in a wave equation simulation becomes a differentiable parameter. 
This natively facilitates the seamless differentiation of simulation outcomes with respect to these geometric initial conditions, as demonstrated in Section~\ref{sec:application:targeted}. 
For one-dimensional simulations, our framework also supports the direct definition of a density field in space, from which initial particle positions are generated using an inverse CDF sampling approach, proving particularly effective for initializing one-dimensional shock problems~\cite{win2023spheric}.

Regarding user interaction and accessibility, \emph{diffSPH} is primarily designed for use within Jupyter notebooks. 
This environment offers an interactive and flexible platform for simulation setup, execution, and analysis, benefiting from Python's visualization and data processing library ecosystem. 
Note that utilizing matplotlib and similar libraries for data analysis and visualization can be computationally more expensive than individual simulation steps.
Moreover, this notebook-centric approach facilitates the easy distribution and execution of simulation scenarios on cloud platforms like Google Colab, enhancing reproducibility and enabling access to the framework's capabilities even without local high-performance compute resources.

\subsection{Neighborhood Search}\label{sec:framework:neighborhoods}
Efficient and accurate neighbor identification is paramount in SPH, as it underpins all particle interactions and significantly impacts computational performance. 
Therefore, the choice and implementation of the neighborhood search algorithm warrants special attention within the \emph{diffSPH} framework.
This becomes even more important due to the variety of simulation schemes implemented in our framework and their vastly different neighborhood requirements.

We base our neighbor search on a compact hashmap-based approach, a technique well-established for its efficiency, particularly in contexts like computer animation~\cite{ihmsen2013implicit,DBLP:conf/vmv/Winchenbach019}.
While we adopt the core principles of this method, several modifications and extensions are necessary to meet the specific demands of a versatile and differentiable SPH framework, diverging in key aspects from prior implementations~\cite{DBLP:conf/vmv/Winchenbach019}. 
Our key modifications to the underlying hashmap approach are a native integration of periodic boundary conditions, a modified search radius update loop, a combined neighbor search with masking instead of separate searches, and adding a Verlet list approach, while building the entire algorithm around a core in PyTorch and differentiability, see Appendix Sec.~\ref{appendix:neighborhoods} for more details.%

\textbf{Differentiability} interacts with neighbor lists in two important ways.
Firstly, the neighbor list is not inherently differentiable as particle adjacency is a binary choice.
However, this does not impede the backpropagation process, as gradient information still propagates between interacting particles during the backward pass. 
Moreover, even if the neighbor list itself were notionally differentiable, the compact support of SPH kernels, with both the kernel and its derivatives being zero beyond the cutoff radius, ensures that any gradient contributions from particles outside this radius would naturally be zero.
Secondly, information flow in a differentiable simulation is bi-directional: in the forward pass, information typically flows from a particle's neighbors to that particle, whereas during the backward pass, gradient information flows from the particle to its neighbors. 
Consequently, the representation of the neighbor list must support this bi-directional flow. 
We achieve this by employing a coordinate list format that explicitly stores interacting particle pairs $(i,j)$, a common practice in frameworks such as Graph Neural Networks~\cite{fey2018splinecnn,Winchenbach2024SFBC} 
This ensures that for every forward interaction, a corresponding path for gradient backpropagation is readily available, which would be much more challenging with compressed formats, such as Compressed Sparse Row formats~\cite{DBLP:journals/cgf/WinchenbachK20}.

\textbf{Periodic Boundary Conditions} are frequently employed in SPH simulations, particularly for validation cases in compressible flows and turbulence studies. 
A canonical approach involves creating ghost particles near domain boundaries, synchronized with their real counterparts. 
However, such a method can introduce complexities for differentiable frameworks, as the dynamic introduction and removal of ghost particles leads to discontinuities in the computational graph, making this approach inappropriate for backpropagation.
We adopt an alternative strategy to circumvent these issues and maintain a consistent particle set. 
Particles maintain their absolute positions throughout the simulation, whereas for neighborhood search and interaction calculations, these absolute positions are mapped into the primary simulation domain using modulo arithmetic. 
Inter-particle distance vectors are also computed using this modulo arithmetic, effectively applying the \textit{minimum image} convention~\cite{deiters2013efficient}. 
It is important to note that C++ and Python conventions for modulo operations can differ, necessitating careful implementation, especially when using mixed-language components via our DSL, to ensure consistent behavior. 
This approach of using unwrapped coordinates with modulo-based distances is physically sound due to the Galilean Invariance of the SPH equations; SPH interactions depend fundamentally on relative positions (and their magnitudes), rather than their absolute coordinates.

\textbf{Ensuring symmetric interactions} becomes a vital step in neighbor search algorithms that deal with varying resolutions ~\cite{DBLP:conf/vmv/Winchenbach019}, where each particle is assigned a, potentially unique, support radius $h_i$.
Consequently, a particle cannot utilize its own support radius to perform a neighborhood search as this would potentially exclude interactions with particles with much larger support radii, i.e., the correct search radius for an interaction between particle $i$ and $j$ should effectively consider $\operatorname{max}\left\{h_i, h_j\right\}$.
To ensure this, we initialize a search radius $s_i$ for each particle with its respective suport radius $h_i$ and then perform a loop over all potentially neighboring particles using the underlying cell datastructure and atomically update the search radius of a neighboring particle $k$ to be $s_k = \max\{s_k, h_i\}$ if $k$ is a neighbor of $i$ but the search radius of $k$ is too small to consider particle $i$ during the neighbor search.
We can then perform the neighbor search using $s_i$, instead of $h_i$, to avoid any potential asymmetries.
Contrary to some prior work~\cite{DBLP:conf/sca/WinchenbachHK16,DBLP:journals/cgf/WinchenbachK20}, we do not limit the maximum support radius of particles to conserve memory, as this introduces unintended numerical artifacts not desirable in CFD.

\textbf{Neighborhood construction} is then performed by searching for all neighbors using a super symmetric SPH scheme, as any other SPH scheme is a strict subset of this set of neighbors.
All particle data is stored in individual contiguous PyTorch tensors, with a dedicated tensor indicating particle type (e.g., fluid, boundary). 
This unified data structure simplifies the search process and allows the neighborhood search to simultaneously identify all potential interactions (e.g., fluid-fluid, fluid-boundary, boundary-boundary).
We then sort the neighborhoods based on the particle kinds, which allows us to efficiently access the correct subset of neighborhoods using tensor slicing operations.

\textbf{Verlet lists} are also readily supported in \emph{diffSPH}, i.e., computing neighborhoods using larger search radii than necessary and reusing them across timesteps.
These filtering operations are straightforward to implement using tensor masking in PyTorch and significantly reduce the overall computational costs.
Note that care needs to be taken with the recomputation as changes in position and support radii simultaneously affect the correctness, see Appendix~\ref{appendix:neighborhoods}.
\subsection{Implementation of Differentiable SPH Operators}\label{sec:framework:operators}

The core of any SPH simulation lies in the discretization of governing PDEs into particle interactions and relies heavily on kernel functions and their derivatives.
In \emph{diffSPH}, these SPH operators are implemented as compositions of differentiable PyTorch functions, ensuring that the entire physics computation chain is amenable to gradient-based optimization and learning.
However, several numerical and computational aspects need to be taken into account when implementing these operations in a differentiable framework.

\textbf{Composability} is a core design principle in \emph{diffSPH}, a choice that significantly influences the structure of the framework. 
While many traditional SPH codes interleave various computational steps for maximal performance, our design prioritizes a modular, extensible, and inherently differentiable process. 
To this end, we provide a set of fundamental SPH operator modules (implementing interpolations, gradients, curl, divergence, and Laplacians) that act on particle data and neighbor information. 
This modular design, where each SPH operator is an independent differentiable component, ensures that gradients can be computed correctly for any composition of these operations.
It also dramatically simplifies the implementation and verification of forward physics computations and their corresponding backward gradient passes. 
Consequently, a vast array of SPH processes, including iterative pressure solvers, can be constructed by simply assembling these fundamental, validated operators.

\textbf{Kernel functions} are a cornerstone of any SPH framework, as they directly model particle interactions.
diffSPH implements a broad set of commonly used kernels, including various B-Spline and Wendland functions~\cite{dehnen2012improving},
as detailed in Appendix~\ref{appendix:operators}. 
%
Traditional SPH codes primarily utilize the kernel, its spatial gradient, and sometimes its derivative with respect to the smoothing length.
A differentiable framework necessitates access to gradients of these quantities for backpropagation. 
For example, computing gradients of SPH gradient operators with respect to particle positions requires the Hessian of the kernel function, which is typically not defined in SPH codes.
While we explicitly implement some of these terms, such as the Hessian, for improved performance and direct verifiability of gradients, AD is utilized to implement any higher-order and mixed terms that might be necessary, e.g., for differentiating multiple times.

A critical special case arises for kernel derivatives when the inter-particle vector ($\mathbf{x}_{ij}$) becomes zero, e.g., for interactions of a particle with itself. 
While the kernel gradient is typically zero (i.e., $\nabla W(\mathbf{0},h)=\mathbf{0}$), the Hessian at this point can be mathematically undefined or singular for some of the standard kernel functions. 
Directly evaluating this case via automatic differentiation (AD) without special handling would yield non-finite values. 
However, a well-defined value for this term is necessary for robust backpropagation, and it is explicitly required by some advanced SPH schemes, such as the implicit particle shifting technique by Rastelli et al.~\cite{rastelli2022implicit}.
We address this by defining the Hessian based on a physically sensible limit or a definition that ensures a continuous extension of the Hessian, for zero separation. 
Assuming kernels are defined by Dehnen and Aly~\cite{dehnen2012improving}, i.e.,
\begin{equation}
    W(\mathbf{x},h) = C_d\frac{1}{h_d}w\left(\frac{|\mathbf{x}|}{h}\right),
\end{equation}
with $C_d$ being a normalization constant dependent on the kernel and dimensionality, and $w$ being the \emph{actual} kernel, 
we can define this modified hessian as
\begin{equation}
    \nabla^2W(\mathbf{x}, h) = A \left[\frac{C_d}{h^{d+2}}w^{\prime\prime}\left(\frac{|\mathbf{x}|}{h}\right)\right] + \left(\frac{\text{Id}}{|\mathbf{x}|+\epsilon h} - \frac{\mathbf{x}\otimes\mathbf{x}}{|\mathbf{x}|^3+\epsilon^3h^3}\right)\left[\frac{C_d}{h^{d+1}}w^\prime\left(\frac{|\mathbf{x}|}{h}\right)\right];\;A=\begin{cases}
    \text{Id},&\frac{|\mathbf{x}|}{h} < \epsilon,\\
        \frac{\mathbf{x}\otimes\mathbf{x}}{|\mathbf{x}|^2+\epsilon^2h^2}, &\text{else},
    \end{cases}
\end{equation}
with $\epsilon$ being a small value to avoid singularities.
It is important to distinguish this use of analytical kernel derivatives within the AD framework from traditional SPH approximations of differential operators like the Laplacian. 
The derivatives utilized during backpropagation are the exact analytical derivatives of the chosen kernel functions and their evaluation in the forward pass. 
Their mathematical consistency, including the defined value at zero separation, is paramount for correct gradient computation. 
Furthermore, these explicit higher-order derivatives are generally not used to evaluate physical field quantities but solely for backpropagation.

\textbf{Gradient Computation and Memory Usage} are primary challenges in applying AD to complex, iterative simulations like SPH due to the potentially large memory footprint. 
Naïve AD applications often store all intermediate variables from the forward pass to efficiently compute gradients during the backward pass. 
For SPH, this would imply storing numerous intermediate terms for each interacting particle pair, which require memory $\mathcal{O}(n\cdot m_\text{avg})$ for each involved quantity in an interaction, with $m_\text{avg}$ being the average number of neighbors.

To mitigate this, we employ a strategy analogous to gradient checkpointing for many SPH operators. 
Instead of storing all intermediate values for materialized pairs, only the fundamental particle properties and the neighbor lists are typically retained from the forward pass. 
Required intermediate components for a specific interaction are then recomputed on-the-fly during the backward pass by indexing into these base arrays. 
%
This significantly reduces the memory footprint associated with AD.

Nonetheless, recomputing the kernel function itself can induce computational overhead, which is especially prominent when the backpropagation involves complex computations, e.g., higher-order derivatives of the base kernel form. 
Therefore, \emph{diffSPH} implements a further optimization: kernel values ($W_{ij}$) and their necessary spatial derivatives (e.g., $\nabla_i W_{ij}$) are computed once per interacting pair $(i,j)$ at the beginning of a simulation step. 
These precomputed values are then directly passed to the SPH operators, instead of the kernel functions. 
During backpropagation, gradients are first accumulated with respect to these precomputed values, and then only a single backpropagation through the actual kernel function is necessary, rather than being repeatedly invoked for every instance where the kernels appear.
This significantly reduces redundant computations in the backward pass and simplifies the effective computational graph for these terms, avoiding a significant number of summation operations.
As we will show below,
this careful combination of recomputing certain values, while precomputing others 
provides a balanced use of compute and memory resources for learning and optimization tasks.

\subsection{Temporal Integration}\label{sec:framework:timestepping}
In our framework, we make several important choices regarding explicit time integration schemes, specific substep and end-of-step operations within complex SPH schemes, and finally regarding optimizations for multi-stage integrators that reduce computational cost and simplify the computational graph for AD.

The selection of an appropriate timestep $\Delta t$ is crucial for numerical stability and accuracy in any explicit numerical scheme and we adopt existing, well studied, Courant-Friedrichs-Lewy conditions for compressible (e.g., considering sound speed and viscous effects~\cite{price2012smoothed}), weakly compressible (e.g., $\delta$-SPH criteria~\cite{antuono2012numerical,dominguez2022dualsphysics}), and incompressible solvers (e.g., related to particle spacing and maximum velocity~\cite{ihmsen2013implicit}), see also Appendix~\ref{appendix:timestepping}. 
For machine learning applications and to simplify data handling and network training, simulations presented in this work utilize a fixed $\Delta t$, chosen conservatively based on these stability criteria, as many neural networks do not natively support adaptive timestepping.

Our framework implements several explicit time integration schemes to advance the particle system. 
These include standard methods such as semi-implicit Euler~\cite{ihmsen2013implicit}, symplectic Euler~\cite{dominguez2022dualsphysics}, and the standard fourth-order Runge-Kutta (RK4). 
Many of these integrators are implemented using their Butcher tableau representations, facilitating straightforward extension to other explicit schemes if required. 
We do not explicitly implement a frozen dissipation scheme~\cite{antuono2012numerical}, but implementing these would be straightforward in \emph{diffSPH}.

Certain advanced SPH schemes, e.g., CRKSPH, require a distinct treatment for substepping and overall temporal integration.
To facilitate these schemes, we implement the respective $\frac{du}{dt}$ updates in each substep~\cite{frontiere2017crksph}, which are also used to compute intermediate states, and then perform the temporal integration of the entire timestep not using the substep's partial updates but the corrected updates.
While this is consistent with the brief description by Frontiere et al.~\cite{frontiere2017crksph}, they only give an explicit formulation for their modified RK2 scheme, however, extending this formulation to arbitrary multi-step schemes is straightforward.
Note that this leads to an inevitable increase in memory as we need to store actual pair-wise interactions both for forward and backwards passes.

For multi-step schemes in general, we chose to apply particle-shifting only at the end of each overall timestep instead of each substep to reduce computational costs~\cite{sun2019consistent}, whereas enforcement of Dirichlet and Neumann boundary conditions is performed during each substep.
To avoid complications of the compute graph, inlet and outlet boundary conditions are also only enforced at the end of each overall timestep to avoid disconnected sub-compute graphs.
In line with prior work~\cite{frontiere2017crksph}, we support using a modified time-integration scheme that re-uses the results of the previous timestep to perform the first substep of the next timestep.
This optimization not only significantly reduces computation, e.g., compute costs are approximately halved for an RK2 scheme, but also significantly simplifies the computational graph for backpropagation.
Note that this does not reduce the order of convergence of the timestepping~\cite{frontiere2017crksph,spheral}.

%% file: 04-00-application.tex
\section{Applications of Differentiability \& Adjoint Capabilities}\label{sec:application}
We now move on to highlight the capabilities of our differentiable framework to solve inverse problems and to aid in the development of new approaches, both for machine learning and classical approaches.
We investigate five different areas of application, i.e., solving inverse problems  in Sec.~\ref{sec:application:inverse}, performing geometric optimization in Sec.~\ref{sec:application:targeted}, building a hybrid neural network and solver approach for correction setups in Sec.~\ref{sec:application:sitl}, and using differentiable physics to find optimal initial conditions highlighting the potential of \emph{diffSPH} to be used for the development of novel particle shifting approaches in Sec.~\ref{sec:application:shifting}.
For the validation of our solver using several well-established validation cases, e.g., the Sod-Shock Tube for compressible simulations and the Taylor-Green Vortex for weakly compressible simulations, see Appendix~\ref{appendix:validation}.
However, before providing details of the different applications, we will discuss the computational scaling of our solver code in the next section.

\input{04-01a-scaling}
\input{04-01-inverse}
\input{04-02-shape}
\input{04-03-solver-in-the-loop}

\input{04-05-shifting}

%% file: 04-01a-scaling.tex
\subsection{Computational Scaling}
\label{sec:applications:scaling}
\begin{figure}
    \centering
    \includegraphics[width=0.8\linewidth]{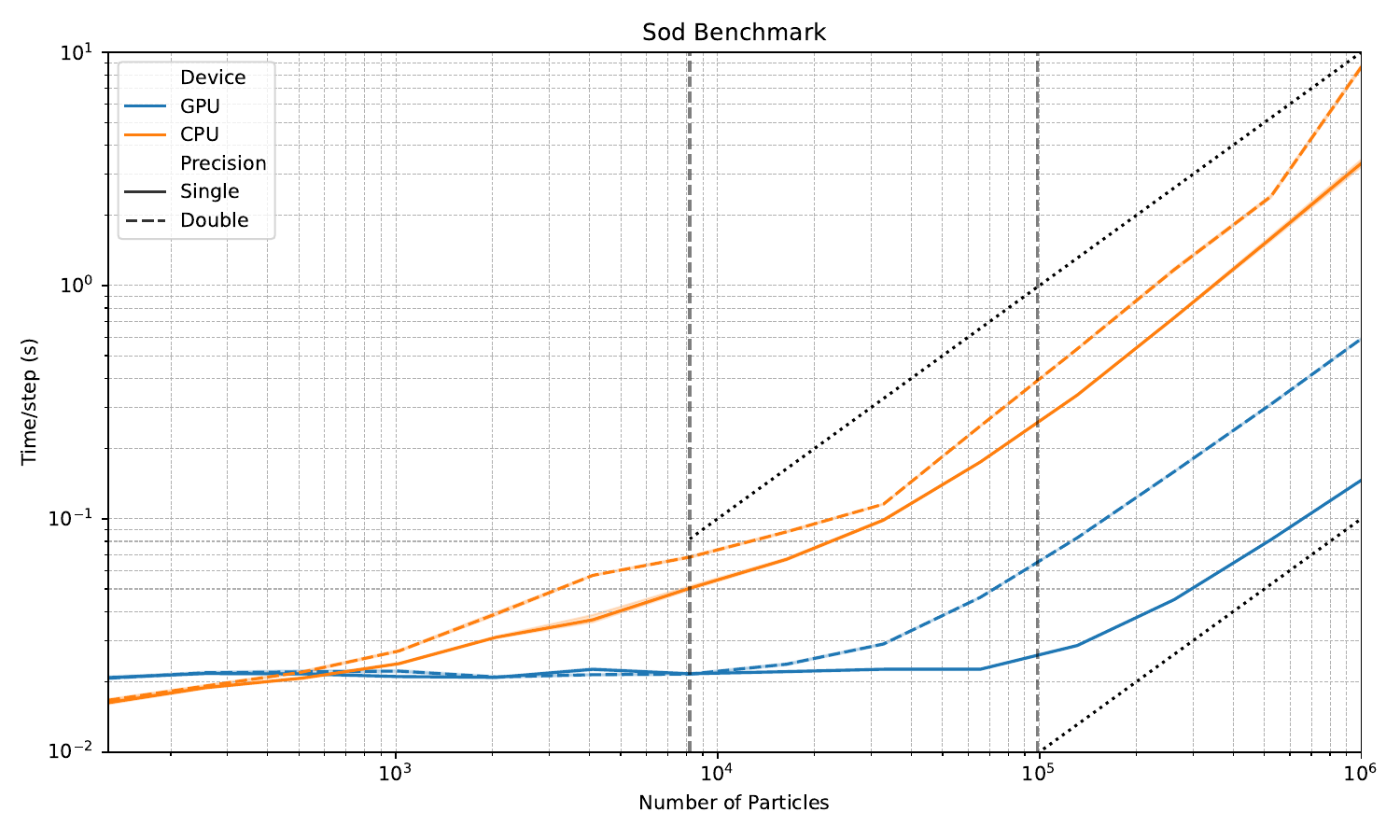}
    \caption{Compute scaling for the Sod Shock Tube in 1D with increasing numbers of particles. The dotted lines indicate $\mathcal{O}(n)$ scaling: whereas the left dashed line indicates the number of CUDA cores on the GPU utilized (8192), the right dashed line indicates the maximum number of threads that can be active simultaneously on the GPU  ($64\times1535=98304$). Results are the average time per timestep over the first $200$ timesteps or at most $30$ seconds of compute.}
    \label{fig:computeBenchmark}
\end{figure}

The computational performance of \emph{diffSPH} directly results from its design philosophy, which prioritizes extensibility and usability via a PyTorch frontend. 
While this introduces overhead compared to low-level languages, our implementation achieves excellent scaling and GPU acceleration, as we demonstrate in this section.
As we implement our solver with a PyTorch front end, each call to a Python function induces a corresponding call overhead.
Typically, the majority of the simulation code is written in Python for extensibility and ease of use. 
However, we implement key aspects of the solvers via custom, carefully optimized C++/CUDA code to ensure that the solver can efficiently handle simulation scenarios with larger amounts of particles. 
Correspondingly, we expect computational performance to be dominated by the calling
overheads for very small problems, while exhibiting scaling according to the underlying algorithms for larger problems.
All our performance measurements were done on a system with a single NVIDIA RTX A5000 GPU with 24GB of VRAM and an Intel Xeon W-2255 CPU with 10 cores and 64GB of system RAM running Ubuntu 22.04 and Cuda 12.8.

\begin{figure}
    \centering
    \includegraphics[width=1.0\linewidth]{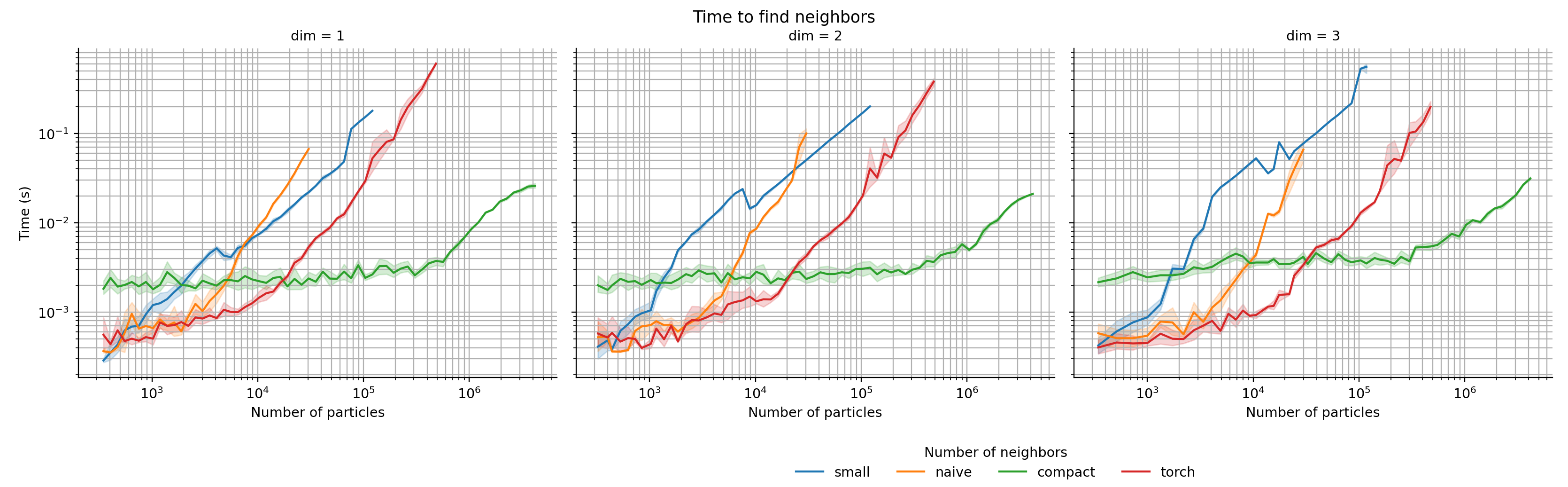}
    \caption{Compute scaling for the neighbor search for increasing numbers of particles in a densely sampled domain for, from left to right, one, two, and three spatial dimensions. The blue line is a Python-based for loop implementation, naïve utilizes an explicit distance matrix with $\mathcal{O}(n^2)$ memory requirements, \emph{torch} refers to the torch-cluster library, and \emph{compact} is our implementation of compact hashing. }
    \label{fig:neighborhoodBenchmark}
\end{figure}

We first investigate computational scaling for the neighborhood search as a key operation for SPH simulations. We evaluate the runtime of our implemented scheme and other alternatives for increasing numbers of particles for fully occupied domains, with random noise added to the particle positions, with the usual neighborhood sizes. We perform these tests in 1D, 2D, and 3D. 
The most straightforward implementation in Python for a neighbor search is a series of for loops, denoted as \emph{small} in the scaling, which shows $\mathcal{O}(n^2)$ scaling and quickly becomes impractical for performance reasons, as shown in Fig.~\ref{fig:neighborhoodBenchmark}.
As an alternative, it is possible to compute a dense distance matrix to find the adjacency of the particles, denoted as \emph{naive}, but this scheme scales with $\mathcal{O}(n^2)$ in memory consumption and performance, making it impractical for systems with more than $4096$ particles.
Neighborhood searches are also a common problem in GNNs, commonly referred to as radius searches in this context, and we chose the torch cluster library as a comparison due to its inclusion in the widely utilized PyTorch Geometric library.
This approach implements a for loop per particle, but manually implements it in CUDA, resulting in good performance for small problems, which are commonly faced in machine learning research.
This results in an inherent $\mathcal{O}(n^2)$ compute cost, making it impractical for large problems and practical SPH simulations.

Our implementation of the compact neighborhood search achieves $O(n\log{n})$ scaling for large problems, due to relying on a sort operation, but remains at a constant compute time for problems below very large scales due to calling overheads that provide a consistent threshold for minimum compute time.
Consequently, for very small problems it is possible to select any of the other search schemes within our framework, whereas for large problems we can rely on the efficient scaling of the compact neighborhood queries. 
Overall, the compute cost of a simulation step can be significantly higher than the cost of the neighborhood search, especially if a velocity Verlet formulation is chosen for the neighborhoods, combined with temporal consistency of the sorting and particle positiong, the computational scaling is oftentimes reduced to $O(n)$, which is the same scaling as the actual SPH simulation.
Due to the usage of the multi-level memory approach~\cite{DBLP:journals/cgf/WinchenbachK20}, our neighborhood search also natively supports neighborhood queries with significantly different support radii, such as those encountered in compressible shock simulations. 

To evaluate the computational cost of a solver scheme implemented in \emph{diffSPH}, we utilized the \emph{compSPH} scheme for a performance benchmark in the Sod-Shock Tube, see Fig.~\ref{fig:computeBenchmark}.
For this benchmark, we evaluated CPU and GPU scaling and single and double precision performance and simulated $200$ timesteps, or at most $30$ seconds of compute per configuration, from $32$ to one million particles.
In general, we expect the performance scaling to be linear, i.e., $\mathcal{O}(n)$ with a constant factor based on the number of neighbors, as the nonlinear scaling of the neighborhood, as discussed before, is negligible for all cases in relative terms.

Regarding the compute costs on the CPU side, we observe the expected scaling of $\mathcal{O}(n)$ for larger problems, i.e., more than $16$K particles, where the performance for smaller problems scales sub-linearly due to the cache size of our CPU.
In this case, the overhead for double precision is approximately a factor of $2$, showing good agreement with the expected computational scaling and costs.
On the GPU side for single precision, we observe constant scaling up to approximately $100$K particles, at which point we have more particles than threads that can be simultaneously in flight on our GPU, leading to the compute cost per thread dominating after this point and a resulting $\mathcal{O}(n)$ scaling.
For double precision, we see this change in scaling behavior earlier, at $8192$ particles, which aligns with the number of CUDA cores on our GPU ($8192$). This indicates that we are strongly limited by the compute cost per particle and that memory operations do not hide these costs.
Overall, we observe a $6\times$ increase in costs for double precision on our GPU, as memory transfers are only affected by a $2\times$ increase compared to a $64\times$ reduction in floating point speed.

Our solver framework focuses on differentiability, and automatic differentiation is a well-studied problem~\cite{BAUR1983317}. 
Consequently, the upper bound for overhead of automatic differentiation is given by a constant factor of $5\times$, with significantly lower factors observed in practice.
We utilized this test case to evaluate the computational cost of the backwards pass and first evaluated the relative computational cost of the forward and backwards pass of a single simulation step.
As a test problem, we compute the summation density of all particles after the step and mark the initial particle positions as requiring gradients.
Based on our observations of the compute scaling, we choose a particle count of $64$K and found that the compute cost of the backwards pass for this single step is half that of the forward pass, i.e., the backwards pass required $32\%$ of the overall compute time, with no significant increase in the cost of the forward pass.
When evaluating a longer rollout, such as the one we will use in Sec.~\ref{sec:application:inverse}, we found that the cost of the backwards pass is similar to the cost of the forward pass in most cases.
These empirical results are significantly below the upper bound of $5\times$, primarily due to using checkpoints and storing intermediate results for the backward pass. 
Furthermore, for short trajectories, the neighborlist is the same for all timesteps and is only computed in the forward pass once, making the backward pass computationally cheaper.
In all of our test cases, the computational costs of the backwards pass never exceeded an overhead of $2\times$.

To summarize, we observe the expected scaling of our SPH code for large particle counts, which are of interest for practical research questions, whereas for small problems, 
call overheads start to dominate.
This can be utilized from a weak scaling perspective 
by running multiple simulations in parallel, i.e., \textit{batched} processing, as our datastructure layout allows for a straightforward merging of particle states across simulations with a batch support in the neighborhood search.
Accordingly, we can run many simulations in parallel for smaller problems, which is, e.g., attractive for training tasks in machine learning scenarios.
Overall, \emph{diffSPH} achieves approximately a $25\times$ speed-up when using a GPU instead of a CPU, and can handle several million particles.

%% file: 04-01-inverse.tex
\subsection{Inverse Problems}\label{sec:application:inverse}

\begin{figure}
    \centering
    \includegraphics[width=0.65\linewidth]{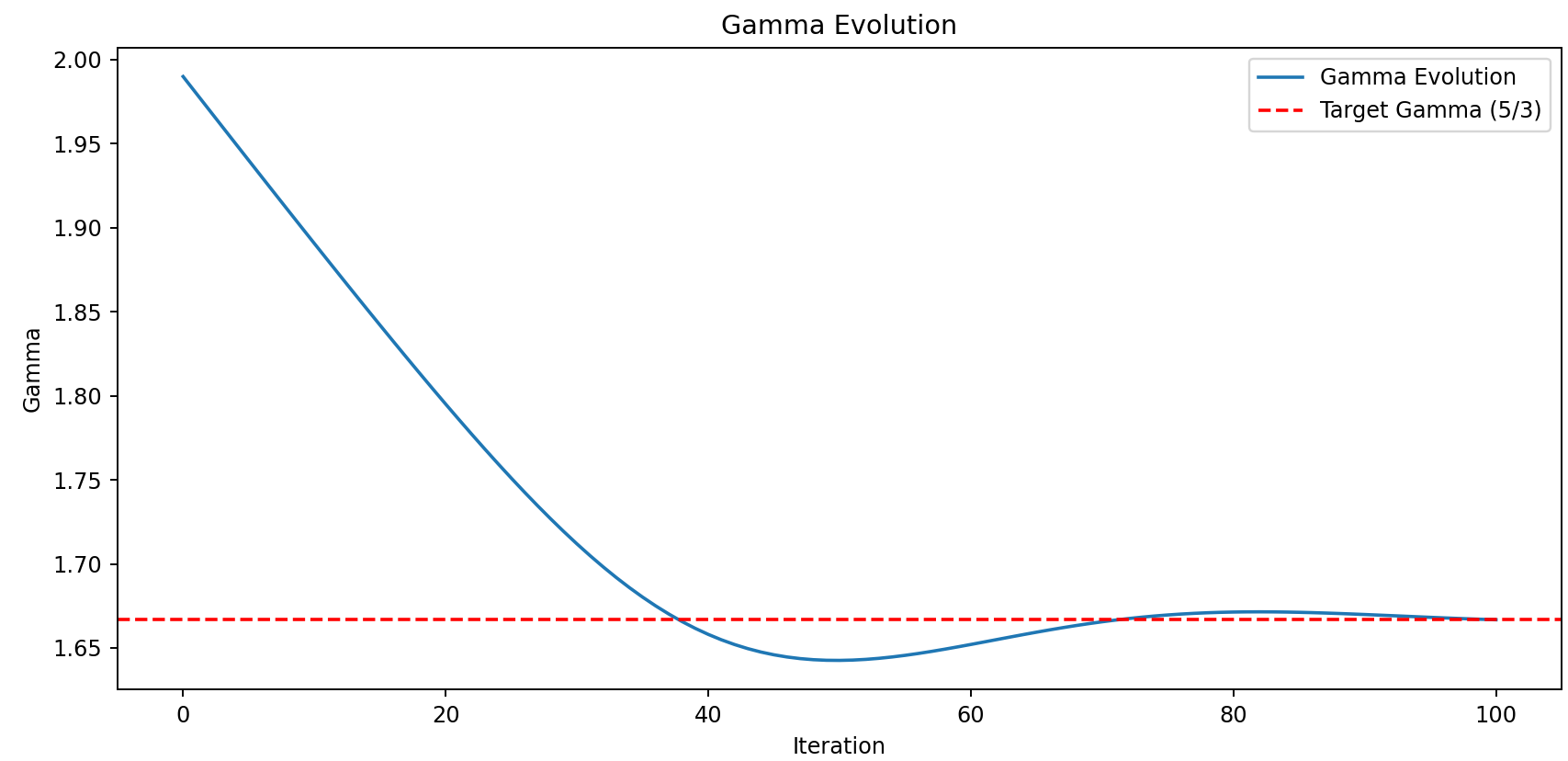}
    \caption{Optimization trajectory of the optimization of the adiabatic constant $\gamma$ for the Sod-Shock tube starting with an initial value of $\gamma=2$ converging to the reference $\gamma=\frac{5}{3}$ using an inverse solve.}
    \label{fig:application:inverse:gamma}
\end{figure}

Inverse Problems are a central area of concern in many practical engineering tasks and involve a large set of potential tasks.
Using traditional non-differentiable frameworks makes solving these problems challenging, as gradient-free optimization techniques in large search spaces tend to be computationally expensive, and implementing specific adjoint optimization paths around existing solvers is difficult in practice.
On the other hand, using a framework that is designed from the ground up to be differentiable makes the realization of these tasks straightforward. 
To demonstrate the capabilities of \emph{diffSPH}, we solve two classic inverse problems: first, denoising initial conditions, and second, finding a physical system parameter from a time evolved state.
We evaluate the performance of the forward pass in Sec.~\ref{appendix:validation:compressible:sod}.

\begin{figure}[t]
    \centering
  \includegraphics[width=\linewidth]{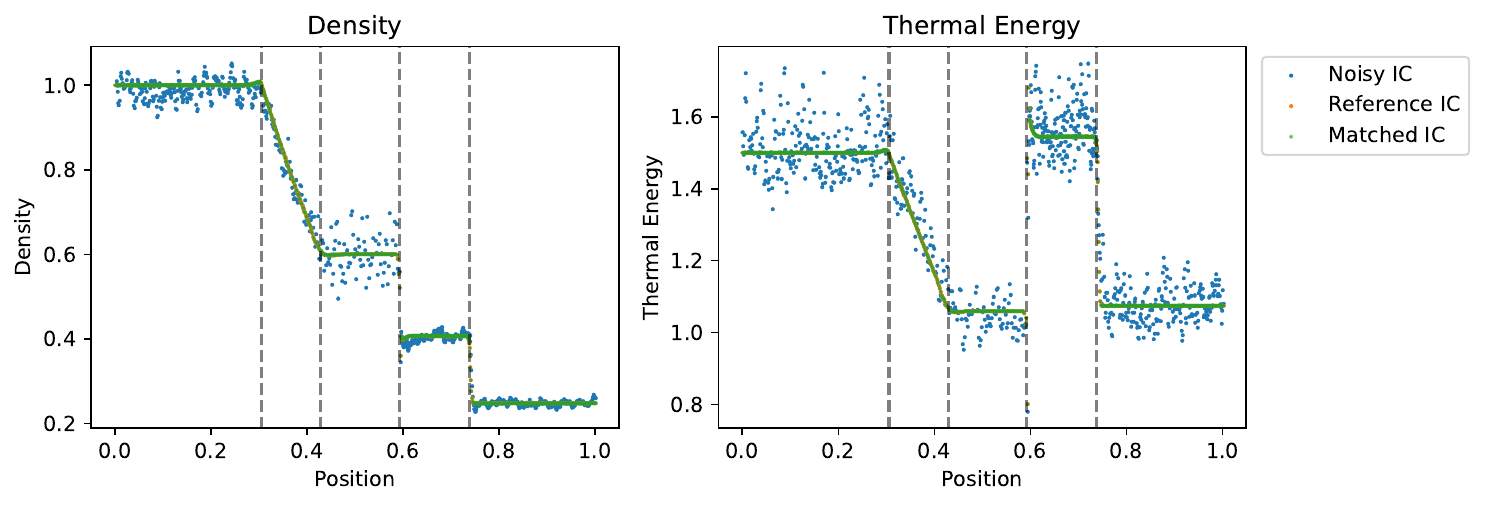}
    \caption{Comparison of the Sod-Shock Tube simulated using the noise free~(orange) and the noisy~(blue) ICs at the reference timepoint $t=0.15$ after 832 timesteps simulated using the compSPH scheme within \emph{diffSPH}. Note that the green points are the result of the optimization of the inverse problem and exactly overlap the results of the noise-free ICs.}
    \label{fig:application:inverse:sod_IC}
\end{figure}

For both tasks, we utilize a reference state after performing several hundred timesteps of a shock-capturing compressible simulation.
To set up this reference state, we use a classic Sod-Shock Tube setup, using the \emph{CompSPH} scheme~\cite{michael2014compatibly} due to its capabilities to handle noisy initial conditions.
For this specific setup, we use a variable-mass setup, i.e., the spatial density of particles is uniform across the domain, with a fixed smoothing length per particle and no viscosity switch to avoid potential instabilities in the forward pass due to the noisy initial conditions. 
Specifically, we choose $[\rho_l,P_l,v_l]=[1,1,0]$ and $[\rho_r,P_r,v_r]=[0.25,0.1795,0]$ for the initial conditions of the left and right half domain with a fixed timestep $\Delta t=1.8\cdot10^{-4}$ as well as the modified RK2 integrator and B7 kernel of Frontiere et al.~\cite{frontiere2017crksph}.
For the reference state, we run the simulation for $832$ timesteps to yield the system state at $t=0.15$ for $n_x=800$ particles, see Fig.~\ref{fig:application:inverse:sod_IC}.

\begin{figure}
    \centering
    \includegraphics[width=\linewidth]{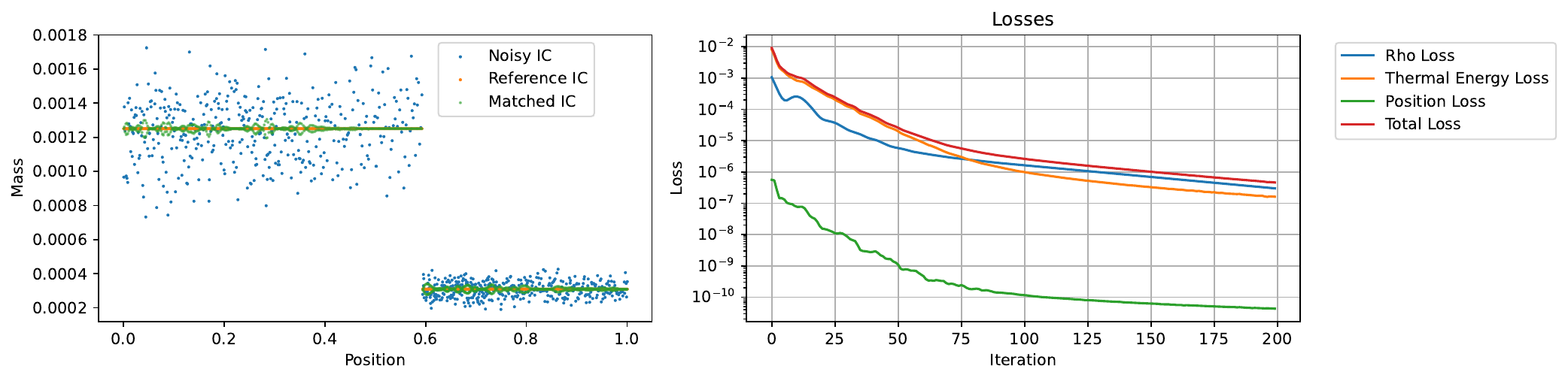}
    \caption{Result of the optimization process showing both the initial noisy conditions and the optimized initial conditions compared to the actual noise-free state, as well as the convergence of the inverse problem} 
    \label{fig:application:inverse:optimizedIC}
\end{figure}
\noindent\textbf{Optimizing Physical Parameters:} 
For this case, we start with the noise-free initial conditions but modify the adiabatic constant $\gamma$ from the value used in the reference trajectory, $\gamma=5/3$, to a higher value $\tilde{\gamma}=2$.
This higher constant leads to significant differences in the resulting state, especially for the velocity magnitude, and we utilize the MSE of the velocity of the particles as the loss function, i.e.,
\begin{equation}
    \mathcal{L} = \frac{1}{n_x}\sum_{i=0}^{n_x}||\mathbf{v}^t_i - \tilde{\mathbf{v}}_i^t||_2^2,
\end{equation}
as the velocity field is only indirectly dependent on $\gamma$, whereas a quantity such as the internal energy would be directly dependent on $\gamma$, leading to a much easier optimization problem.
We then mark $\gamma$ as requiring a gradient in PyTorch and use the Adam optimizer with a learning rate of $10^{-2}$, where this large learning rate is possible as the influence of the adiabatic constant is smoothed over the number of particles and not a per particle quantity, leading to well-behaved gradients.
Running the optimization for $100$ steps, see Fig.~\ref{fig:application:inverse:gamma}, shows the expected convergence to the reference value, with the under and overshooting behavior being emblematic of the used Adam optimizer.

\noindent\textbf{Optimizing Initial State:} 
To generate the noisy initial conditions we add a normal distributed noise using a standard deviation of $1\%$ of the maximum mass of a particle in the system, i.e., $\mathcal{N}(0,0.01\operatorname{max}_i m_i)$, onto the initial mass of every particle and perform a forward simulation using these ICs, see Fig.~\ref{fig:application:inverse:sod_IC}.
To compute the difference between the reference state and the current state, we use the Mean Squared Error (MSE) of the positions, i.e.,
\begin{equation}
    \mathcal{L} = \frac{1}{n_x}\sum_{i=0}^{n_x}||\mathbf{x}^t_i - \tilde{\mathbf{x}}_i^t||_2^2,
\end{equation}
where $\mathbf{x}^t_i$ and $\tilde{\mathbf{x}}_i^t$ denote the position of particle $i$ at timepoint $t$ in the current and the reference simulation.
We then use the Adam optimizer~\cite{Kingma2014AdamAM} using a learning rate of $10^{-5}$ for $100$ iterations to optimize the noisy initial conditions to match the reference trajectory.
Performing this optimization in \emph{diffSPH} only requires a few lines of code to set up, i.e., we  mark the initial state as differentiable, utilize an existing optimizer, such as Adam~\cite{Kingma2014AdamAM}, run the simulation normally and then the backwards pass on the error, in total requiring less than ten lines of code. 
We find that, see Fig.~\ref{fig:application:inverse:optimizedIC}, the optimization reliably converges towards the expected noise-free initial conditions, with the residual noise depending on the number of optimization steps. 

%% file: 04-02-shape.tex
\subsection{Geometry Optimization}\label{sec:application:targeted}

Shape optimization is another classic problem in CFD and engineering challenges and involves optimization of the geometry of a domain to yield specific desired outcomes.
Oftentimes, one of the main challenges in such a scenario is the parameterization of the initial conditions with respect to the geometry~\cite{li2021learnable}; however, our usage of a combination of Signed Distance Fields with Constructive Solid Geometry allows for a direct parameterization of the flow geometry.
Note that this description can also yield derivative quantities of the boundary geometry, e.g., analytic surface normals required for boundary sampling~\cite{english2022modified}.

\begin{figure}
    \centering
    \includegraphics[width=0.95\linewidth]{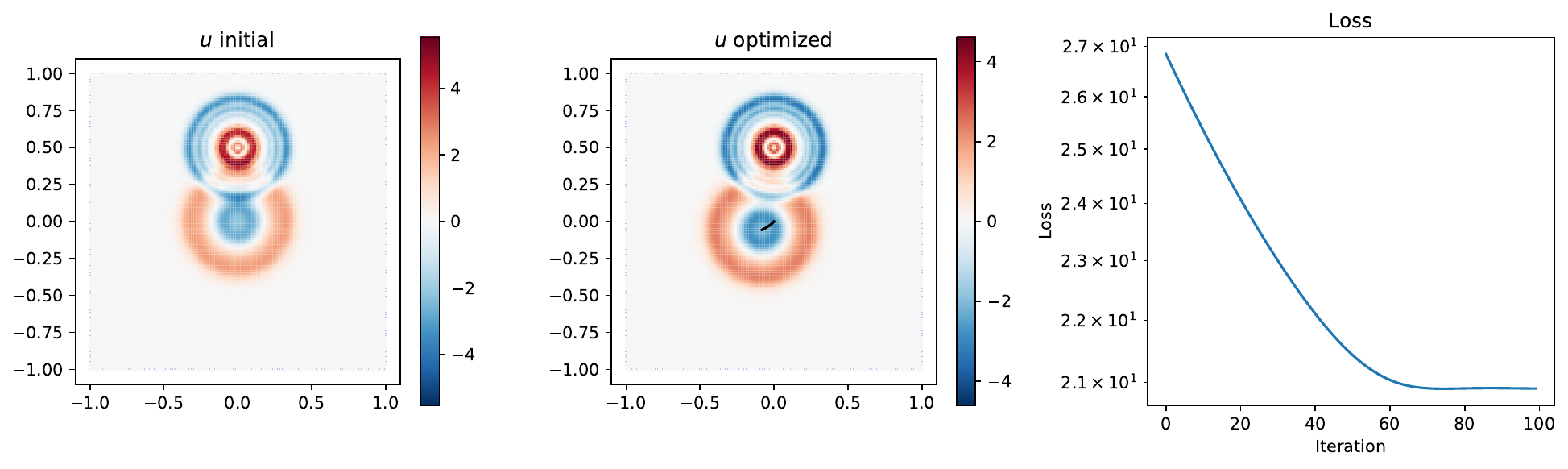}
    \caption{Result of the simulation after $250$ timesteps using the initial geometry and the results for the optimized initial conditions to maximize the wave intensity at the target location $[0,0.25]$. The black line indicates the position of the wave emitter we optimized over iterations. The right figure shows the convergence of the loss we minimized.}
    \label{fig:application:shape:final}
\end{figure}

As this task focuses on the geometry's differentiability, we utilize a wave equation simulation instead of a Navier-Stokes formulation.
We implement the wave equation by splitting the second order differential equation into two first order differential equation and using a Runge Kutta 4~(RK4) scheme as the time integrator, with $\Delta t = 0.001$, with no evolution of the particle positions over time, with uniform speed of sound $c=1$ and no damping in the domain $[-1,1]^2$ with $n_x=n_y=256$ particles, resulting in $65536$ particles in total.
As initial conditions, we utilize two circular areas of high absolute magnitude $u_\text{mag}=10$, at locations $[0,0]$ and $[0,0.5]$ with radius $r=1/8$, where the source at $[0,0]$ is the target for optimization. 
We then simulate $500$ timesteps and evaluate the magnitude of the field $u$ at $[0,0.25]$ as the target point in space.
We optimize the location of the first source with respect to this target using the Adam optimizer and a learning rate of $10^{-3}$.
As we want to maximize the magnitude of the $u$ field at a location, we define the loss based on the difference to a maximum value, i.e.,
\begin{equation}
    L = \left[5 - \sum_j \frac{m_j}{\rho_j} u_j W(\mathbf{x}-\mathbf{x}_j,h)\right]^2,
\end{equation}
with an arbitrary choice of $5$ as the reference value as no point in the simulation domain has a magnitude greater than $5$ after the target time period.
Performing this optimization for $100$ steps yields the results shown in Fig.~\ref{fig:application:shape:final}, where the magnitude at the target point has been maximized from an initial value of $-0.18$ to $0.43$.
It is important to note here that this process finds a potential local maximum and is not guaranteed to find the global maximum due to the periodic nature of the wave equation.
Despite these ambiguities, the optimization with \emph{diffSPH} reliably finds the closest local minimum via the gradient computed through the $500$ simulation steps.

%% file: 04-03-solver-in-the-loop.tex
\subsection{Learned Solver Corrections}\label{sec:application:sitl}

\begin{figure}[t]
    \centering
\begin{subfigure}{.5\textwidth}
  \centering
  \includegraphics[width=\linewidth]{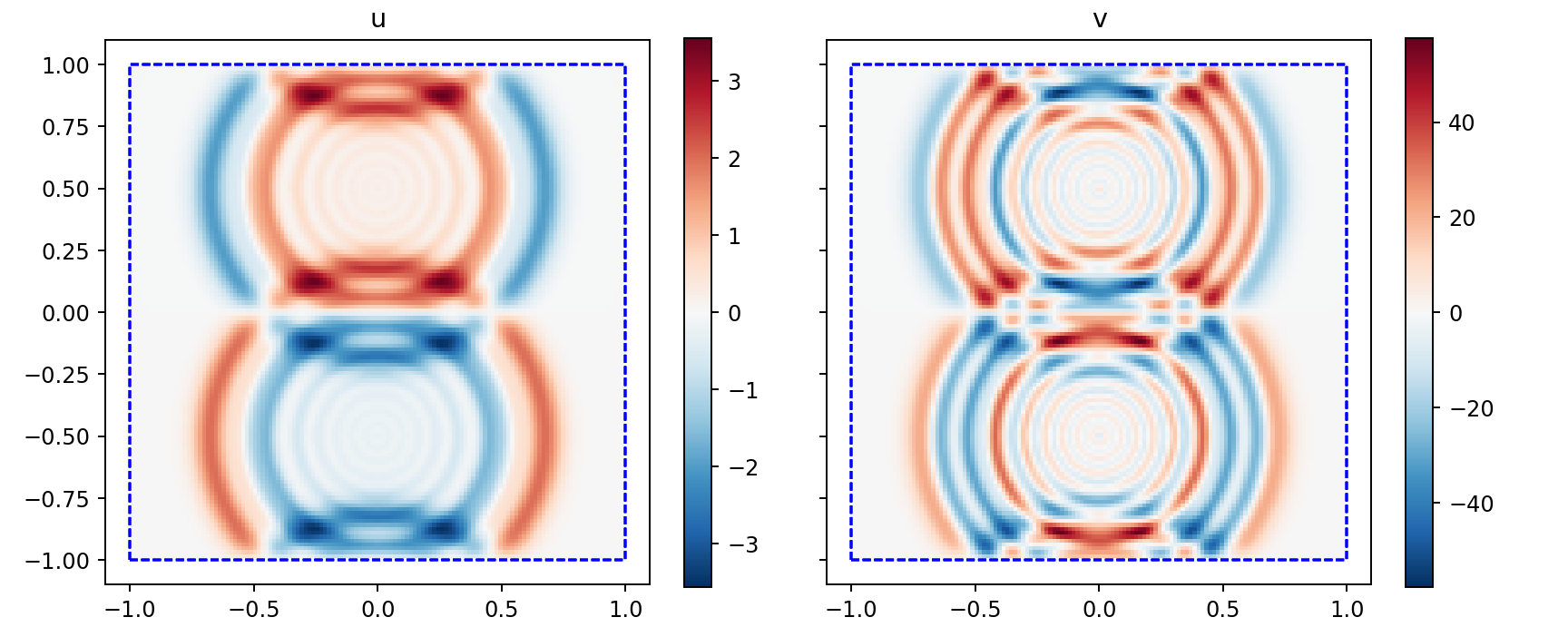}
  \caption{Runge Kutta 4}
  \label{fig:application:sitl:rk4}
\end{subfigure}%
\begin{subfigure}{.5\textwidth}
  \centering
  \includegraphics[width=\linewidth]{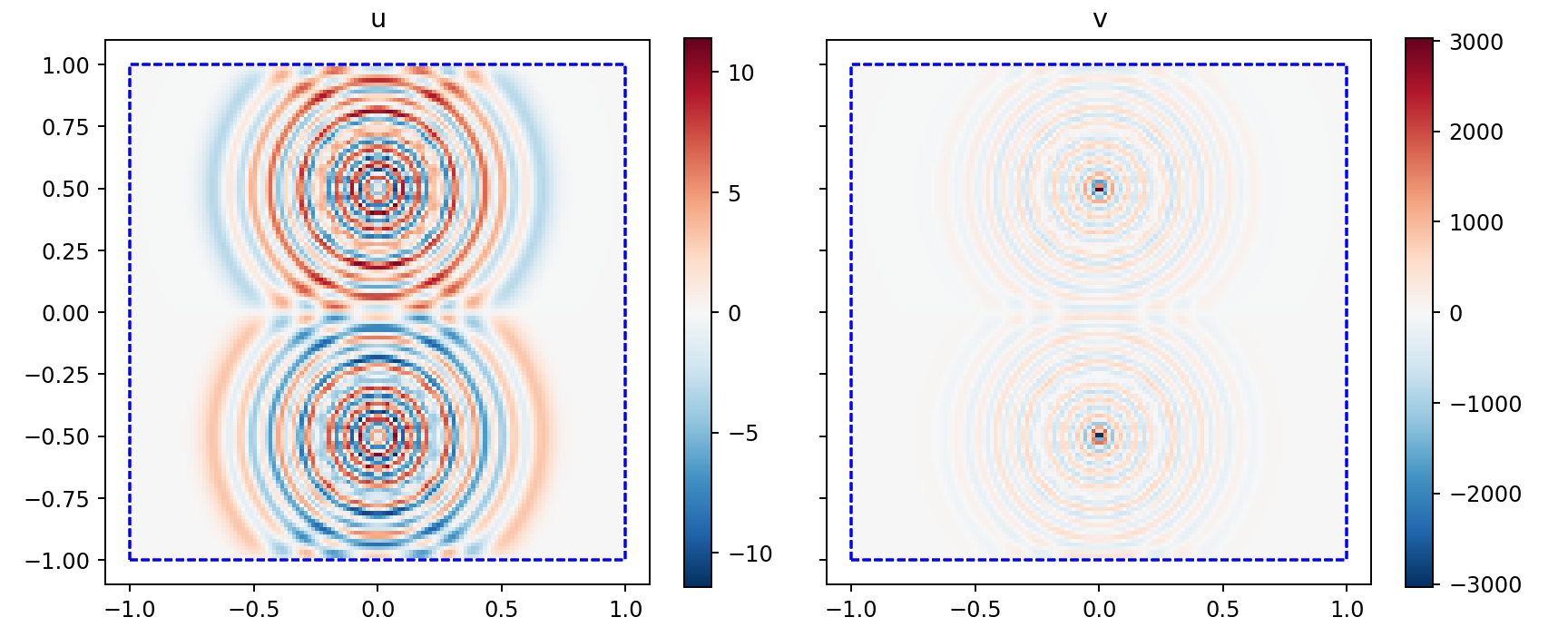}
  \caption{Explicit Euler}
  \label{fig:application:sitl:eeuler}
\end{subfigure}%
    \caption{Comparison of the wave equation setup from Sec.~\ref{sec:application:sitl}, evaluated for $250$ timesteps using both an RK4 integrator (left) and an explicit Euler integrator(right), showing the instability due to a low order numerical scheme that we aim to correct with a Neural Network.}
    \label{fig:application:sitl:reference}
\end{figure}

\begin{figure}[t]
    \centering
\begin{subfigure}{.5\textwidth}
  \centering
  \includegraphics[width=\linewidth]{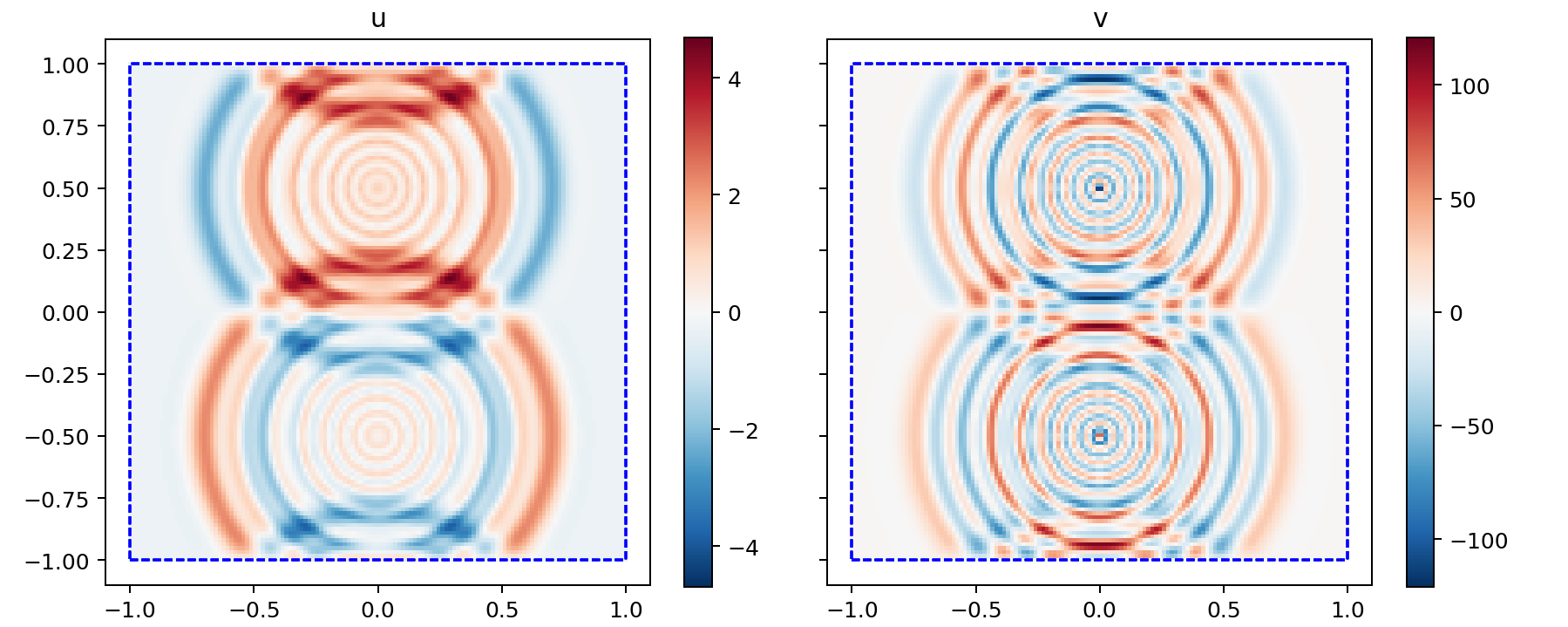}
  \caption{Results without temporal unrolling}
  \label{fig:application:sitl:nounroll}
\end{subfigure}%
\begin{subfigure}{.5\textwidth}
  \centering
  \includegraphics[width=\linewidth]{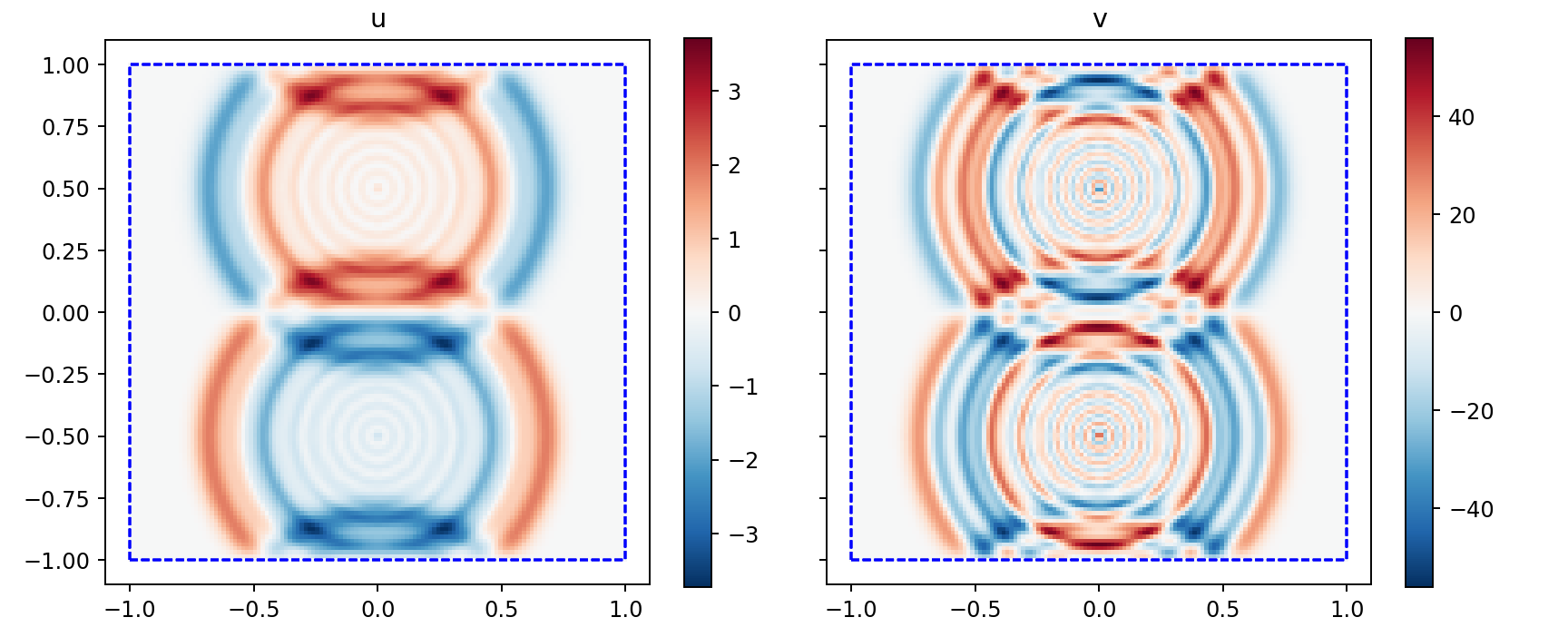}
  \caption{Results with temporal unrolling}
  \label{fig:application:sitl:wunroll}
\end{subfigure}%
    \caption{The results of training a solver-in-the-loop approach for the wave equation using no temporal unrolling during training (left) and using temporal unrolling during training (right), showcasing significant differences and better performance for temporal unrolling, which is only possible with a differentiable framework.}
    \label{fig:application:sitl:trained}
\end{figure}

\begin{figure}
    \centering
\begin{subfigure}{.5\textwidth}
  \centering
  \includegraphics[width=\linewidth]{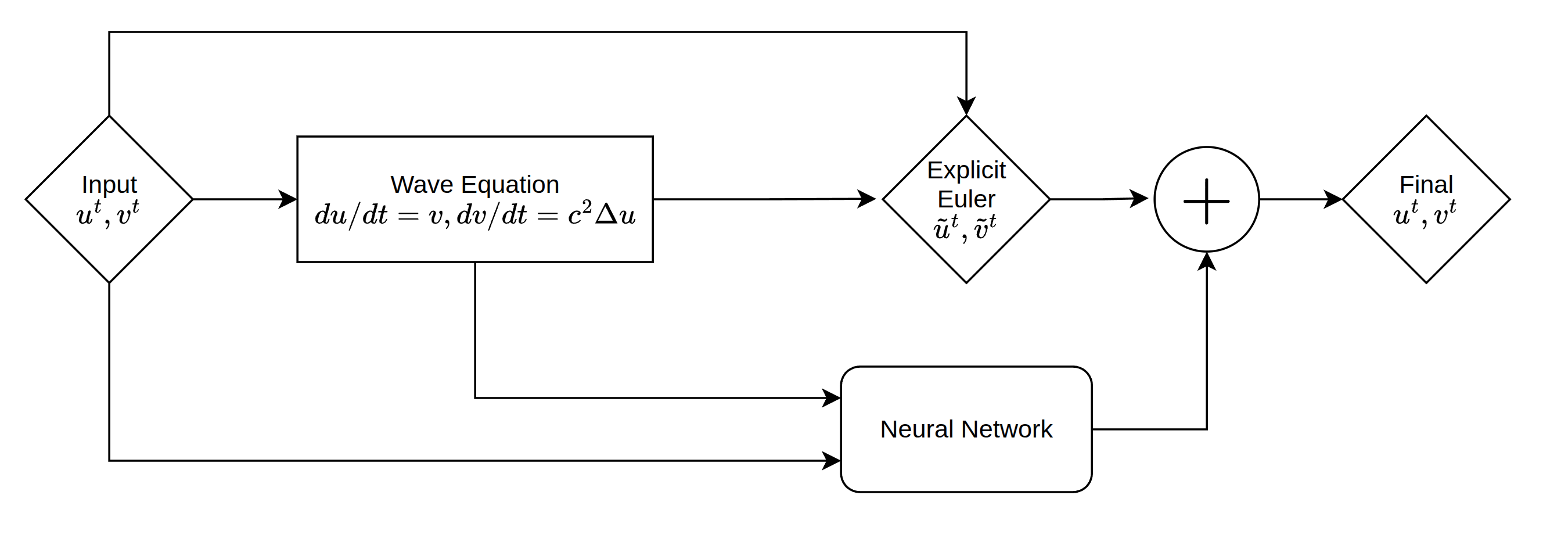}
  \caption{The overall architecture of the solver-in-the-loop approach}
  \label{fig:application:sitl:setup}
\end{subfigure}%
\begin{subfigure}{.5\textwidth}
  \centering
  \includegraphics[width=\linewidth]{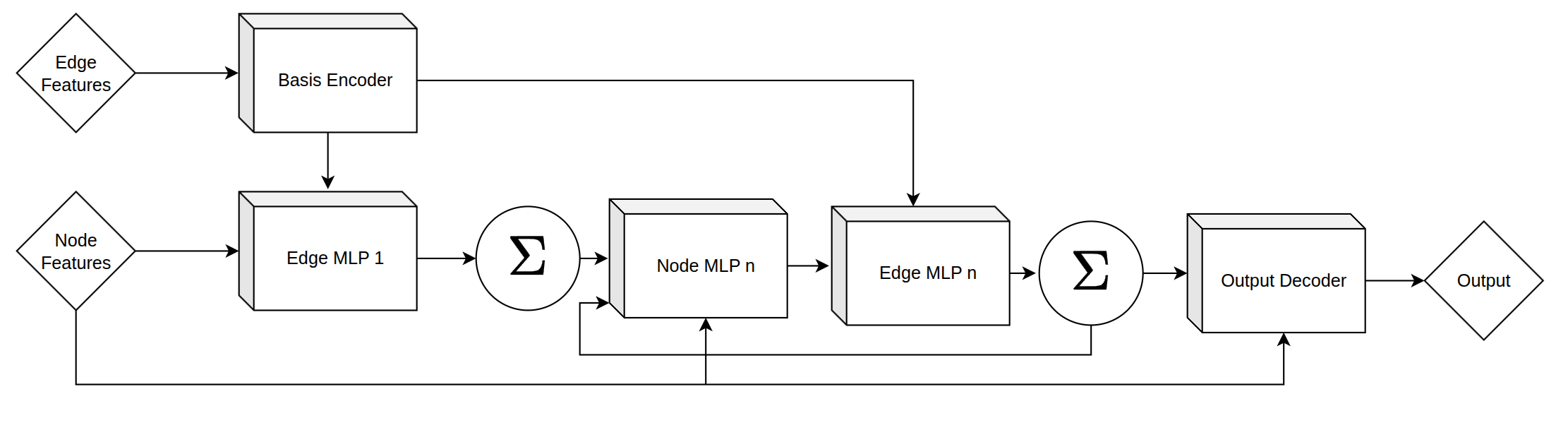}
  \caption{The internal architecture of the GNN used}
  \label{fig:application:sitl:gnn}
\end{subfigure}%
    \caption{Schematic representations of the solver-in-the-loop architecture (left) and the internal architecture of the GNN.}
    \label{fig:application:sitl:schemas}
\end{figure}

In addition to solving classic inverse problems, \emph{diffSPH} is also designed to integrate tightly with ML-based applications, especially regarding hybrid network-solver combinations.
A noteworthy example of these approaches is the solver-in-the-loop approach that combines a low-order numerical scheme with a neural network corrector to match a higher-order numerical scheme~\cite{um2021solverintheloop,kochov2021}.
To demonstrate the capabilities of such hybrids within \emph{diffSPH}, we chose a setup similar to the one in the previous section, see Sec.~\ref{sec:application:targeted}, with slightly modified initial conditions as we now place two symmetric sources at $[0,\pm0.5]$.
To build our solver-in-the-loop application, we combine an explicit Euler integration scheme with a neural network and aim to match the behavior of the RK4 scheme.
Using an explicit Euler scheme with no correction, see Fig.~\ref{fig:application:sitl:eeuler}, leads to a highly unstable and inaccurate simulation, whereas an RK4-based integration, see Fig.~\ref{fig:application:sitl:rk4}, leads to a smooth and stable solution, which provides an ideal basis for a solver-in-the-loop approach.
The prior definition of the wave equation leads to an explicit Euler scheme as
\begin{equation}
    u^{t+1}=u^t+\Delta t v^t;\;v^{t+1}=v^t+ c^2\Delta t\Delta u^t.
\end{equation}
To build the neural network component, we chose a simple message passing graph Neural Network~\cite{brandstetter2022message} as the basis, where we input several physical parameters into each node, i.e., the particle mass, density, and support radius, as well as the input variables of the wave equation, see Fig.~\ref{fig:application:sitl:setup}.
The neural network itself is based on combining an input encoder for the edge features, i.e., the relative distances of particles within each particle's support radius, using a Fourier basis encoder~\cite{tancik2020fourier}, and then performing three message passing steps followed by a per-particle MLP output decoder, see Fig.~\ref{fig:application:sitl:gnn}, where each MLP consists of $2$ hidden layers with $64$ neurons, see Fig.~\ref{fig:application:sitl:setup}.

As the training dataset, we then utilize a single trajectory of the simulation containing the first $1000$ timesteps and train the network for $4000$ iterations using Adam as optimizer and a learning rate of $10^{-3}$ that is reduced by $\frac{3}{4}$ every $500$ iterations.
To evaluate the performance of the network, we then provide only the initial state at $t=0$ and recurrently invoke the solver-in-the-loop setup for $250$ timesteps to evaluate the ability of the network to generalize to unseen trajectories, as the network has never been called recurrently during training.
The resulting network, see Fig.~\ref{fig:application:sitl:nounroll}, shows a significantly stabilized behavior when compared to just the explicit Euler results, but still exhibits undesirable high-frequency components in both the $u$ and $v$ components with significant errors on the $v$ component.

A classic approach in ML to avoid some of these issues is training the network using \textit{unrolling}, i.e., invoking the solver-network combination recurrently during training, which requires a solver that is fully differentiable~\cite{list2025differentiability}.
To realize this approach, we can simply provide the outputs of the previous invocation to the solver, where we initially train with no unrolling for the first $250$ training iterations and then increase the unroll length by $1$ every $250$ iterations afterwards.
This training results in a much more stable performance, see Fig.~\ref{fig:application:sitl:wunroll}, which closely matches the expected ground truth.

%% file: 04-05-shifting.tex
\subsection{New Applications: Differentiable Particle Shifting}\label{sec:application:shifting}

In addition to the previous examples, differentiability also provides a new perspective on existing problems that are traditionally difficult to solve.
One such problem is particle shifting, which is a central component of ensuring the accuracy and convergence of SPH~\cite{Vacondio2021Grand}, as particle disorder can lead to significant issues, e.g., $\nabla1\neq0$~\cite{price2012smoothed}.
Several schemes exist in traditional SPH methodology to implement these particle shifting techniques, e.g., using explicit~\cite{sun2019consistent} or implicit~\cite{rastelli2022implicit} schemes, however, they are often challenging to implement or require significant tuning of heuristic parameters.
These schemes are especially challenging near free surfaces and boundaries, which are also areas of the biggest concern when requiring accurate and robust simulations.
Furthermore, generating optimal initial conditions for SPH has been an ongoing area of research since its inception, especially regarding shock-capturing and astrophysics, see for example Diehl et al.~\cite{Diehl_Rockefeller_Fryer_Riethmiller_Statler_2015} for an overview of various sampling techniques.

On a conceptual level, particle shifting aims to reduce the error of SPH operators by adjusting particle positions, i.e., these schemes minimize a loss term, defined by an SPH operator, with respect to particle positions.
This problem description naturally lends itself to being addressed by a differentiable physics approach. In this section, we demonstrate how \emph{diffSPH} can be directly employed to reduce such physical losses to generate optimal initial conditions.
We do this by first demonstrating how to optimize the particle positions of a 1D SPH simulation to yield a desired density distribution, e.g., to generate optimal initial conditions for shock-capturing simulations, where our differentiable approach yields results that are at least on par with traditional inverse CDF approaches, see Sec.~\ref{sec:application:shifting:1D}.
Next, we demonstrate how we can implement a differentiable shifting approach in 2D based on the errors in SPH operators, combined with regularization terms, to find optimal particle distributions, including proper treatment of boundary particles, see Sec.~\ref{sec:application:shifting:2D}.
Finally, we demonstrate how this can be utilized in 3D scenarios, see Sec.~\ref{sec:application:shifting:3D}.

\subsubsection{Generating Optimal Initial Conditions}
\label{sec:application:shifting:1D}
\begin{figure}[t]
    \centering
  \centering
  \includegraphics[width=\linewidth]{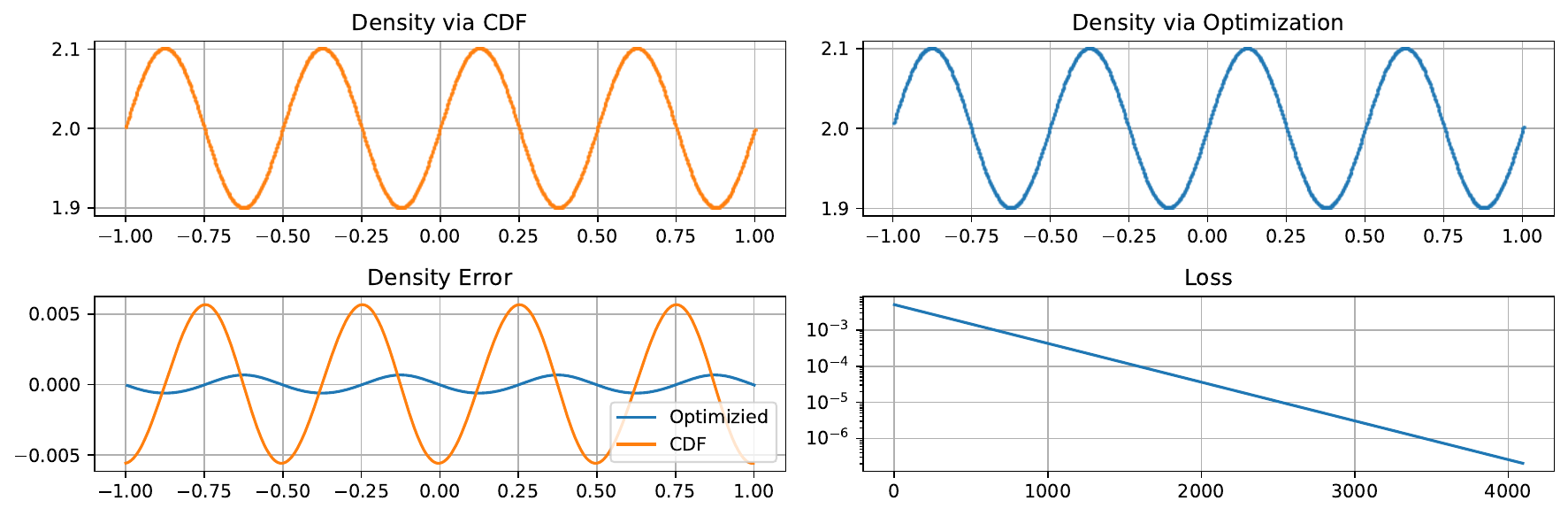}
    \caption{Results for generating initial conditions for a sinusoidal density profile $\rho(\mathbf{x}) = 2 + 0.1\sin{\left(4\pi\mathbf{x}\right)}$. The top left shows the inverse CDF sampling, the top right shows the gradient optimization sampling, the bottom left shows the relative errors to the desired density profile, and the bottom right shows the convergence of the gradient-based sampling.}
    \label{fig:application:differentiable:1D_sin}
\end{figure}

\begin{figure}[t]
    \centering
  \centering
  \includegraphics[width=\linewidth]{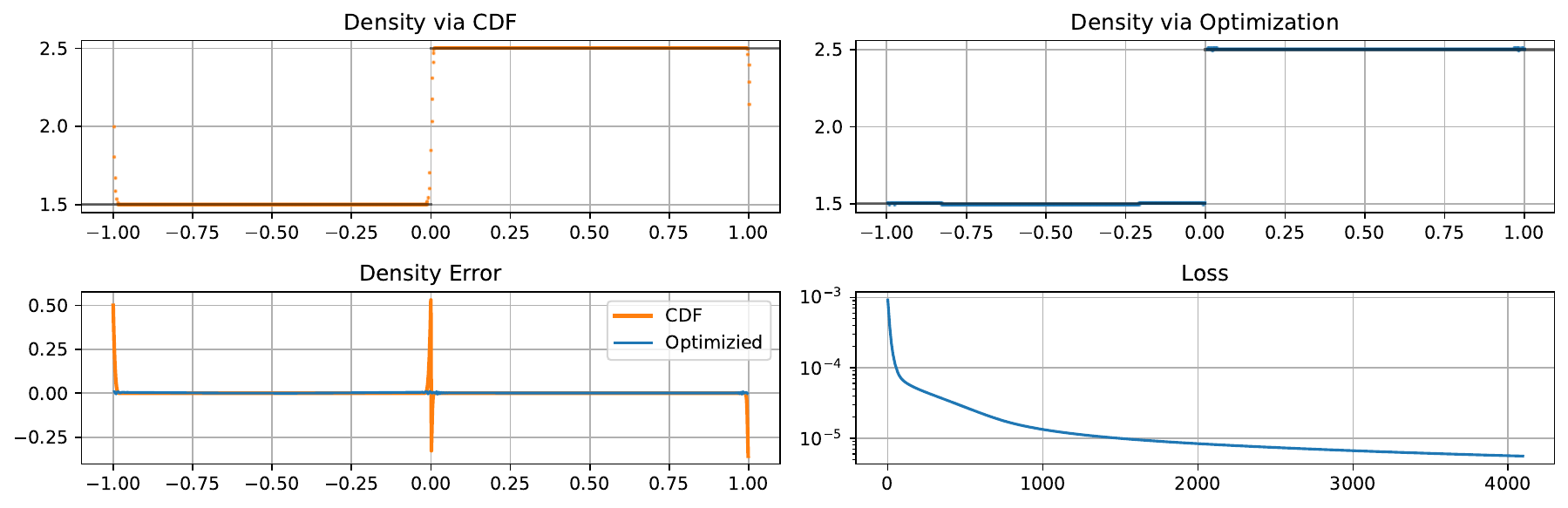}
    \caption{Results for generating initial conditions for a sinusoidal density profile $\rho(\mathbf{x}) = 2 + \frac{1}{2}\operatorname{sgn}{\mathbf{x}}$. The top left shows the inverse CDF sampling, the top right shows the result of using gradient optimization sampling, the bottom left shows the relative errors to the desired density profile, and the bottom right shows the convergence of the gradient-based sampling.}
    \label{fig:application:differentiable:1D_step}
\end{figure}

For 1D simulations a straightforward and, barring numerical errors, optimal strategy is using the close relationship of SPH with its foundations in statistical science and treating a desired density field as a probability density function~(PDF), integrating the PDF numerically into its cumulative density function~(CDF) and numerically inverting the CDF to find the inverse CDF that matches the given PDF~\cite{Diehl_Rockefeller_Fryer_Riethmiller_Statler_2015}.
Due to the properties of the inverse CDF, a regularly spaced sampling in the inverse CDF will yield a particle distribution with the given PDF, making it very useful in generating initial conditions.
However, this only works for certain simulation scenarios, e.g., sharp interfaces with symmetric SPH formulations can yield errors on the interfaces due to numerical errors inherent to SPH.

A potential solution to avoid these errors, which we are using here, is to instead define the density field as a target and optimize the particle positions with respect to the error on the density field.
This loss-based formulation gives a straightforward process, i.e., starting with an initially regularly sampled distribution of particles, the particle density $\rho_i$ is evaluated using a summation approach, the difference to the target density field $\hat{\rho}(\mathbf{x}_i)$ is computed, and then backpropagated via the MSE to the positions. We thus minimize
\begin{equation}
    \mathcal{L} = \frac{1}{n}\sum_{i}\left[\sum_{j\in\mathcal{N}_i}m_jW_{ij} - \hat{\rho}(\mathbf{x}_i)\right]^2.
\end{equation}
This results in the distribution and error shown in Fig.~\ref{fig:application:differentiable:1D_sin} for a simple sinusoidal density profile $\rho(\mathbf{x}) = 2 + 0.1\sin{\left(4\pi\mathbf{x}\right)}$, starting from a regular particle distribution.
While this indicates that the gradient-based optimization matches the results of the CDF, albeit with a convergence based on iteration count rather than sample points, the target density profile is smooth and easy to represent using SPH.
If we instead use a step function as a density profile, i.e., 
\begin{equation}
    \rho(\mathbf{x}) = \begin{cases}
        1.5, &\mathbf{x} < 0,\\
        2.5, &\mathbf{x} > 0,
    \end{cases}
\end{equation}
using the inverse CDF sampling yields a smoothed out density profile on the step, see Fig.~\ref{fig:application:differentiable:1D_step}, as the process is defined independently of SPH and does not take the kernel smoothing into account, nor a dependence of support radius on density.
The optimization process, however, is based on an automatic differentiation of the actual SPH operators and can take these effects into account.
Using the CDF-based sampling as the starting point and then optimizing using gradient descent using AD allows us to reduce the total absolute error from $4.02$ to $1.83$, a reduction by $2.2\times$, while reducing the maximum density error from $0.53$ to $0.01$, a reduction by $53\times$.

These results demonstrate that loss-based approaches using differentiability can achieve results comparable to or better than traditional approaches with minimal effort, i.e., implementing this process only takes around ten lines of code.
Note that the loss function used here might not be suitable in all cases, and using a regularization term, as often done in loss-based approaches, can significantly stabilize the optimization process.
Regularization terms, e.g., minimizing the difference between a particle's spacing and its direct neighbors compared to the mean particle spacing, are a potential choice we will explore in the two-dimensional case, where the loss landscape is more challenging.

\begin{figure}[t]
    \centering
\begin{subfigure}{.45\textwidth}
  \centering
  \includegraphics[width=\linewidth]{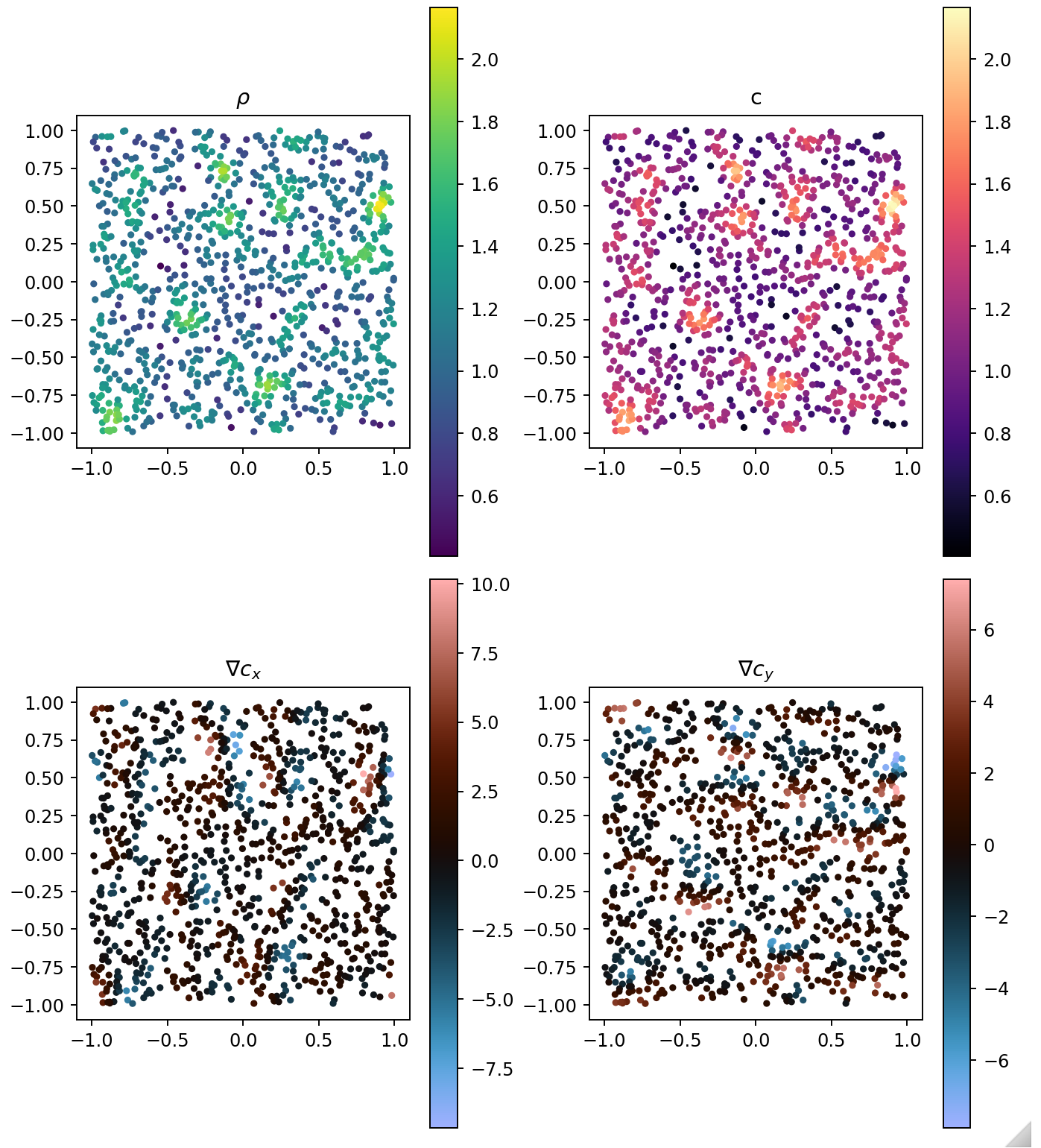}
  \caption{Initial Conditions}
  \label{fig:application:differentiable:initial}
\end{subfigure}%
\begin{subfigure}{.45\textwidth}
  \centering
  \includegraphics[width=\linewidth]{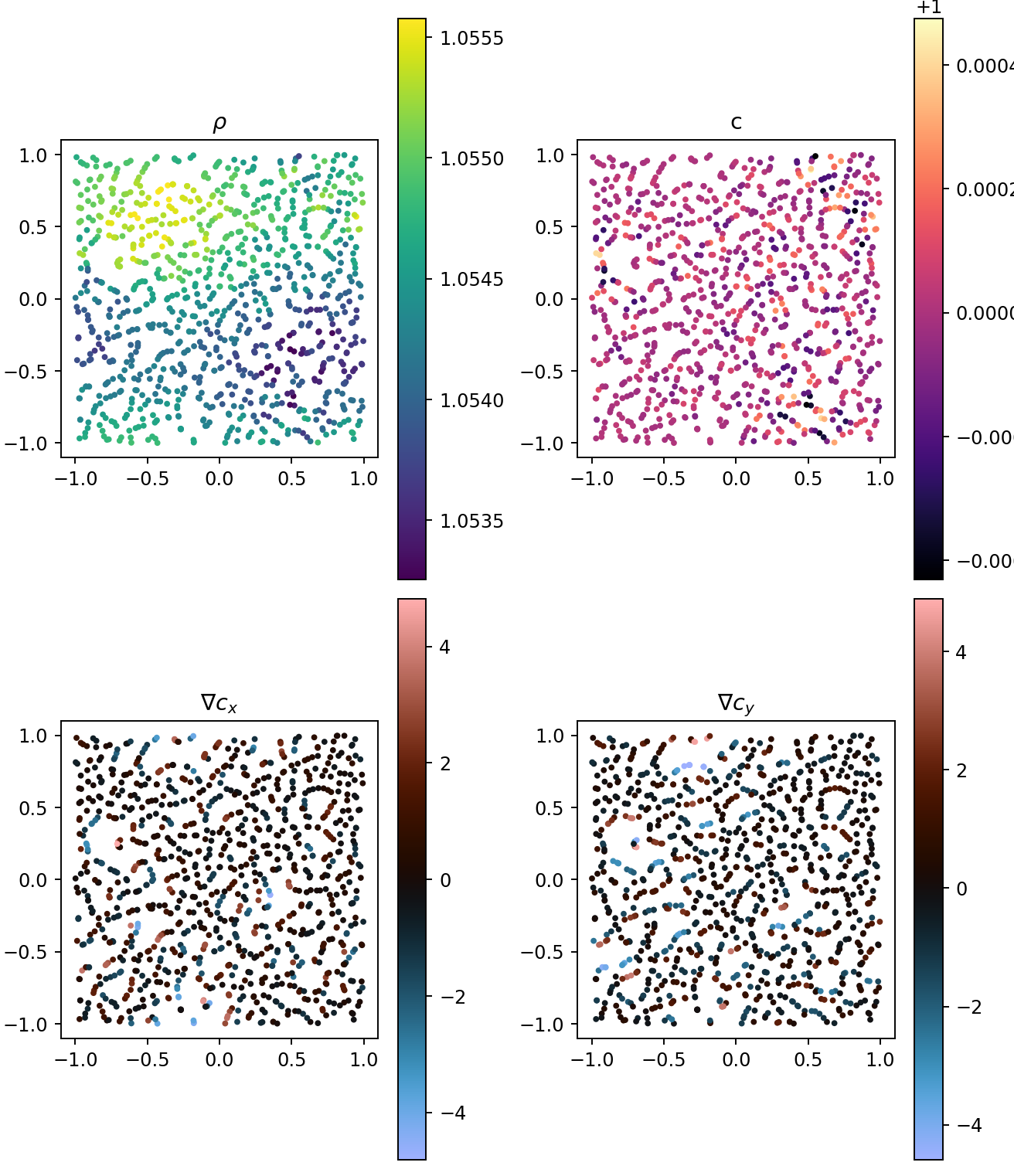}
  \caption{Optimized using MSE of color field}
  \label{fig:application:differentiable:color}
\end{subfigure}%
    \caption{This figure shows the initial random configuration of particles used to highlight the applicability of our framework for particle shifting. The right figure shows the result of using solely the MSE of $\langle1\rangle$, which leads to pairing instabilities.}
    \label{fig:application:differentiable:initial_color}
\end{figure}

\begin{figure}[t]
    \centering
\begin{subfigure}{.45\textwidth}
  \centering
  \includegraphics[width=\linewidth]{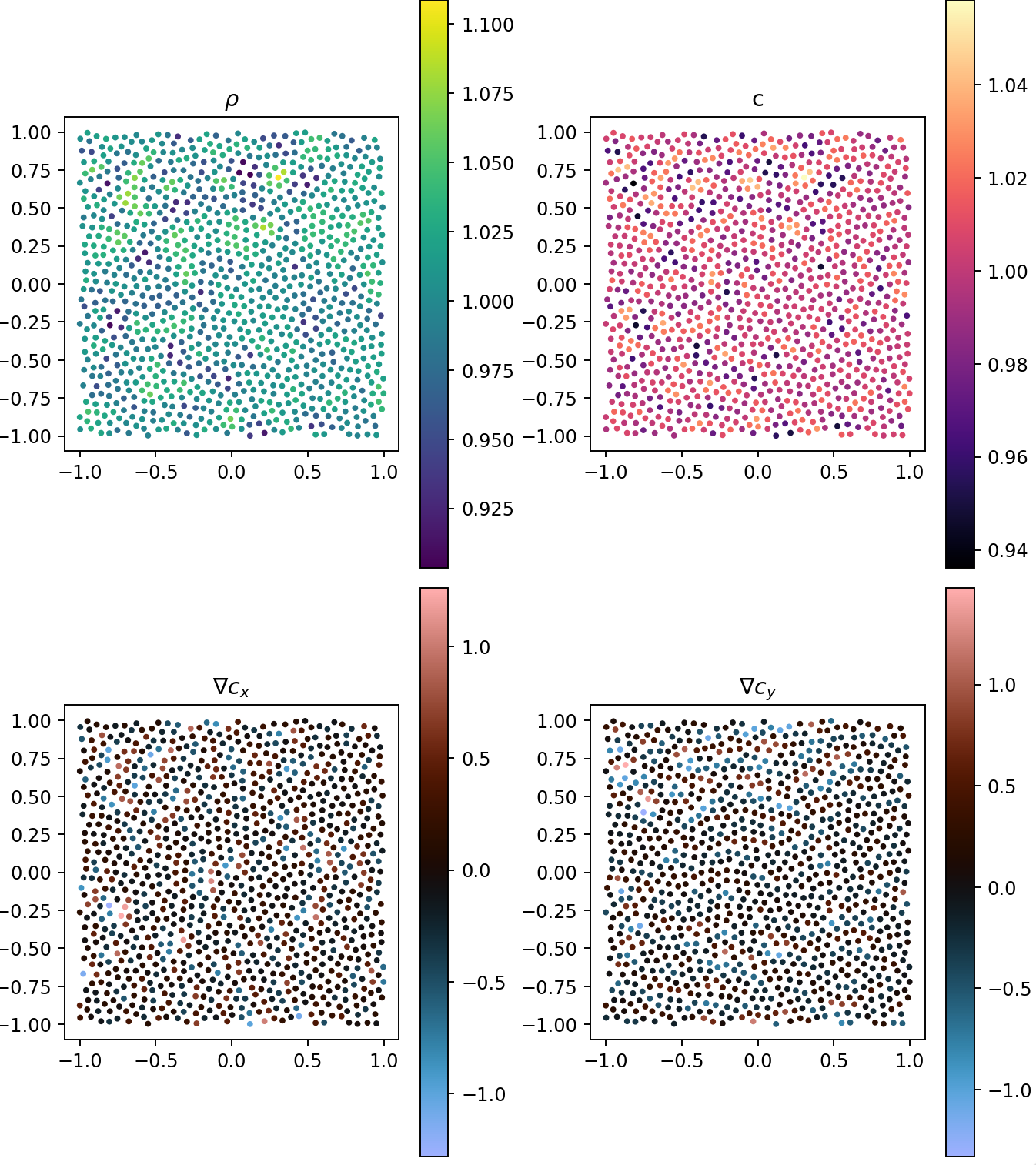}
  \caption{Using $\langle1\rangle$ and $\langle\nabla1\rangle$}
  \label{fig:application:differentiable:gradient}
\end{subfigure}%
\begin{subfigure}{.45\textwidth}
  \centering
  \includegraphics[width=\linewidth]{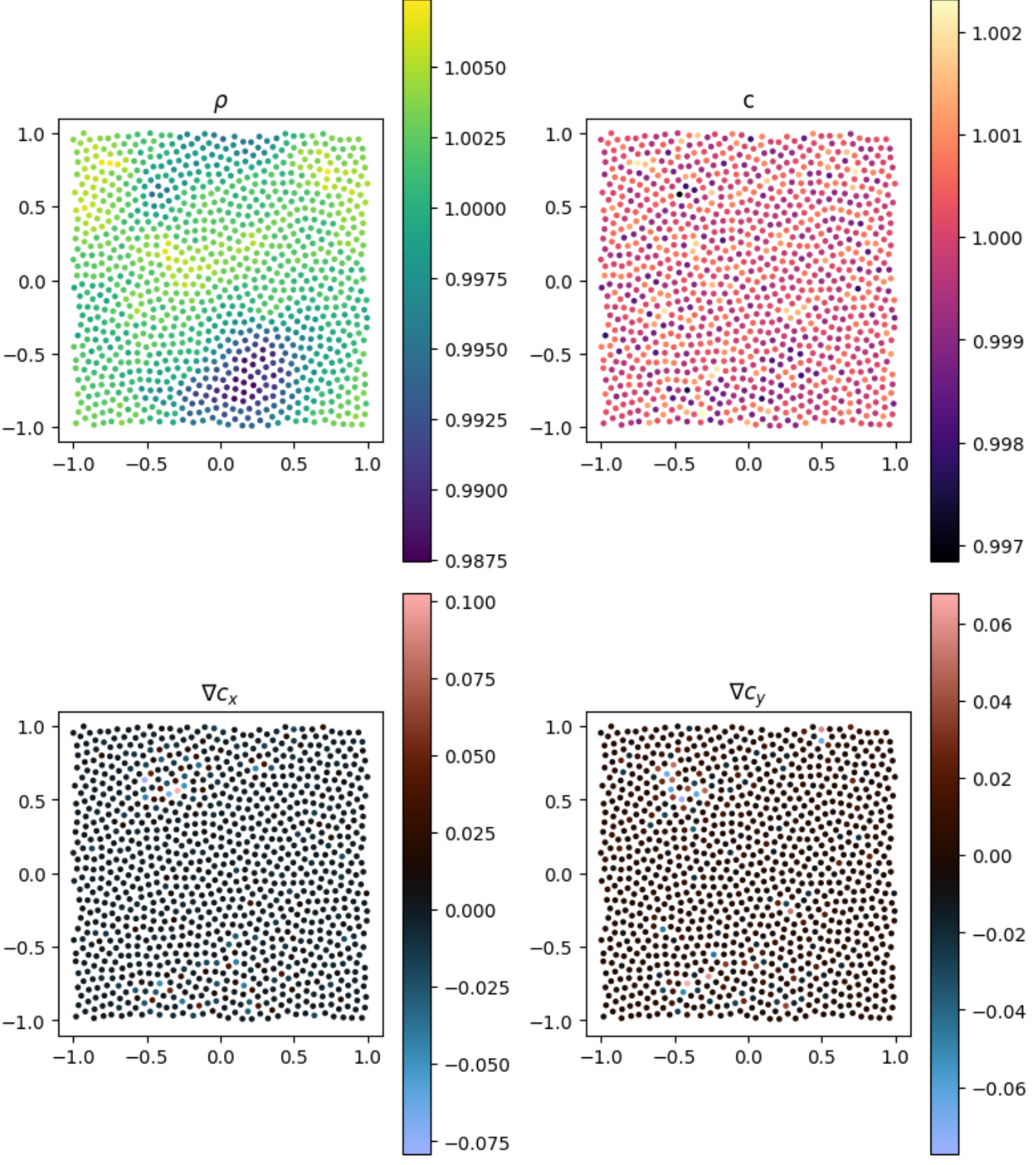}
  \caption{Using an additional regularizer}
  \label{fig:application:differentiable:regular}
\end{subfigure}%
    \caption{This figure shows a comparison of using the gradient of the color field as an additional loss term (left), as well as adding a regularization term (right). This highlights the benefit of regularization terms, which are 
    non-trivial to include in a 
    linear solve.}
    \label{fig:application:differentiable:2Dfinal}
\end{figure}

\begin{figure}[t]
    \centering
    \includegraphics[width=0.9\linewidth]{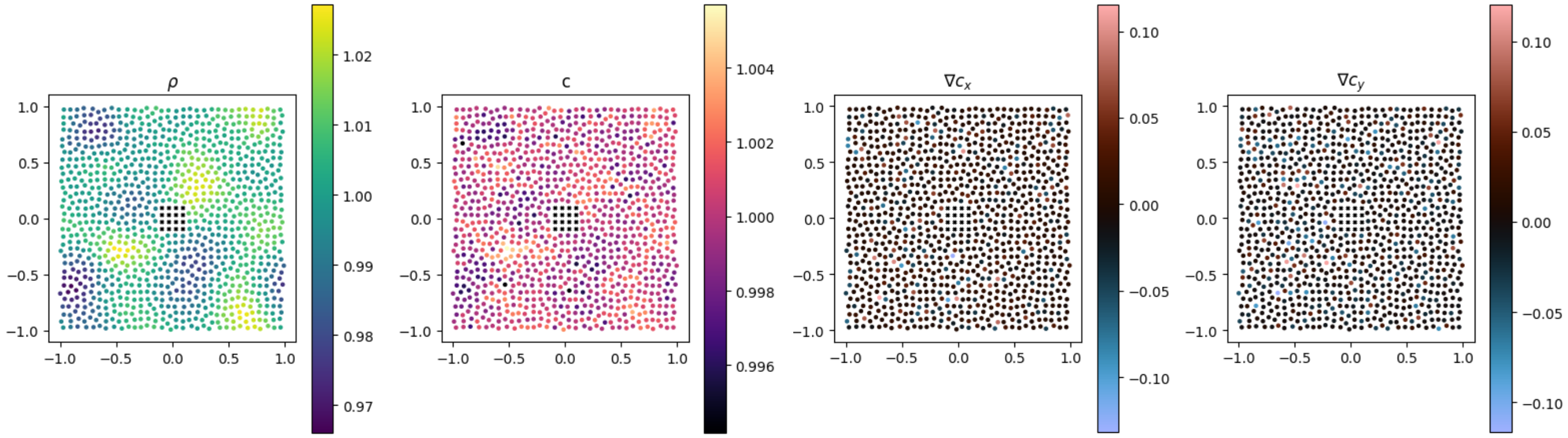}
    \caption{This figure shows the result of adding a boundary obstacle to the domain, where our loss-based approach can be easily modified to support this task.}
    \label{fig:application:differentiable:boundary}
\end{figure}

\subsubsection{Particle Shifting via Differentiable Physics}
\label{sec:application:shifting:2D}
To demonstrate how physical loss terms can be used to find optimal initial conditions and how the choice of loss term influences the resulting distribution, we focus on a straightforward task in 2D.
For this problem, we initially sample the domain $[-1,1]$ with $n=32^2$ particles of equal mass, see Fig.~\ref{fig:application:differentiable:initial}, using uniform random sampling. 
A naïve loss formulation to obtain the initial distribution would be to minimize error on the SPH interpolation of a constant field, i.e., $\langle1\rangle=1$, also known as the color field $c_i = \langle1\rangle$. This can be evaluated in a two-step process, i.e., we first compute a summation density for each particle $\rho_i$ and then evaluate the MSE loss
\begin{equation}
    \mathcal{L} = \frac{1}{n}\sum_i \left[1-\sum_j\frac{m_j}{\rho_j}W_{ij}\right]^2,
\end{equation}
and then minimize this loss term using the particle positions.
However, doing this results in a non-optimal distribution, see Fig.~\ref{fig:application:differentiable:color}, due to an issue commonly referred to as particle pairing~\cite{ihmsen2013implicit,muller2003particle,dehnen2012improving}, where particles that are initialized very close to each other will be close enough to where the gradient magnitude is not sufficient to separate the particles, leading to a pseudo-merging of particles~\cite{DBLP:journals/tog/WinchenbachHK17}.
We can avoid this issue by not just including the SPH interpolation of the quantity, but also optimizing the particle positions to lead to a zero gradient, i.e., $\nabla1=0$, yielding:
\begin{equation}
    \mathcal{L} = \frac{1}{n}\sum_i \left[1-\sum_j\frac{m_j}{\rho_j}W_{ij}\right]^2 + \frac{1}{n}\sum_i\left[\sum_j\frac{m_j}{\rho_j}\nabla_iW_{ij}\right]^2,
\end{equation}
which gives the result shown in Fig.~\ref{fig:application:differentiable:gradient}.
While this result is a notable improvement over the previous term, it is not optimal.
Adding a regularization term, i.e., by adding a penalty term if particles are closer than a fraction of the initial particle spacing $\Delta x$ (if the sampling had been uniform), and using only the gradient of the color field, results in a much more optimal sampling, see Fig.~\ref{fig:application:differentiable:regular}.
Using only the gradient is closely related to the implicit particle shifting technique of Rastelli et al.~\cite{rastelli2022implicit}. However, instead of requiring a custom, pre-conditioned linear solve, the \emph{diffSPH} functionality allows for solving this problem with a very short and simple program.

Furthermore, while it is possible to add boundary handling to the implicit approach~\cite{rastelli2022implicit}, it tends to be complicated to account for boundary particles, as this requires modifying the linear system.
Contrarily, implementing boundary particles in the \emph{diffSPH} optimization involves the masking of the gradient of boundary particles, which can be implemented in a single line of code. This results in the expected optimal distributions, see Fig.~\ref{fig:application:differentiable:boundary}.

\subsubsection{Extension to 3D}
\label{sec:application:shifting:3D}
\begin{figure}
    \centering
\begin{subfigure}{.4\textwidth}
  \centering
  \includegraphics[width=\linewidth]{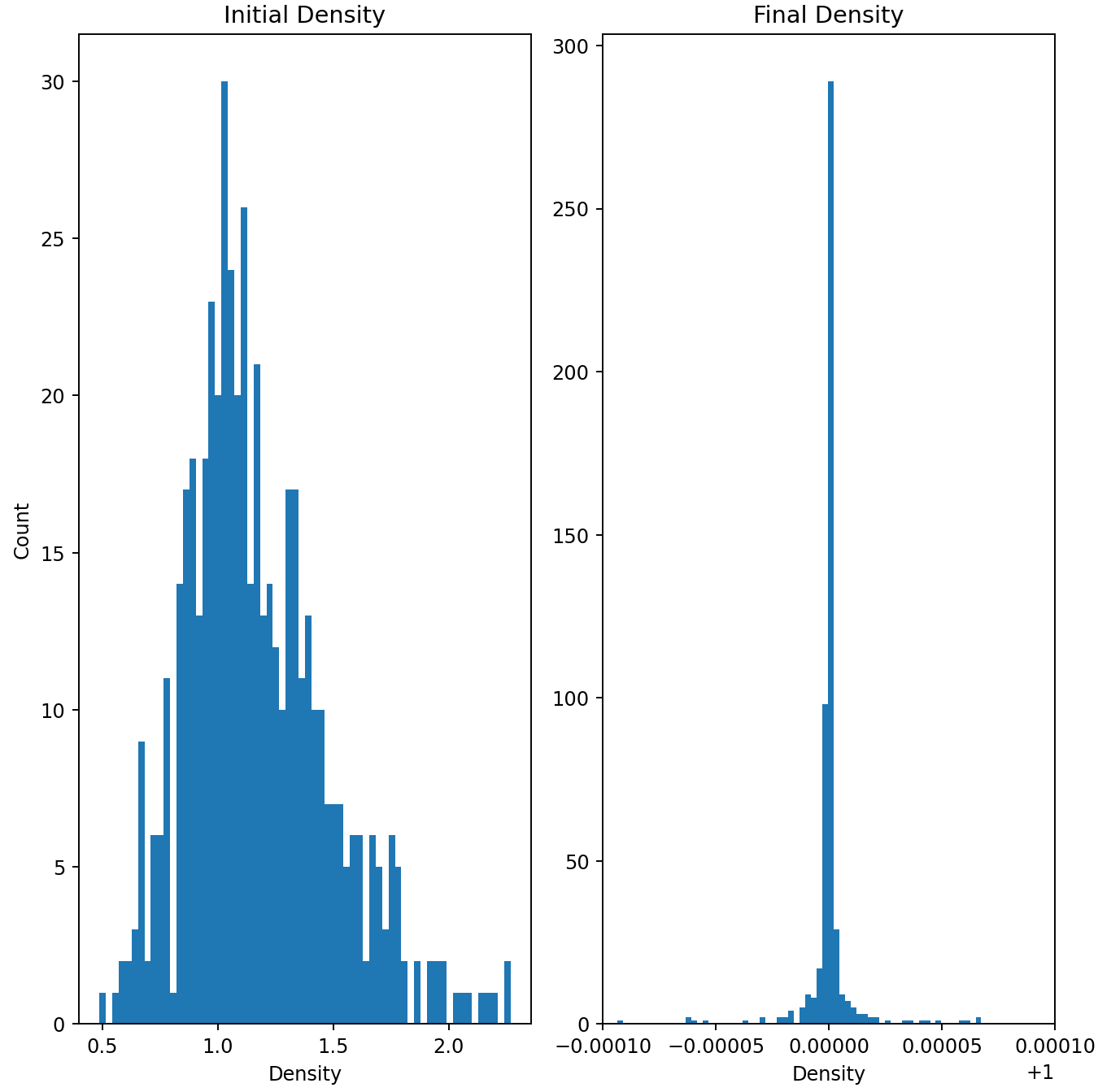}
  \caption{Density Distribution before and after optimization}
  \label{fig:application:differentiable:3Ddensity}
\end{subfigure}%
\begin{subfigure}{.4\textwidth}
  \centering
  \includegraphics[width=\linewidth]{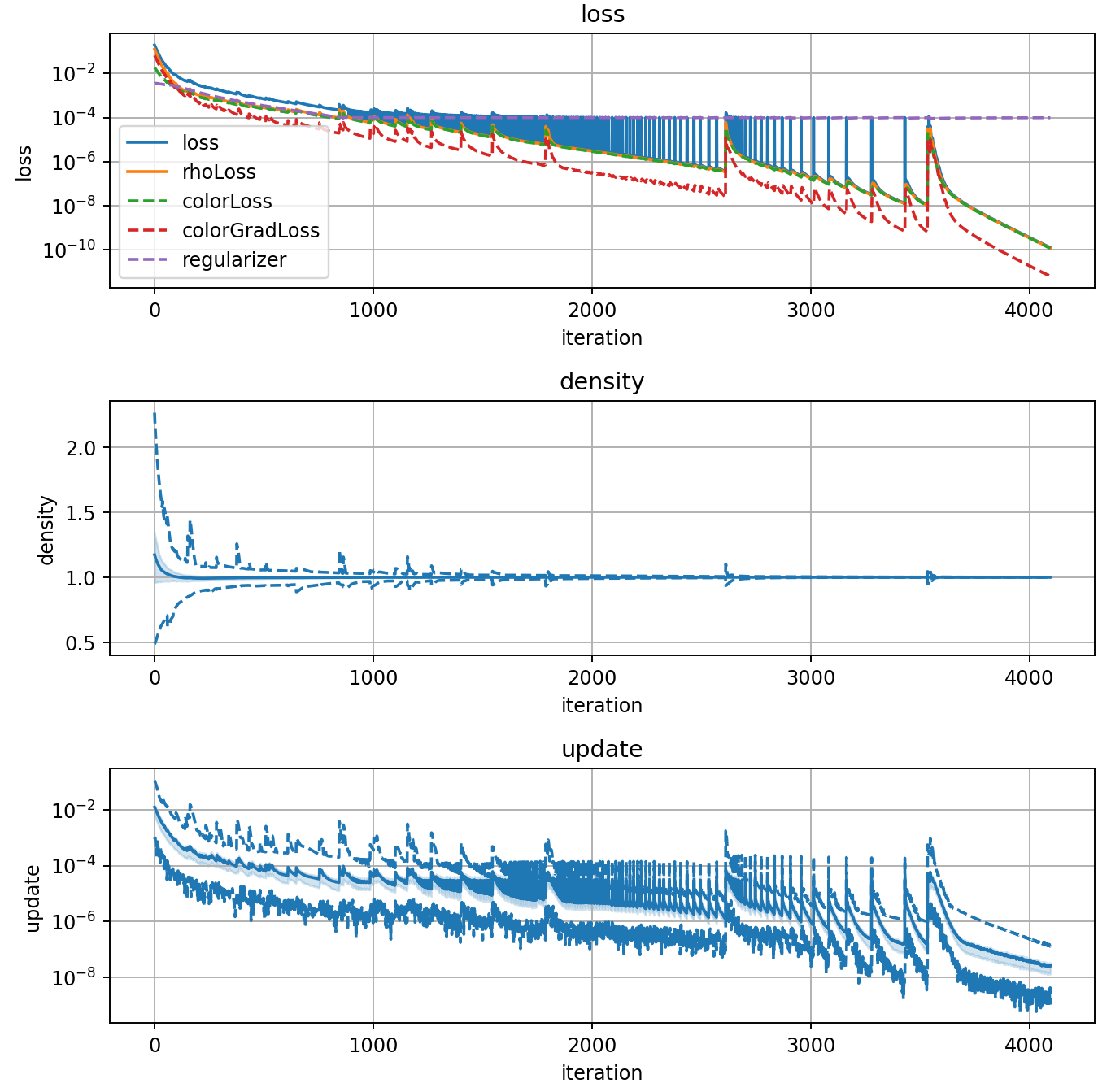}
  \caption{Convergence of the optimization process}
  \label{fig:application:differentiable:3Dconv}
\end{subfigure}%
    \caption{This figure shows the result of performing the optimization task in 3D by highlighting the achieved density error (left) as well as the convergence of the optimizer (right).}
    \label{fig:application:differentiable:3D}
\end{figure}

As a final application of our framework, we extend the previous example to 3D using the same loss terms and regularization without any specific changes for this case.
while the initial particle sampling had a density variation from $0.5$ to $2.5$, the optimized sampling has a variation of less than $\pm 0.00005$, 
as shown in Fig.~\ref{fig:application:differentiable:3Ddensity}.
On the other hand, the convergence rate of the process, see Fig.~\ref{fig:application:differentiable:3Dconv}, highlights that the convergence of this approach is not optimal and requires a larger number of iterations to converge. 
While convergence could be improved with extensions such as adaptive step sizes, these examples already highlight the potential of differentiable approaches to utilize different and novel problem formulations, e.g., using error metrics that are difficult to formulate with classic solvers, and their ability to match the quality of computationally more expensive approaches, highlighting an exciting area of future research.

%% file: 05-discussion.tex
\section{Conclusion}\label{sec:discussion}
In this paper, we have presented our \emph{diffSPH} framework and its applicability to a broad 
range of ML-related tasks in the context of SPH simulations.
Our framework implements a wide set of capabilities to simulate problems using compressible, weakly compressible, and incompressible schemes and can be readily extended to support other schemes, as demonstrated by the wave equation cases.
For each sub-field of SPH, our solver provides a wide range of schemes and operators, and can be easily extended to implement other SPH schemes.
Due to the inherent design choices, the \emph{diffSPH} framework can also be applied to schemes beyond traditional SPH, e.g., using Eulerian SPH-like approaches, as demonstrated in the wave equation example.

In addition to these forward schemes, \emph{diffSPH} was designed from the ground up around differentiability, e.g., for the neighborhood search and choice of integration schemes.
This provides a strong foundation for inverse problems, such as optimization of physical parameters to match reference trajectories, shape optimization, training of hybrid neural network models, and the development of novel loss-driven simulation schemes.
We have also demonstrated how our solver can be used to differentiate through hundreds of timesteps, e.g., up to 832 steps in Section~\ref{sec:application:inverse} and up to $65536$ particles in Section~\ref{sec:application:targeted}, showcasing the potential of our solver for a wide range of application scenarios, especially in the ML area.

Overall, our framework enables the simulation of millions of particles, using compressible and weakly compressible simulation schemes, with backpropagation through trajectories over time. This is provided via an extensible, flexible, high-level language with full GPU support and differentiability. 
Based on the results presented above, several interesting avenues for future research can be identified.
One is to further reduce the overhead from high-level abstractions in Python for small-scale problems. 
This could potentially be addressed by just-in-time compilation and additional custom operators for bottleneck operations.
In addition, optimizing general, non-linear inverse problems in SPH settings poses interesting challenges in terms of loss formulations and optimization techniques.
In this context, our results show that \emph{diffSPH} framework can provide an essential building block for developing solutions.

%% file: A0-appendix.tex
\input{A1-operators}
\input{A2-neighborhoods}

\input{A3-timestepping}

\input{A4-DSL}
\input{A5-00-validation}

%% file: A1-operators.tex
\section{SPH Kernel Operators}\label{appendix:operators}

This appendix provides the detailed mathematical formulations of the fundamental SPH operators implemented within the \emph{diffSPH} framework. 
The core of SPH lies in the approximation of field quantities and their derivatives through kernel-based interpolation over neighboring particles. The specific forms of the kernel functions, their spatial derivatives, and the resulting SPH discretizations for gradient, divergence, curl, and Laplacian operators are crucial for both the accuracy of the physical simulations and the correct propagation of gradients in the context of automatic differentiation. 
The following sections detail these formulations used in our framework for completeness and to centralize definitions from across the various SPH subfields our framework touches upon.

\subsection{Kernel Functions and Their Derivatives}

Kernel functions are at the core of SPH and are an important aspect to the numerical accuracy and usability of the overall scheme~\cite{dehnen2012improving}.
Accordingly, many different kernel functions have been proposed in prior research, with the respective subfields of research tending to use different kernel functions.
However, as noted before, several different approaches exist to define kernel functions and their terminology in SPH, and this subsection will describe the conventions and kernels we used.
In general, we follow the kernel definition of Dehnen and Aly~\cite{dehnen2012improving}, where a kernel function is defined as
\begin{equation}
W(\mathbf{r},h)=\frac{C_d}{H^d}\hat{w}\left(\frac{|r|}{H}\right),
\end{equation}
where $\mathbf{r}\in\mathbb{R}^d$ denotes the distance, $h$ being the smoothing scale, $C_d$ a normalization constant and $\hat{w}$ being the actual smoothing kernel.
Here $H$ denotes the support radius, i.e., the distance the kernel becomes $0$, where $h$ and $H$ are related by a kernel-dependent scaling factor $\frac{H}{h}$.
For the usage of our framework and the derivation of SPH schemes, the specific conventions do not affect the outcome; however, when implementing an SPH framework, it is important to keep the choices consistent.
For memory and compute efficiency, we directly store $H$ for each particle as the \emph{support}, where we can back relate this to $h$ through the \emph{kernel scale} parameter $H/h$, and following conventions from Computer Animation~\cite{ihmsen2013implicit,DBLP:journals/tog/WinchenbachA020} and machine learning~\cite{toshev2023lagrangebench}, we chose to denote the support itself as $h$ from this point onwards.

While our framework implements a broad set of kernel functions, the most commonly used ones for the schemes described in this paper are the Wendland-4 Kernel and the B7-Spline Kernel.
The Wendland kernels are a family of polynomial, non-piece-wise, kernel functions with the Wendland-4 kernel given in 1D~\cite{dehnen2012improving} as
\begin{equation}
    \hat{w}^{1}_{W4}(q)=\left(1-q\right)^5_+(1+5r+8r^2),
\end{equation}
with $\left(\cdot\right)_+=\max\{\cdot,0\}$, and in 2D and 3D as
\begin{equation}
     \hat{w}^{2,3}_{W4}(q)=\left(1-q\right)^6_+\left(1+6r+\frac{35}{3}r^2\right),
\end{equation}
with normalization constants in 1D, 2D and 3D as $\frac{3}{2}$, $\frac{9}{\pi}$ and $\frac{495}{32\pi}$, respectively, and kernel scales $1.936492$, $2.171239$ and $2.207940$.
The B7 kernel is defined as~\cite{hopkins2015new,frontiere2017crksph}
\begin{equation}
\hat{w}_{B7}(q) = 56 \left(q-\frac{1}{4}\right)_-^7 - 28\left(q-\frac{1}{2}\right)_-^7 + 8 \left(q-\frac{3}{4}\right)_-^7 - \left(q-1\right)_-^7,
\end{equation}
with $\left(\cdot\right)_-=\min\{\cdot,0\}$, with normalization constants in 1D, 2D and 3D as $\frac{4096}{315}$, $\frac{589824}{7435\pi}$ and $\frac{1024}{105\pi}$, respectively, and kernel scales $2.121321$, $2.158131$ and $ 2.195775$, with the kernel scales being adapted from the quintic B-spline kernel~\cite{dehnen2012improving}.

For each kernel function in \emph{diffSPH} we then define the first through third derivative terms explicitly, e.g., $\frac{\partial}{\partial q}\hat{w}(q)$, with a general wrapper around each kernel function that computes the kernel function itself, as well as the kernel gradient
\begin{equation}
    \nabla W(\mathbf{r},h)=\hat{\mathbf{r}} \frac{C_d}{h^{d+1}}\hat{w}^\prime\left(\frac{|r|}{h}\right),
\end{equation}
the kernel Laplacian and Hessian, based on the definition in Sec.~\ref{sec:framework:operators}, as well as the derivative with respect to the support radius
\begin{equation}
\frac{d}{dh} W(\mathbf{r},h)=-\frac{C_d}{h^{d+2}}\left[hd\hat{w}\left(\frac{|r|}{h}\right) + \frac{|r|}{h}\hat{w}^\prime\left(\frac{|r|}{h}\right)\right],
\end{equation}
with several more terms, e.g., the spatial third-order derivative, a $d^3$ tensor, defined for backpropagation.
Note that all of these functions can be evaluated using jit script by passing the desired kernel type as an enum to the kernel wrappers.

Setting the support radius of a particle can then be done either by relating the number of neighbors to the SPH density~\cite{monaghan2005smoothed}, or by utilizing the kernel moment~\cite{owen2010asph}, where the latter scheme is popular in the Astrophysics community~\cite{michael2014compatibly,frontiere2017crksph}, but beyond the scope of what we can cover here.
Relating the particle density to the number of neighbors, however, can be done straightforwardly by relating the apparent volume $V_{\text{SPH}}=\frac{m}{\rho}$ of a particle to the volume of the support radius, e.g., $V_H=\frac{4}{3}\pi H^3$ in 3D, where the number of neighbors is their ratio, i.e.,
\begin{equation}
    N_H = \frac{V_H}{V_\text{SPH}},
\end{equation}
alternatively one can also use the initial particle sampling density $\Delta x$ to compute the apparent volume $V_\text{SPH}=\Delta x^d$, or the rest density and initial mass of a particle $V_\text{SPH}=\frac{m}{\rho_0}$.
This relationship can then be refactored to yield $H$ with respect to the desired number of neighbors, e.g., for 2D, we find 
\begin{equation}
    H = \Delta x\sqrt{\frac{N_H}{\pi}},
\end{equation}
using the initial particle sampling.
For the specific choice of $N_H$ see the discussions by Dehnen and Aly~\cite{dehnen2012improving}.
To ensure consistency across schemes, we adopt the neighborhood sizes of Owen~\cite{owen2010asph}, aiming for $N_H=8$ in 1D, $N_H=50$ in 2D, and $N_H=100$ in 3D, which results in $H\approx4\Delta x$ in 2D, which aligns with the common choice of $H/h=2$ and $h=2\Delta x$.
Note that choosing a different formulation, i.e., $V_{\text{SPH}}=\frac{m}{\rho}$, creates an iterative problem where the choice of support means that the density and support radius of a particle influence each other, which can be resolved either through fix point iteration or a Newton-Raphson iteration~\cite{monaghan2005smoothed}, with $f=h-h(\rho)$ and $f^\prime=\Omega$, with~\cite{monaghan2005smoothed}
\begin{equation}
\label{eqn:omega}
    \Omega_i = 1+ \frac{h}{\rho d}\sum_jm_j \frac{d}{dh} W(\mathbf{x}_{ij},h).
\end{equation}
Note that in our differentiable framework, if the support radius $h$ is dynamically updated based on density, this iterative process and its dependence on particle properties are incorporated into the computational graph, allowing gradients to flow through the support radius adaptation. 
This is handled automatically by AD when using a fixed-point solver, or can be accounted for explicitly by differentiating through the Newton-Raphson update.

\subsection{SPH Interpolation of Field Quantities}

Based on the Kernel functions, it is then straightforward to compute an SPH interpolation of a field quantity as described before in Sec.~\ref{sec:governing}, giving the fundamental SPH operator
\begin{equation}
\langle A_i\rangle =\sum_j\frac{m_j}{\rho_j}A_j W_{ij},    
\end{equation}
with $W_{ij}$ being determined as $W(\mathbf{x}_{ij},h_i)$ for Gather-SPH, $W(\mathbf{x}_{ij},h_j)$ for Scatter-SPH, $W(\mathbf{x}_{ij},0.5(h_i+h_j))$ for Symmetric-SPH and $\frac{1}{2}(W(\mathbf{x}_{ij},h_i) + W(\mathbf{x}_{ij},h_j))$ for the Super Symmetric form that is commonly used in compressible SPH~\cite{frontiere2017crksph}.
A related formulation to the SPH interpolant has been utilized by PESPH~\cite{hopkins2015new} and the Computer Animation community~\cite{band2018mls}, where $\frac{m_j}{\rho_j}$ is replaced by the apparent volume $V_j$ of a particle, which can be computed in different ways, e.g., by relating the apparent volume to the kernel moment~\cite{hopkins2013general}.
From a naïve perspective, this SPH interpolant requires gradients for backpropagation with respect to mass, density, field quantities, positions and support radii, i.e., $m_j$, $\rho_j$, $A_j$, $\mathbf{x}_{i}$, $\mathbf{x}_j$, $h_i$ and $h_j$, which are straight forward to compute.
However, as mentioned before in Sec.~\ref{sec:framework:operators}, this can be significantly simplified by precomputing $W_{ij}$, which simplifies the backpropagation step significantly as only a single backpropagation through the kernel $W$ is necessary and all gradient terms for an SPH interpolant are simple products of particle quantities.
Finally, the summation density of a particle can be evaluated either directly by inserting $\rho$ as the field quantity, i.e.,
\begin{equation}
    \langle \rho_i\rangle =\sum_j m_j W_{ij},  
\end{equation}
or by using a Shepard filter~\cite{english2022modified}, as
\begin{equation}
    \langle \rho_i\rangle = \frac{\sum_j m_j W_{ij}}{\sum_j\frac{m_j}{\rho_j} W_{ij}}.
\end{equation}

\subsection{SPH Gradient Operators}

Based on the SPH operator for a field quantity, evaluating a naïve gradient term is straightforward in SPH  yields
\begin{equation}
    \langle \nabla_i A_i\rangle =\sum_j\frac{m_j}{\rho_j}A_j \nabla_i W_{ij},
\end{equation}
however, due to particle disorder and numerical effects, this term is not numerically accurate, e.g., computing the naïve gradient of a constant field does not result in a zero gradient.
However, this can be resolved by explicitly evaluating the gradient of $1$ using the product rule, i.e., $\nabla_i (1\cdot A_i)$, yielding the difference formulation.
Several other gradient formulations exist that also ensure symmetric interactions, and \emph{diffSPH} supports the following formulations:
\begin{align}
    \langle \nabla_i A_i\rangle &=\sum_j\frac{m_j}{\rho_j}\left(A_j-A_i\right) \nabla_i W_{ij}& \text{(difference)},\\
    \langle \nabla_i A_i\rangle &=\sum_j\frac{m_j}{\rho_j}\left(A_j+A_i\right) \nabla_i W_{ij}& \text{(summation)},\\
    \langle \nabla_i A_i\rangle &=\rho_i\sum_jm_j\left(\frac{A_i}{\rho_i^2}+\frac{A_i}{\rho_j^2}\right) \nabla_i W_{ij}& \text{(symmetric)},\\
\end{align}
which can also be directly utilized to compute the curl and divergence of vector fields, e.g., $\nabla\cdot\mathbf{A}$.
Regarding divergence terms, \emph{diffSPH} also supports computing the divergence of higher order fields using either the first or last dimension of the tensor field, and computing the so called consistent divergence~\cite{antuono2021delta}, which uses $\frac{m_j}{\rho_i}$ instead of the usual $\frac{m_j}{\rho_j}$.
Note that the specifics of choosing a kernel formulation are beyond our scope here, and we refer the reader to the respective SPH schemes implemented for details on the choice of formulations, as we chose to adopt the respective terms from the literature.
Furthermore, certain terms, such as the divergence of the velocity field, are central to the evolution of an SPH scheme but may require very different formulations for different SPH schemes, e.g., when using a momentum formulation~\cite{marrone2011delta} or when using the divergence as part of a viscosity switch~\cite{cullen2010inviscid,hopkins2015new}.
Concerning backpropagation, all terms here function identically to the SPH operators and are straightforward to implement when precomputing the kernel gradients.

\subsection{SPH Laplacian Operators}
Analogously to the gradient operators, a naïve Laplacian operator is straightforward to define, but is numerically highly unstable~\cite{price2012smoothed} and is rarely utilized in an SPH scheme, except in special cases, such as the implicit particle shifting scheme~\cite{rastelli2022implicit}.
Instead, a difference of gradients is utilized, which uses the relative particle pair distance to compute a finite-difference gradient combined with an analytic first-order SPH gradient.
This formulation was initially popularized by Brookshaw~\cite{brookshaw1985method}, and has been widely adopted since with several other derivative formulations, see~\cite{price2012smoothed} for an overview.
The Brookshaw formulation is given as
\begin{equation}
\langle\Delta A_i\rangle = \sum_j \frac{m_j}{\rho_j} \left[A_j-A_i\right]\left(\frac{\mathbf{x}_{ij}\cdot\nabla_iW_{ij}}{|\mathbf{x}_{ij}|^2+\epsilon^2h^2}\right),
\end{equation}
where $\epsilon$ is a small value used to avoid singularities.
This term also has practical advantages in terms of backpropagation and derivative terms, as this avoids the explicit requirements for a third-order derivative of the kernel function, which can be computationally expensive and numerically intractable.
Note that the Laplacian terms are commonly utilized in the computation of viscosity terms, e.g., $\nu\Delta u$, but the specifics of the choice of viscosity term are beyond our scope here.
However, at this point, it is noteworthy that instead of providing the per-particle quantities $A_i$ and $A_j$ to the SPH operators, we can also directly provide a pairwise quantity $A_{ij}$ to the SPH operator.
This can be utilized for the computation of viscosity effects by computing the viscosity term $\Pi{ij}$~\cite{monaghan2005smoothed} separately and passing it along to the SPH operator, which also means that for backpropagation the SPH operator itself is unaffected, as the $\Pi_{ij}$ term appears as a direct input and AD ensures proper execution of the chainrule.

\subsection{Kernel and Gradient Correction Terms}

In addition to the explicit SPH operators defined thus far, \emph{diffSPH} supports a variety of correction terms that can improve the numerical properties of an implementation.
Specifically, and in addition to the Shepard filter that can be naïvely implemented for the density computation, \emph{diffSPH} supports grad-H correction terms, gradient renormalization matrices, and the CRKSPH correction scheme.

\textbf{grad-H} terms occur in a broad range of contexts in SPH and are a direct result of coupling the SPH density, i.e., via summation evaluation, to the support radius of a particle. 
Due to this coupling, the derivative of the support radius is not zero and needs to be explicitly accounted for. However, this process becomes challenging when non-gather formulations are utilized due to the complexity of tracking the derivatives in traditional SPH codes.
Note that it would be possible to utilize AD to automatically track these derivative terms exactly in \emph{diffSPH}, but it is beyond the scope of our discussions here.
For a gather SPH formulation, the corrective term was already described before in Eqn.~\ref{eqn:omega} and is applied inversely to any gradient quantities, e.g., the evolution of energy over time becomes
\begin{equation}
    \frac{du_i}{dt} = - \frac{p_i}{\Omega_i\rho_i^2}\sum_j m_j(\mathbf{v}_j - \mathbf{v}_i)\cdot\nabla_iW_{ij}.
\end{equation}

\textbf{Kernel renormalization} terms appear commonly in free-surface problems where the lack of explicit air-phase particles leads to an incomplete support for particles near the free-surface.
This is generally resolved by utilizing a correction matrix $L_i$ and applying it to the kernel, e.g., the corrected gradient would be~\cite{marrone2011delta}
\begin{equation}
  \langle \nabla_i A_i\rangle =\sum_j\frac{m_j}{\rho_j}\left(A_j-A_i\right) L_i\nabla_i W_{ij}.   
\end{equation}
This correction matrix $L_i$ is based on the inverse of the covariance of the local particle positions, i.e., it directly compensates for the sampling deficiencies of the support, where $L_i$ is computed as
\begin{equation}
L_i = \left[\sum_j\frac{m_j}{\rho_j}\left(\mathbf{x}_j-\mathbf{x}_i\right) \nabla_i W_{ij}\right]^{-1},
\end{equation}
where the matrix inversion is commonly performed as a pseudo-inverse for numerical stability.
Our \emph{diffSPH} framework implements an explicit analytic pseudo-inverse in 2D and utilizes the built-in batch pseudo-inverse of PyTorch for 3D and 1D and fully supports backpropagation through this operation as well.

\textbf{CRKSPH} utilizes the geometric moments (and their derivatives) of the kernel~\cite{frontiere2017crksph} to ensure exact reconstruction of linear problems.
Evaluating these terms requires the computation of two corrective terms, denoted as $A\in\mathbb{R}$ and $B\in\mathbb{R}^d$ in~\cite{frontiere2017crksph}, and their spatial gradients, requiring a total of $d^2+2d+1$ values to be stored per particle, and furthermore involves a more expensive kernel computation as it necessitates the computation of two kernel gradients per corrected kernel evaluation.
The exact equations for these terms can be found in~\cite{frontiere2017crksph}.
We do not precompute the corrected kernels and gradients; instead, we pass the CRK terms to each SPH operator.
This does increase the complexity per SPH operator, but these kernels are sparsely utilized, and storing the corrected kernels requires significantly more memory.
By re-computing them on-the-fly from the base correction terms $A$ and $B$, we trade a small amount of computation for a significant reduction in memory, which would be a significant bottleneck in larger-scale differentiable simulations.

\subsection{Boundary Conditions}
Boundary handling is a quintessential part of any SPH simulation, and in the past, a broad range of boundary handling schemes have been proposed for SPH~\cite{english2022modified,chiron2019fast,akinci2012versatile,leroy2014unified,DBLP:journals/tog/WinchenbachA020}.
The most widely utilized approach in CFD, and the one we chose to adapt, is using boundary particles sampled densely into the boundary domain, which are then reflected into the fluid domain as ghost particles for estimations of fluid quantities on the boundary.
Note that we can directly sample these boundary and their respective ghost particles using the underlying description of the boundary as SDFs, allowing easy use of level-sets for sampling in a differentiable and versatile manner.
This extrapolation is usually~\cite{english2022modified,band2018mls} performed using a first-order Taylor series approximation using the fluid field at the position of the ghost particles. 
This can be done using a linear system per particle~\cite{Liu2010}, where the linear system is defined as
\begin{equation}
\begin{bmatrix}
\langle1\rangle & \langle\mathbf{x}_{ij}\rangle^T\\
\langle \nabla 1\rangle & \langle\nabla\otimes\mathbf{x}_{ij}\rangle
\end{bmatrix} \cdot \begin{bmatrix}
f\\f^\prime
\end{bmatrix}= \begin{bmatrix}
\langle q\rangle\\
\langle \nabla q\rangle
\end{bmatrix},
\end{equation}
which yields a system of equations $A\mathbf{x} = b$, where the most widely used solution uses a pseudo-inverse of $A$.
Note that several special cases must be accounted for, e.g., to handle rank deficiency and particles with few neighbors.
Computing the individual terms of the matrix can be done straightforwardly using the previously described SPH operators, where we can use the SPH operators' ability to utilize pairwise information to efficiently compute the gradients and interpolations of positions.
Overall, this system is a $d+1\times d+1$ system where we can then extrapolate the quantity on the boundary node using the distance from the boundary node to its corresponding ghost node, $\mathbf{x}_{bg}$ as 
\begin{equation}
q_b = f + \mathbf{x}_{bg}\cdot f^\prime,
\end{equation}
which yields a direct, first-order, extrapolation from the fluid field into the boundary.
Note that for reflecting boundary conditions, we can directly utilize $f$ for the boundary node, and these extrapolation terms are also utilized for inflow and outflow regions to prevent instabilities from incomplete support~\cite{negi2020improved}.
Handling Dirchlet boundary conditions in \emph{diffSPH} is straightforward as we can define regions, using the same SDF+CSG approach used for the boundaries and fluid descriptions, to model regions that directly impose quantities on fluid particles either before the evaluation of the Equation of State used in the simulation and/or after the evaluation of the updates at the end of a timestep.

%% file: A2-neighborhoods.tex
\section{Neighborhood Searches}\label{appendix:neighborhoods}

Efficient and accurate neighbor identification is the computational backbone of any SPH simulation, often consuming most of the runtime. 
This appendix describes the neighborhood search implementation within the \emph{diffSPH} framework. Our approach is founded on a well-established compact spatial hashing scheme, which we have tailored specifically to the demands of a fully differentiable and extensible simulation environment. 
The following sections will elaborate on the key components of this system: the core hashing algorithm for cell indexing, the seamless integration of periodic boundary conditions via modulo arithmetic, the iterative process for handling adaptive particle support radii, and the strategy for implementing and updating Verlet lists to further reduce computational overhead.

For the overall data structure and algorithm for neighborhood searches, we utilize the approach of~\cite{DBLP:journals/cgf/WinchenbachK20}, which was initially designed for spatially adaptive SPH simulations in computer animation, which aligns well with the requirements of our framework, especially considering the adaptive support radius for compressible SPH simulations.
The overall approach works in a three-step process that can be briefly summarized as 
\begin{enumerate}
    \item Divide the simulation domain into cells of some size $c$ and assign each particle to a cell
    \item Construct a compact hash map to enable queries from physical space into the list of occupied cells
    \item Use the hash map to perform a neighbor query on adjacent cells for each particle
\end{enumerate}
When combined, this procedure results in a data structure with memory consumption that, at worst, is $\mathcal{O}(n)$, where $n$ is the number of particles, and allows for neighborhood queries in constant time, due to the hash map.
Furthermore, this approach utilizes an efficient space-filling curve at its core, enabling good memory locality and performance. It also allows for the fast construction of multiple variants of the hash map to enable fast access for spatially adaptive simulations.
However, as initially proposed, this procedure does not work for compressible SPH simulations, does not naïvely support periodic boundary conditions, and was computationally expensive when multiple kinds of particles were used, e.g., fluid and ghost particles, as each kind of particle required a neighborhood query for each other kind.

The primary issue regarding compressible simulations is the usage of different SPH formulations for different parts of the SPH simulation, e.g., using a gather formulation for the adaptive support radius but a scatter formulation for pressure forces.
Resolving this involves an additional step, which can be combined with the first loop that is part of the neighborhood construction process, that assigns a search radius $s$ to each particle using atomic operations to ensure that $s_i$ for every particle is as large as $h_k$, where $k$ is the furthest particle from $i$ that would have $i$ as a neighbor in a scatter formulation.
We then construct the neighborlist for all particles, regardless of kind, in a single neighborhood query using $s_i$, and then sort the neighborlist based on the kinds of particles interacting, i.e., fluid-fluid, fluid-boundary, boundary-fluid, boundary-boundary, and fluid-ghost interactions.
This sorting enables tensor slicing operations to select the appropriate subset of neighbors for any SPH operation.
Furthermore, we can create masks, either on-demand or a priori, that select an appropriate subset of neighbors using a given SPH formulation.
This masking and slicing can also be utilized with a velocity verlet approach, where the neighborhood query is run with a larger support radius per particle to avoid recomputations in every sub-timestep.

Supporting periodic boundary conditions is more important as this requires a significant change in the simulation's design.
Two options are widely utilized: either particles are mirrored explicitly as ghost particles, or a minimum image convention is used.
The former approach is straightforward and flexible, but requires significant bookkeeping of ghost particles, including a varying number of total particles over time, and makes tracking gradients over time more challenging as some ghost particles become real particles and vice versa.
Similarly, in a ghost particle approach, the position of a particle is periodic, e.g., when a particle moves out of the right border of a simulation, its x-coordinate wraps around to the value of the left border.
This consequently means that the change of position over time is strongly decoupled from the particle velocity, which is a significant limitation for machine learning tasks.

Instead, for a minimum image convention-based approach, particles have an absolute position in space and perform all interactions in a modulo space.
This makes computations of distances between particles more expensive, as they now require several modulo operations, but avoids any explicit tracking of \emph{real} and \emph{ghost} particles.
Note that, as physics generally are translation invariant and all SPH operations are performed solely on relative particle positions, and not using absolute positions, this approach is straightforward to implement in an SPH simulation.
However, if certain SPH schemes are implemented, e.g., surface detection schemes using covariance matrices, the modulo distances must be utilized and not the absolute distances of positions.
To enable efficient implementation of such schemes, we store a list of all particle distances while constructing the neighborhood lists, similar to storing the kernel values, and make these values accessible during the simulation.
While this increases the simulation framework's overall memory consumption, it significantly simplifies implementing additional SPH approaches.

Finally, to reduce the computational cost of the neighborhood queries, we utilize Verlet lists to compute a neighbor list that is larger than necessary for an individual timestep.
Traditionally, these are utilized as velocity-Verlet lists, where the search radius for the neighborhood query is increased based on the CFL condition of the simulation, i.e., if due to the chosen CFL number a particle can move at most $0.4\times$ its support radius in one timestep, then performing a neighborhood query with $1.8h$ instead of $h$ for all particles ensures that in the next timestep the list of actual neighbors is a subset of the Verlet list.
Consequently, the neighbor list needs to be updated if any particle moves more than half of the skin-width, where it is important to compute this update with respect to the positions used during the original neighborhood query and not the previous timestep.
Note that the tracking of total particle motion is significantly simplified for periodic BCs when using the minimum image convention within our SPH framework.

In addition to changes in particle position, changes in the support radius of a particle can also lead to an invalidation of the neighborhoods.
Accounting for these changes can be done analogously to changes in particle position, i.e., when a single particle changes its support radius by a factor greater than half the skin-width, the neighborhood might have changed.
These two conditions, i.e., position and support radius changes, compound together multiplicatively and allow for a straightforward detection of a potential invalidation of the neighborhood of a particle.
Note that instead of recomputing the neighborhood of the potentially problematic particle, we fully recompute the neighborhoods for all particles due to the internal layout of the data structures and to remove potential iterative updates.
Several optimizations could be applied to this process, e.g., only updating the neighborhoods if there are particles moving towards each other, as parallel motion won't invalidate the neighborhoods, but by using a GPU-based neighborhood query, the computational gains would not be significant for the cases we consider in our studies.

%% file: A3-timestepping.tex
\section{Timestepping}\label{appendix:timestepping}

The chosen temporal integration scheme governs the evolution of the SPH system through time. 
This appendix details the time-stepping procedure implemented in \emph{diffSPH}, covering both the selection of a stable timestep, $\Delta t$, via various Courant-Friedrichs-Lewy (CFL) conditions, and the explicit sequence of operations that constitute a single simulation step. 
Our framework supports several standard explicit integrators, such as Symplectic Euler and multi-stage Runge-Kutta methods, which are implemented modularly using their Butcher tableau representations. 
This design facilitates straightforward extension to other schemes and ensures that the entire simulation loop is composed of differentiable operations, enabling end-to-end gradient propagation for optimization and learning tasks.
Note that the overall structure of our framework and the various pre- and post-simulation step processes align closely with the formal description of Pahlke and Sbalzarini~\cite{pahlke2023unifying}.
Consequently, extending our framework to the simulation of other PDEs and particle-based problems is straightforward, as demonstrated by implementing a wave equation using Eulerian particles.
Regarding the choice of timestep, we utilize three different approaches based on the specific fluid mechanics problem that we are using \emph{diffSPH} for.
For incompressible simulations, we utilize a CFL condition based solely on the maximum speed of any particle in the simulation $\mathbf{v}_\text{max}$ and the smallest support radius of any particle in the simulation $h_\text{min}$ as~\cite{ihmsen2013implicit}
\begin{equation}
    \Delta  t = \operatorname{min}\left(\text{CFL} \frac{H_i}{|\mathbf{v}_i|}\right),
\end{equation}
with a CFL constant of $0.4$.
For weakly compressible simulations, we utilize the canonical triplet of timestep conditions based on viscosity, acoustic effects, and the particle acceleration as 
\begin{align}
\Delta t_\nu &= \frac{1}{8} \frac{h^2}{\nu},\\
\Delta t_c &= \text{CFL} \frac{H_i}{c_s},\\
\Delta t_a & = \frac{1}{4}\sqrt{\frac{h_i}{\operatorname{max}_j \frac{d\mathbf{v}_j}{dt}}},
\end{align}
where the acoustic term using the speed of sound relates closely to the CFL condition for the incompressible simulation as, for weakly compressible simulations, $c_s\geq10|\mathbf{v}_\text{max}|$.
For compressible simulations, we utilize only the acoustic constraint~\cite{price2012smoothed}, where it is important to note that there is no direct relation between the speed of sound and the maximum velocity, i.e., there may be high Mach number effects that significantly limit the timestep size.

Regarding temporal integration schemes, we focus on explicit timestepping schemes that can be represented in the form of Butcher tableaus, e.g., Runge Kutta schemes, as well as Verlet schemes, such as the widely utilized Symplectic Euler scheme~\cite{dominguez2022dualsphysics}.
Similar to the formal description of Pahlke and Sbalzarini~\cite{pahlke2023unifying}, we utilize a particle state dictionary $\mathcal{P}$ that contains all persistent particle information, e.g., positions and velocities, as well as a global state dictionary $\mathcal{C}$ containing configuration parameters, e.g., the magnitude of kinematic viscosity.
Each simulation scheme now defines an update function $f:\mathcal{P}\times\mathcal{C}\rightarrow\mathcal{U}$ that returns the temporal derivatives of all quantities to be updated, e.g., $\frac{d\rho}{dt}$.
The neighborhood function $g:\mathcal{P}\times\mathcal{C}\times\mathcal{N}\rightarrow\mathcal{N}^\star$ is a part of the update function and implemented as described in Appendix Sec .~\ref {appendix:neighborhoods}, whereas the stopping function is implemented externally to the timestepping function.

Each timestepping scheme first performs a pre-processing step, which can be used to reset temporary state variables, and then creates a temporary particle state $\mathcal{P}^\prime$, implemented using tensor views with copy-on-write to avoid excess memory consumption, which is used to call $f$.
The results of $f$ are then utilized to integrate an intermediate state, and if a multi-step integrator is used, they are then utilized to evaluate $f$.
After all necessary substeps have been performed, the overall temporal integration is performed, followed by a post-processing step.
This post-processing step is where we evaluate particle-shifting, similar to the design choice in DualSPHysics~\cite{dominguez2022dualsphysics}, and update the internal energy for compatibly differenced schemes~\cite{michael2014compatibly,frontiere2017crksph}.
Note that to implement the modified temporal integration scheme of Owen~\cite{michael2014compatibly}, we return the update values of the last substep from the time integration function and re-utilize them as the update values for the first substep of the next timestep.

Overall, the temporal integration process performs the physical update part of the evolve function~\cite{pahlke2023unifying}, but does not update the global parameter state or handle inlets and outlets.
These steps are, generally, independent of the subtimesteps and are evaluated outside of the temporal integration scheme, i.e., the evolution of global parameters is performed after the temporal integration scheme is called.
Consequently, we utilize an overall update function that performs the evolution of the particle state $e$, combining the temporal integration scheme and inlet/outlet handling, and the evolution of the global parameter state $\mathring{e}$.
Overall, this design enables the implementation of SPH schemes and any general particle scheme and could be readily expanded to handle other simulation approaches such as Discrete Element Methods.

%% file: A4-DSL.tex
\section{C++ Extensibility}\label{appendix:DSL}

While \emph{diffSPH} is designed primarily in PyTorch to maximize usability and leverage its native automatic differentiation capabilities, certain computationally intensive, per-particle operations can become performance bottlenecks. 
Operations that are challenging to vectorize efficiently or where existing tools like torch.vmap have limitations for complex particle-wise computations necessitate an alternative path for performance optimization. To address this, we have developed a lightweight Domain-Specific Language (DSL) that significantly simplifies the integration of custom, high-performance C++ and CUDA components. 
The primary purpose of this DSL is to abstract away the repetitive and error-prone boilerplate code associated with creating Python bindings, a process that is often tedious to implement manually. Crucially, our DSL is designed with differentiability, providing a straightforward mechanism for users to define a corresponding backward pass for their custom kernels, ensuring seamless integration with PyTorch's autograd engine. 
This extensibility layer provides cheap GPU acceleration via CUDA for custom modules, yet it is intentionally kept simple and can be completely bypassed, allowing users to remain entirely within the PyTorch ecosystem if they so choose.
While a significant portion of this extensibility is a software engineering problem, there are practical cases where using compiled rather than interpreted code can be important.

The first case involves numerically important steps, i.e., in an interpreted and JIT compiled version of a numerical algorithm, there is no explicit and direct control over the operations used. In contrast, with compiled code, these aspects can be directly controlled.
The most direct example of this would be the utilization of fused multiply-add (FMAD) instructions that can preserve higher numerical accuracy for combined operations but are not something that an interpreted can safely utilize, due to the influence these instructions have on the results, and different JIT compilations may chose different operations to fuse.
On the other hand, with manually compiled code, there is complete control over such options, e.g., one may choose to utilize FMAD instructions only for very specific components to preserve numerical stability.
This kind of usage is straightforward with our DSL, as we can define an operation that performs an FMAD for three input tensors and provide appropriate gradients for this operation, which can then be utilized as part of other Python operations.

The second case involves computationally intractable operations in Python, with the most common example being iterative loops.
While PyTorch does offer some tools to speed up the performance of loops, such as vmap, these tools are often very restrictive in the code they allow and are not straightforward to utilize in practice.
However, iterative loops, such as over particles for a neighborhood search, are very easy to implement in C++/CUDA and can yield significant performance benefits on the order of many orders of magnitude.
In our framework, we utilize custom C++/CUDA backends for these operations in the neighborhood search, as these code components are crucial for performance, and do not require gradients (as the neighborhood adjacency is not differentiable by nature).
This can also be utilized to include legacy implementations, if no gradients are required, and can often yield significant performance benefits, e.g., implementing semi-analytic wall boundaries~\cite{DBLP:journals/tog/WinchenbachA020} leads to a 70000$\times$ speedup when using C++/CUDA instead of Python.
However, it is important to note that the core SPH operations, and using the SPH operators of our framework, can be done regardless of the implementation backend, and these extensions are for specific, narrow cases where computational performance is critical.

%% file: A5-00-validation.tex
\section{Framework Validation}\label{appendix:validation}

This appendix provides verification and validation for the Smoothed Particle Hydrodynamics schemes implemented within the \emph{diffSPH} framework. 
The primary goal of this section is not to re-validate the well-established numerical properties of schemes such as CompSPH or $\delta$-SPH, which have been extensively documented in the literature. 
Instead, this appendix serves two principal purposes. 
First, it demonstrates the breadth of physical problems that can be accurately simulated with our framework, spanning compressible, weakly compressible, and incompressible regimes. 
Second, and more importantly, it serves as a verification that our specific implementations reproduce the expected behaviors and accuracies.
This baseline is fundamental to the core contribution of our paper, as it ensures that the gradients obtained through automatic differentiation correspond to a correct and well-behaved underlying physical model. 
For each case presented, we will briefly define the problem setup, specify the relevant numerical parameters used in \emph{diffSPH}, and provide qualitative and quantitative comparisons against established results.

\input{A5-01-linear}
\input{A5-02-sodshock}
\input{A5-03-Rayleigh-Taylor}
\input{A5-04-TGV}
\input{A5-05-LDC}
\input{A5-06-Droplet}

%% file: A5-01-linear.tex
\subsection{Linear Wave Propagation}
\label{appendix:validation:compressible:linear}

\begin{figure}
    \centering
    
  \centering
  \includegraphics[width=\linewidth]{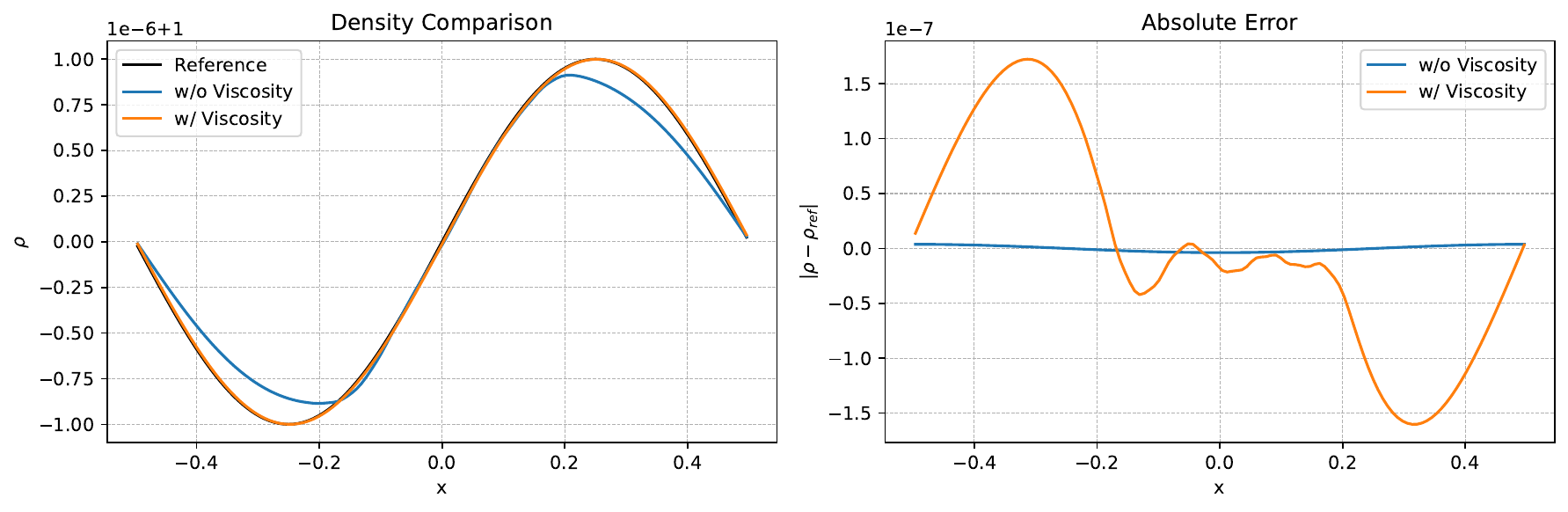}  \caption{Linear Wave Propagation test after $5$ wave periods. The left figure shows the absolute density values, whereas the right figure shows the error with respect to the initial conditions. Note that we utilize the standard viscosity formulation of Monaghan~\cite{monaghan2005smoothed}, which results in an asymmetric error for the propagation using viscosity.}
  \label{fig:LinearWavePeriod}
\end{figure}

The correct propagation of acoustic waves, and thus information, is an essential component to accurately simulating several phenomena, especially for compressible simulations.
To verify the correct propagation of acoustic waves, we utilize the linear wave setup described by Frontiere et al.~\cite{frontiere2017crksph}, where an acoustic wave of small magnitude is propagated through a periodic domain.
For this test case, the magnitude of the wave is chosen to be small to limit the effects to a purely linear propagation and requires high accuracy, i.e., double precision, to be simulated appropriately.
The basic setup of this case involves a periodic domain $[-1,1]$ sampled with particles in regular spacing with an initial wave amplitude $A=10^{-6}$, resulting in a pressure field of
\begin{equation}
P(\mathbf{x}) = c_s^2 \frac{\rho_0}{\gamma} + A \sin{\left(2\pi \mathbf{x}\right)},
\end{equation}
where $\delta_i = A\sin {\left(2\pi\mathbf {x}_i\right)}$ is the wave offset and the left-hand side is a background pressure term based on the chosen Equation of State.
These initial conditions result in a density field of $\rho_i = \rho_0 + \delta_i$ and a per-particle mass of $\tilde{m}_i = m_0 + \delta_i$.
However, this per-particle mass is not the exact mass that would yield the given density field, due to particle sampling and resolution, and thus we chose to employ a fixed-point iteration around 
\begin{equation}
    m_i^{n+1} = m_i^n \left(1 - \sum_j m_j W_{ij} + \rho_0 - \delta_i\right),
\end{equation}
which converges after a few iterations.
Based on these initial conditions, we can then also compute the internal energy of each particle.
The wave propagation then proceeds with the system's speed of sound $c_s$, chosen as $1$, and we compare the simulation state after a time period $T = 5 c_s$, which should exactly match the initial conditions in the absence of viscosity.
However, introducing viscosity into this simulation limits the wave propagation and results in errors specific to the viscosity formulation, which results in only first-order convergence~\cite{price2012smoothed}.
Utilizing a viscosity switch for this simulation~\cite{hopkins2015new} removes artificial viscosity in the absence of shocks, i.e., for this simulation, the artificial viscosity is completely turned off due to the linear behavior.
However, this still introduces some initial artifacts as most viscosity switch formulations initially have some viscosity and converge to some minimal level of viscosity, $\alpha_\text{min}$, instead of zero viscosity.
Without any viscosity, the expected result would be a second-order convergence due to the basic convergence property of SPH~\cite{Vacondio2021Grand}.

\begin{figure}
    \centering
    
  \centering
  \includegraphics[width=0.8\linewidth]{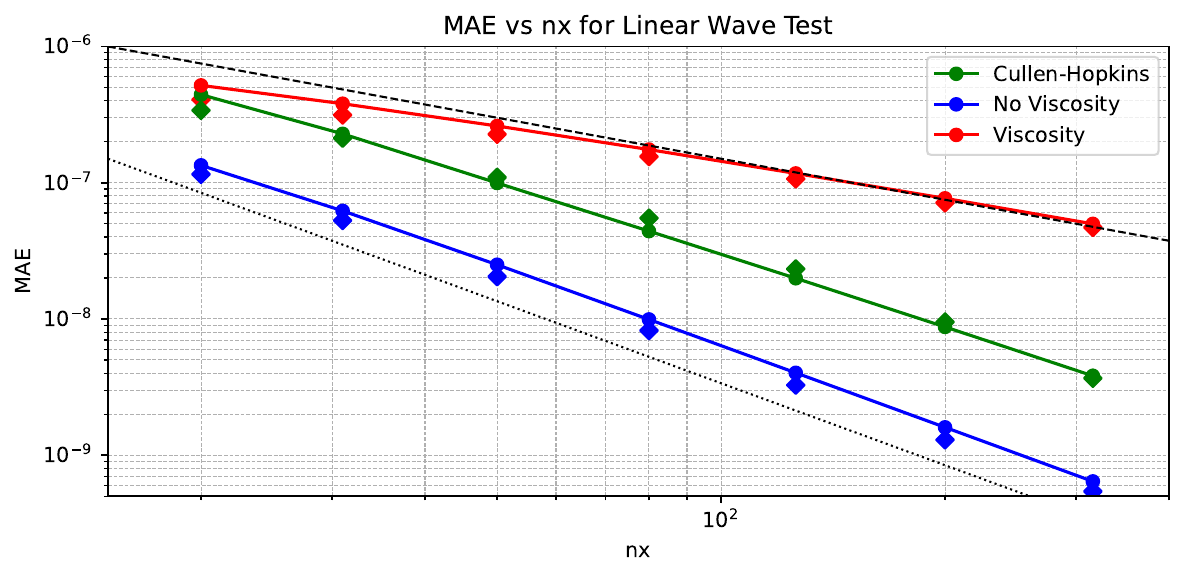}
  \caption{Convergence of the Linear Wave Propagation test for increasing resolution for a simulation with the standard Monaghan viscosity (red), the Cullen-Hopkins switch (green), and no viscosity (blue), with MAE denoting the mean absolute error across all particles. Circles are results obtained using our framework, diamonds indicate values taken from Frontiere et al.~\cite{frontiere2017crksph}, with the dashed line indicating $\mathcal{O}(n)$ convergence and the dotted line indicating $\mathcal{O}(n^2)$ convergence.}
  \label{fig:LinearConvergence}
\end{figure}

We now evaluate this validation case when using the standard artificial viscosity of Monaghan~\cite{monaghan2005smoothed}, using the viscosity switch of Hopkins~\cite{hopkins2015new}, based on the original Cullen and Dehnen switch~\cite{cullen2010inviscid}, and a formulation using zero viscosity, see Fig.~\ref{fig:LinearConvergence}.
With artificial viscosity enabled, the wave propagation is severely distorted, see Fig.~\ref{fig:LinearWavePeriod}, which limits the overall achievable accuracy.
Note that we utilize the standard Monaghan switch that turns off artificial viscosity for separating particle pairs~\cite{monaghan2005smoothed}, which leads to the observed asymmetric errors.
Without any viscosity, the wave after five time periods closely matches the initial wave.
Considering the order of convergence, see Fig.~\ref{fig:LinearConvergence}, we observe a close to linear convergence for the standard viscosity formulation and second order convergence for both the viscosity switch and viscosity free formulations.
Note that all results closely match the reported results from~\cite{frontiere2017crksph}.

%% file: A5-02-sodshock.tex
\subsection{Sod-Shock Tube}
\label{appendix:validation:compressible:sod}
\begin{figure}
    \centering
    \includegraphics[width=0.85\linewidth]{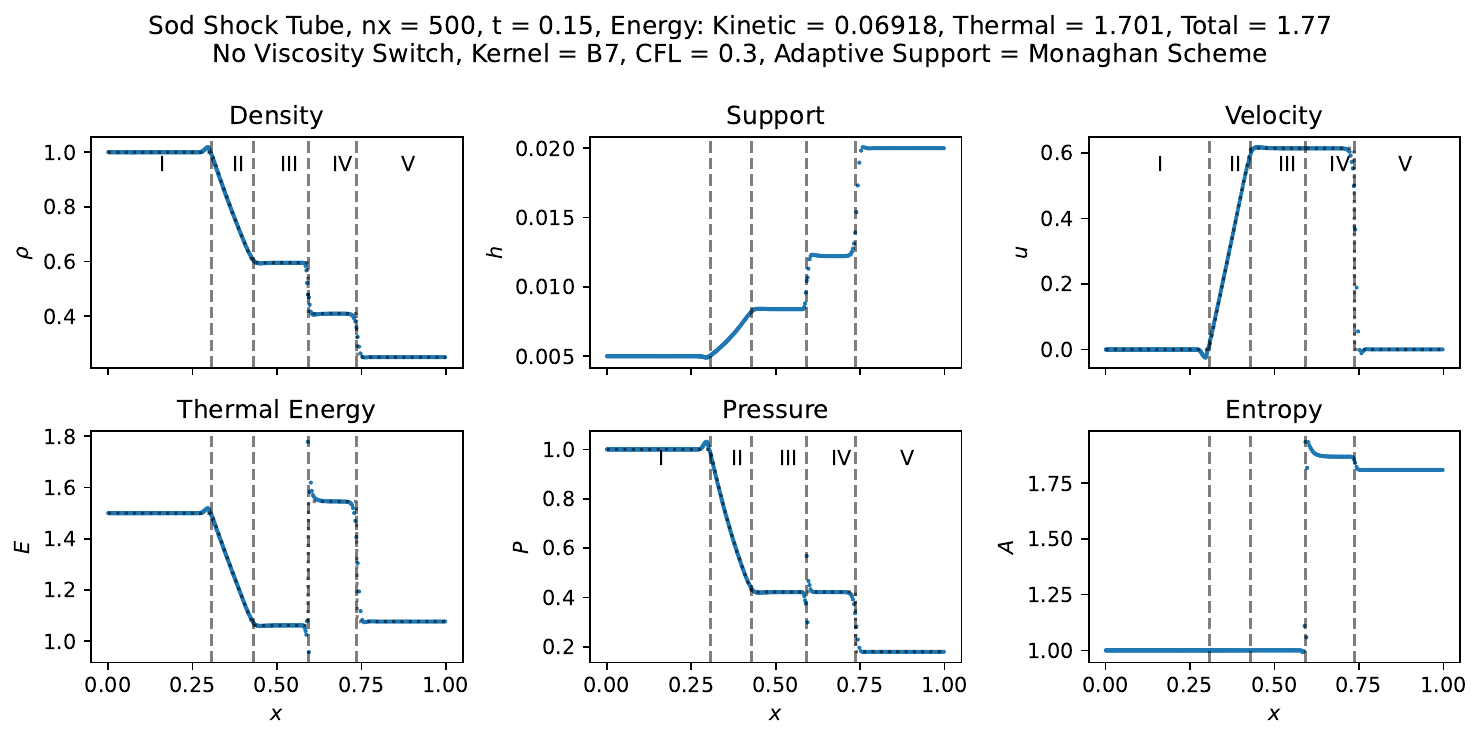}
    \caption{Results of running the Sod-Shock tube as described in Sec.~\ref{appendix:validation:compressible:sod}, for the compSPH scheme without a viscosity switch. The dotted line represents a reference solution obtained on a high-resolution grid, with the roman numerals indicating the different regions of the solution of the underlying Riemann problem.}
    \label{fig:Sod_Shoc}
\end{figure}

To verify the ability of our framework to handle shock capturing, we utilize the standard Sod Shock Tube setup that has been utilized in a wide variety of prior work~\cite{price2012smoothed,frontiere2017crksph,monaghan1994simulating}.
The problem represents a simple one-dimensional Riemann problem with two states, commonly referred to as the left and right initial conditions, with a discontinuity across the interface.
For our implementation, we utilize the initial left and right states
\begin{align}
    [P_l, \rho_l, \mathbf{v}_l] &= \left[1,1,0\right],\\
    [P_r, \rho_r, \mathbf{v}_r] &= \left[0.1795,0.25,0\right],
\end{align}
and an adiabatic constant of $\gamma = \frac{5}{3}$, in line with the description by Owen~\cite{michael2014compatibly}.
Within the SPH community, there are two choices for this case as either utilizing particles of equal mass and varying the initial particle spacing in each phase, or utilizing particles of varying mass but equal spacing.
For our evaluation, we utilize particles of equal mass and varying spacing, similar to Monaghan~\cite{monaghan2005smoothed}, as this also allows us to verify the correct handling of variable spacing and the resulting varying support radius in space, which results in a spacing of $\frac{1}{4}\Delta x$ for the right part of the simulation.
Some prior work on this case in SPH utilized smoothed initial conditions~\cite{price2012smoothed, michael2014compatibly}, but we chose to leave the discontinuity unsmoothed~\cite{monaghan2005smoothed}.
Running this case, see Fig.~\ref{fig:Sod_Shoc}, results in the expected behavior of the utilized scheme and is in close agreement with prior work.

%% file: A5-03-Rayleigh-Taylor.tex
\subsection{Rayleigh-Taylor Instability}
\label{appendix:validation:compressible:rayleigh}
\begin{figure}
    \centering
    \includegraphics[width=\linewidth]{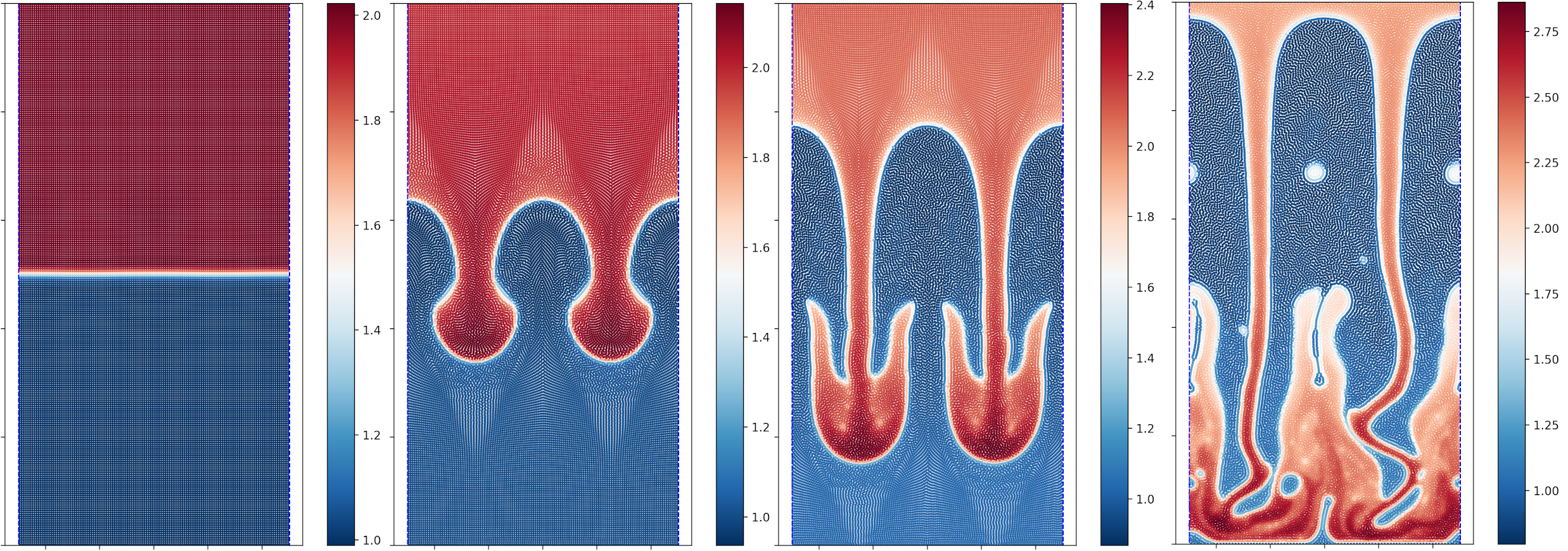}
    \caption{Results of the Rayleigh Taylor Instability using the compSPH scheme in \emph{diffSPH} at timepoints $t=[0,3,5,8]$ (from left to right). Color indicates per-particle density.}
    \label{fig:RayleightTaylor}
\end{figure}
As a final example of our framework for compressible fluid mechanics, we demonstrate its applicability to the simulation of the Rayleigh-Taylor~(RT) instabilities.
The instabilities in an RT instability grow from an interface between two fluids of different density in opposite layering, as would be expected due to gravity, i.e., the denser phase is on top of the less dense phase.
In the classic RT simulation setup~\cite{frontiere2017crksph}, the instabilities grow from an initial set of perturbations, usually due to a perturbation in the initial y-velocity field, that act as seed points for the instabilities.
In contrast to the Sod Shock Tube, we utilize smoothed initial conditions, i.e., the initial density is exponentially interpolated between the two phases at the interface to avoid any spurious shock effects from occurring.
The initial velocity field is then set up to result in two initial perturbations in the x-direction, leading to a symmetric simulation with periodic boundary conditions.
For the top and bottom of the fluid, we utilize Dirichlet boundary conditions that enforce the initial conditions as fixed values to model the infinite extent of the phases.
For the exact details on the setup, see Frontiere et al.~\cite{frontiere2017crksph} and their respective results.
For this case, see Fig.~\ref{fig:RayleightTaylor}, the results achieved with our framework align closely with those presented in related work~\cite{frontiere2017crksph}, and highlight the proper evolution of the instabilities.

%% file: A5-04-TGV.tex
\subsection{Taylor Green Vortices}
\label{appendix:validation:wcsph:TGV}
\begin{figure}
    \centering
    \includegraphics[width=\linewidth]{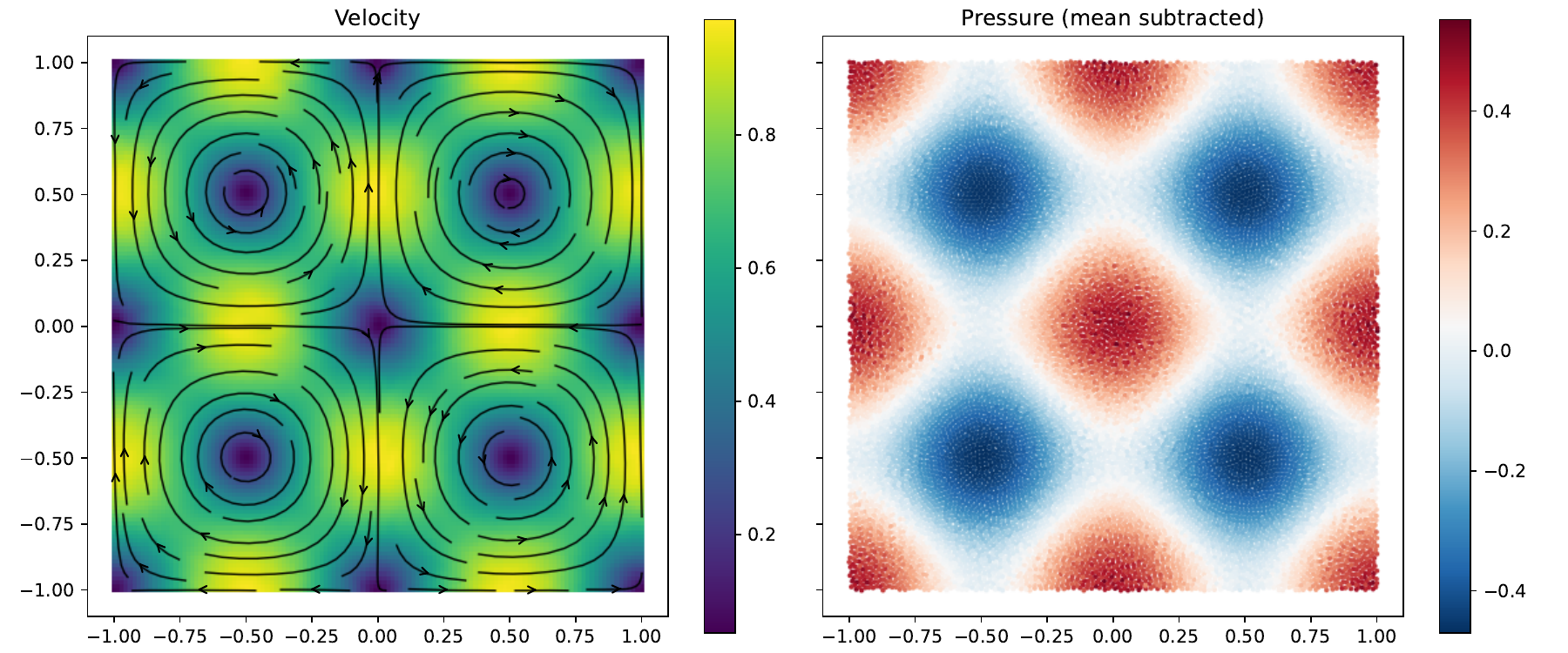}
    \caption{Results of running the Taylor Green Vortex simulation with $nx=128$ particles for four seconds. On the left, we visualize the particle velocity, which is remapped to a regular Cartesian grid to enable the visualization of streamlines. On the right, we visualize the pressure field with respect to the mean density. Due to the periodic boundary conditions of the simulation, the pressure is only determined up to a constant, and we, thus, remove any constant offset from the solution for visualization purposes.}
    \label{fig:TGV_wcsph}
\end{figure}

\begin{figure}
    \centering
    \includegraphics[width=0.5\linewidth]{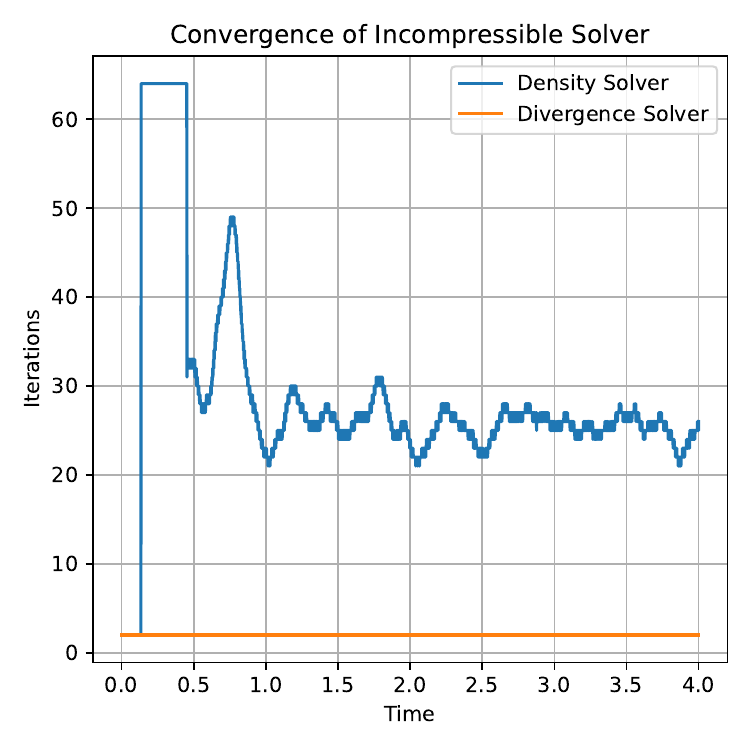}
    \caption{Visualization of the per-timestep number of solver iterations to solve the incompressible and divergence-free problems for the Taylor Green Vortex case when simulated using DFSPH~\cite{bender2015divergence}, using a maximum per-timestep iteration count of $64$. Note that due to a lack of impacts or other strong effects, the solver behavior in this case is dominated by the incompressible problem.}
    \label{fig:TGV_incomp}
\end{figure}

The Taylor Green Vortex~(TGV) case is a classic test case for weakly compressible and incompressible fluid mechanics and can be utilized to estimate the numerical viscosity and stability of a scheme without solid boundaries or free surfaces.
The initial conditions of the TGV case are a periodic domain $[-1,1]^2$ with particles at rest, density, and velocity
\begin{align}
\mathbf{v}_x = \mathbf{v} \cos \left(\frac{k\pi x}{2} + \theta\right) \sin\left(\frac{k\pi y}{2}\right),\\
\mathbf{v}_y = -\mathbf{v} \sin \left(\frac{k\pi x}{2} + \theta\right) \cos\left(\frac{k\pi y}{2}\right),
\end{align}
with $\theta$ being a phase offset set to $\frac{\pi}{2}$ for even $k$ to keep the simulation symmetric in our domain.
These initial conditions then give a total initial kinetic energy of the simulation $E_k^0=\sum_i \frac{m_i}{\rho_0}\rho_i |\mathbf{v}_i|^2$.
This energy then exponentially decays over time with a rate of $F(t) = e^{-4k^2\nu t}$, which can conversely be utilized to estimate the kinematic viscosity $\nu$ using the ratio of current kinetic energy to initial kinetic energy.

For our specific setup of speed of sound, resolution, and neighborhood size, the default choice of $\alpha=0.01$ for the artificial viscosity of the $\delta$-SPH model~\cite{antuono2021delta} results in a Reynolds number of $1853$, whereas our simulation after $4$ seconds achieves a Reynolds number of $1875$.
This achieved viscosity is close to the expected value and indicates no excessive viscosity present in our implementation, as excessive numerical dissipation would lower the measured kinematic viscosity.
This result is also strongly influenced by the particle shifting scheme utilized~\cite{sun2019consistent}, as without particle shifting, this simulation would become unstable due to anisotropic particle distributions, whereas a particle shifting technique can resolve these instabilities.
However, this particle shifting can also introduce excessive dissipation if the flow quantities are not properly handled after particle shifting, which makes this case a good test for the numerical dissipation due to particle shifting.

Note that the Reynolds number here is used solely to describe the initial decay of the TGV case, and that this measurement excludes any potential vortex decay towards the later stage of the simulation.
Qualitatively, our implementation, see Fig.~\ref{fig:TGV_wcsph}, shows a stable vortex location over time, which is influenced by the speed of sound utilized, and good agreement with the pressure magnitude that we expect from this simulation.
Note that we measure the relative pressure variation here as pressure in a periodic system is only determined up to a constant and the absolute pressure magnitude can drift over time in a weakly compressible simulation due to a drift in the momentum equation, however, this problem is not new or unique to our implementation and an effect of the underlying fluid mechanics.

We also utilized our implementation of the Divergence Free SPH scheme by Bender and Koschier~\cite{bender2015divergence}, with the modified source terms of Cornelis et al.~\cite{cornelis2019optimized}, that solves the incompressible Navier Stokes equations by projecting the velocity field to be divergence free and utilizes an incompressibility constraint as a particle shifting technique to stabilize the simulation.
In this case, the overall velocity field is divergence-free during the initial conditions and is temporally smooth and consistent.
Consequently, the projection of the velocity field to be divergence-free is straightforward, whereas the particle shifting problem due to the periodic nature of the flow is significantly more challenging.
In this case, see Fig.~\ref{fig:TGV_incomp}, we observe that the divergence-free step stops after two iterations, the minimum number of iterations~\cite{bender2015divergence}, whereas the incompressible step requires approximately $25$ iterations on average.
This behavior is expected from a shifting approach~\cite{rastelli2022implicit}, as the initial particle constellation is optimal, which is then perturbed until a threshold of perturbance is reached that the shifting approach can reduce, after which the amount of shifting required per timestep becomes consistent.

%% file: A5-05-LDC.tex
\subsection{Lid-Driven Cavity}
\label{appendix:validation:wcsph:LDC}

\begin{figure}
    \centering
    \includegraphics[width=\linewidth]{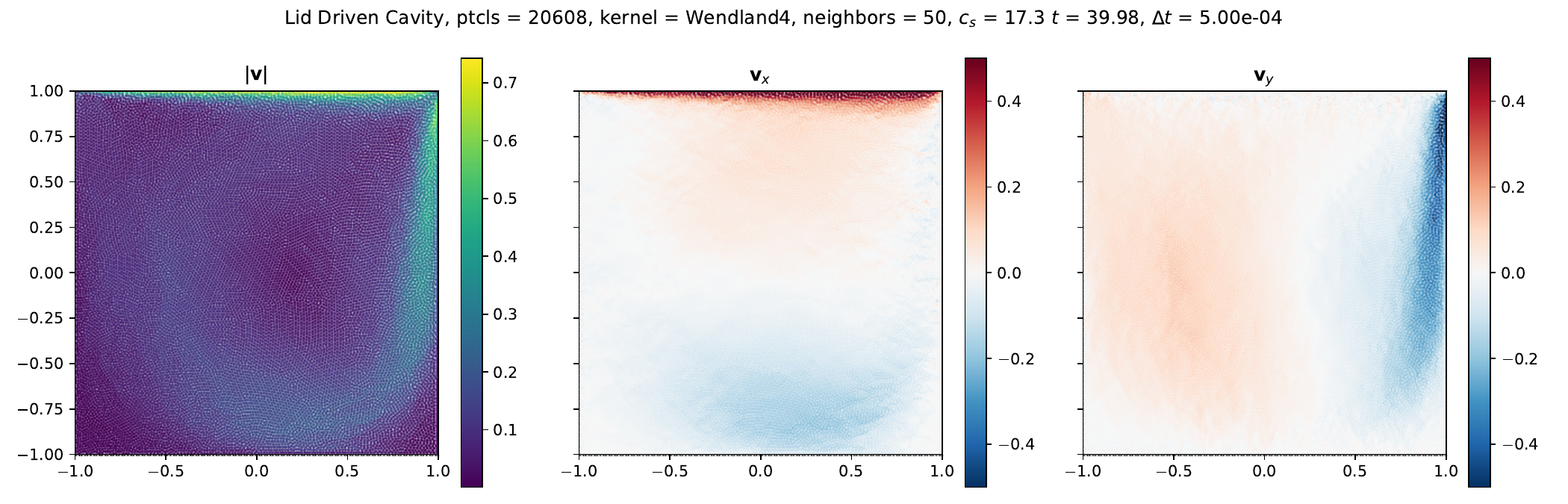}
    \caption{Result of the Lid-Driven Cavity simulation using $\delta$-SPH for $nx = 128$ fluid particles after $40$ seconds. From left to right, the particle velocity magnitude, velocity in x, and velocity in y-direction are visualized.}
    \label{fig:LDC}
\end{figure}
To validate the implementation of the boundary handling scheme for weakly compressible SPH, we utilize the classic Lid Driven Cavity simulation, see Fig.~\ref{fig:LDC},
In this case, a fully closed domain $[-1,1]^2$ is perturbed by a driven lid with a constant velocity $\mathbf{v}=(1,0)$, which creates a vortex in the fluid domain.
This case can be particularly challenging for SPH schemes due to the fully closed domain and the orthogonal drive into the non-driven walls.
However, in our framework, we do not observe notable numerical artifacts and find fluid behavior consistent with prior work~\cite{band2018mls}.

%% file: A5-06-Droplet.tex
\subsection{Oscillating Droplet}
\label{appendix:validation:wcsph:Droplet}
\begin{figure}
    \centering
    \includegraphics[width=\linewidth]{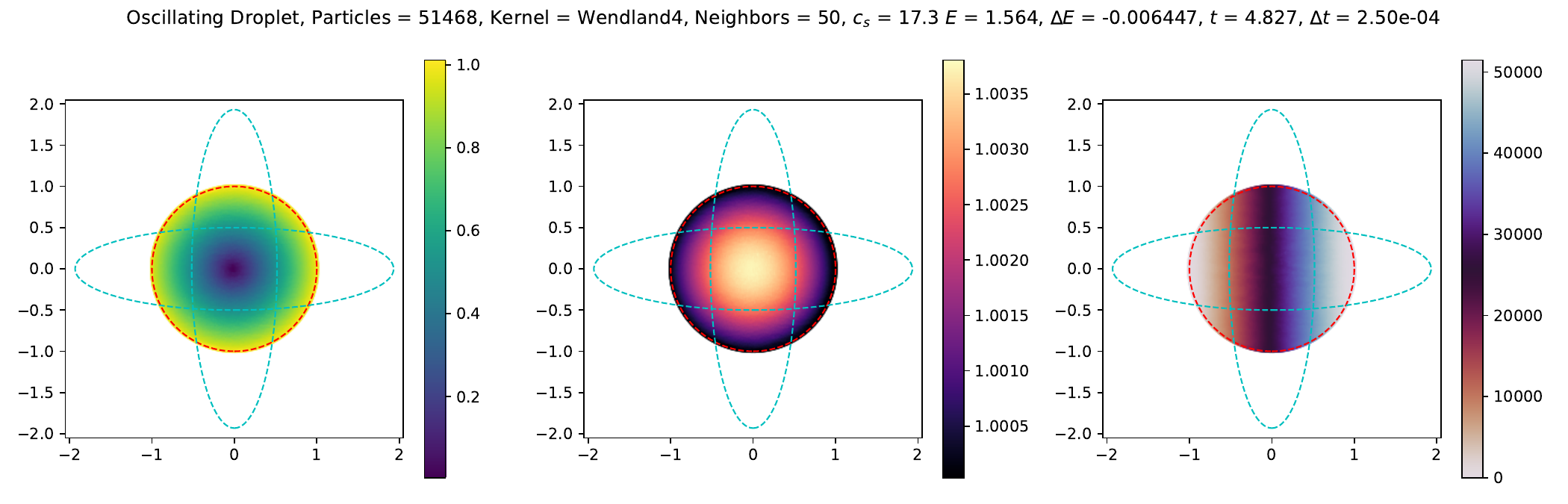}
    \caption{Visualization of the particle state after one period of oscillation for the Oscillating Droplet case. From left to right, we visualize the current particle velocity magnitude $|\mathbf{v}|$, the current particle density $\rho$, and the particle index to track particle motion over time.}
    \label{fig:DropletValidation}
\end{figure}
As the final validation case for our framework, we utilize the Oscillating Droplet case~\cite{sun2019consistent}.
In this case, an initial circle of particles with radius $R=1$ is subject to an initial velocity field
\begin{align}
    \mathbf{v}_x = A x,\\
    \mathbf{v}_y = -A y,
\end{align}
with $A$ being the initial velocity magnitude set to $A=1$ here.
The particles are then subjected to a potential field 
\begin{equation}
    f_\text{potential}(\mathbf{x}) = -B^2 \mathbf{x},
\end{equation}
with $B$ being the potential field magnitude set to $B=1$.
This results in the initially circular droplet oscillating into an ellipsoid with an aspect ratio of $1.931843:1$, with an oscillation period of $T=4.827A$.
This oscillation results in a transfer of potential to kinetic energy and back, where this case can be utilized to measure excess numerical dissipation.
Any issues in handling free surfaces or particle shifting near the free surface will also yield dissipation.

\begin{figure}
    \centering
    \includegraphics[width=0.8\linewidth]{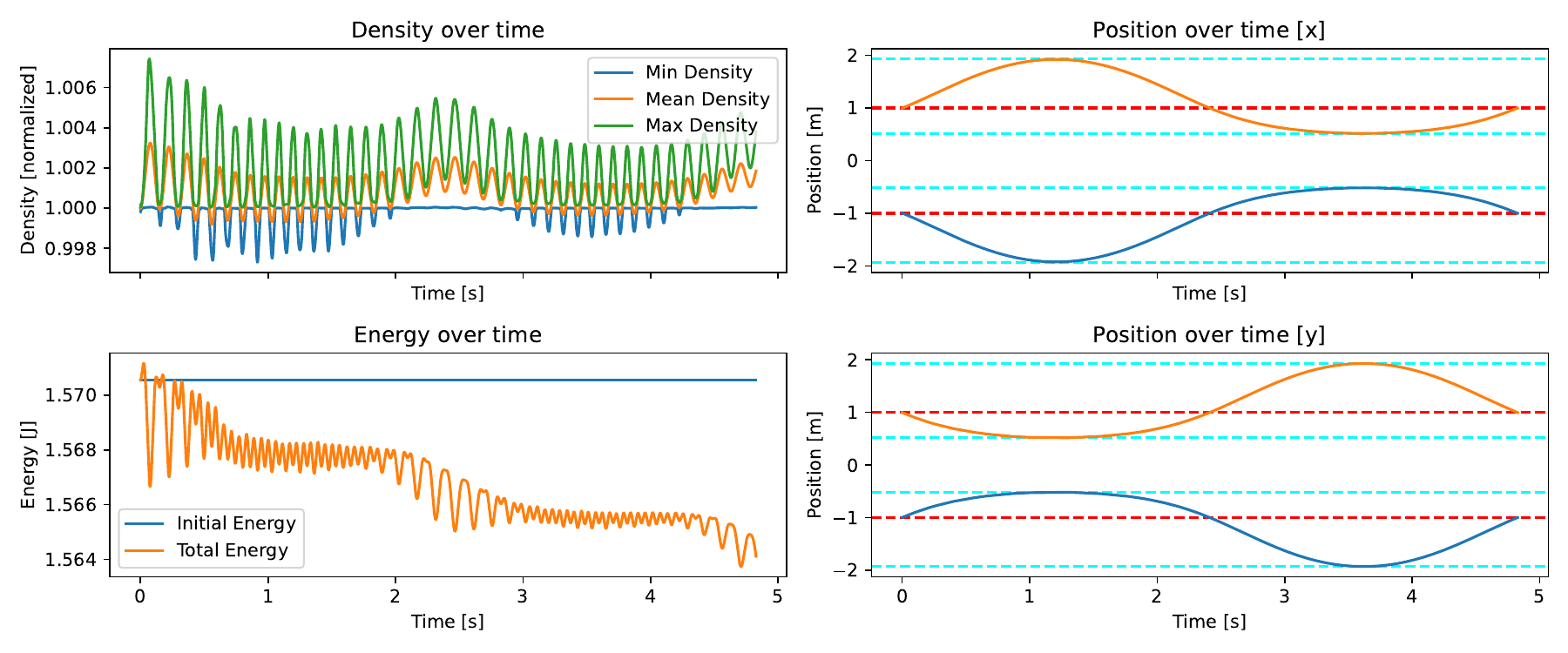}
    \caption{Density, Energy, and position evolution of the oscillating droplet for one period. Note that the high-frequency component of the density oscillation is based on the system's speed of sound. The horizontal lines for the position indicate the analytic solution for the minimal and maximal extent of the ellipsoid, as well as the initial radius of the sphere (red). The orange and blue lines denote the maximum and minimum particle positions on this axis.}
    \label{fig:DropletEnergy}
\end{figure}

Considering the qualitative results, see Fig.~\ref{fig:DropletValidation}, after one oscillation period, our method shows close agreement with the initial particle shape, indicating that the free-surface treatment does not lead to excessive surface distortions.
Considering the evolution of density over time, see Fig.~\ref{fig:DropletEnergy}, we observe an oscillatory effect based on the chosen speed of sound, i.e., the velocity magnitude of the fluid is approximately one in this case, with a speed of sound $c_s=17.5$, which results in acoustic waves that travel significantly faster than any fluid phenomena.
While undesired, these acoustic waves are an inherent property of weakly compressible SPH~\cite{SUN201725}.
We also observe close agreement between the maximum particle position in each axis and the shape of the ellipsoid over time, see Fig.~\ref{fig:DropletEnergy}.
Finally, the overall energy dissipation over one period of oscillation is approximately $0.4\%$, which agrees with the results of Sun et al.~\cite{SUN201725} for this resolution of particles.

